\documentclass[prd,aps,showpacs,twocolumn,superscriptaddress,floatfix,nofootinbib]{revtex4-1}
\usepackage[utf8]{inputenc}
\usepackage{bm}
\usepackage{bbold}
\usepackage{mathtools}
\usepackage[toc,page]{appendix}
\usepackage{dsfont}
\usepackage{combelow}
\usepackage[colorlinks=true,linkcolor=blue,citecolor=blue, urlcolor=blue]{hyperref} 
\usepackage[a4paper,top=2.4cm,bottom=2.4cm,right=1.0cm,bindingoffset=-1.2cm]{geometry}
\usepackage{tikz}
\usepackage{graphicx}
\usepackage{subfigure}
\usepackage{scalerel,amssymb}
\DeclareMathAlphabet{\pazocal}{OMS}{zplm}{m}{n}

\newcommand{\xT}{\mathbf{x_\perp}}
\newcommand{\yT}{\mathbf{y_\perp}}
\newcommand{\zT}{\mathbf{z_\perp}}
\newcommand{\rT}{\mathbf{r_\perp}}
\newcommand{\bT}{\mathbf{b_\perp}}
\newcommand{\dT}{\mathbf{d_\perp}}

\newcommand{\pT}{\mathbf{p_\perp}}

\newcommand{\kT}{\mathbf{k_\perp}}

\def\deltan{\hat\delta}

\begin{document}
\title{Rapidity dependence of initial state geometry and momentum correlations\\ in p+Pb collisions}
\author{B.~Schenke}
\affiliation{Physics Department, Brookhaven National Laboratory, Bldg.\,510A, Upton, NY 11973, USA}
\author{S.~Schlichting}
\affiliation{Fakultät für Physik, Universität Bielefeld, D-33615 Bielefeld, Germany}
\author{P.~Singh}
\email{prasingh@jyu.fi}
\affiliation{Fakultät für Physik, Universität Bielefeld, D-33615 Bielefeld, Germany}
\affiliation{Department of Physics, P.O. Box 35, 40014 University of Jyv\"askyl\"a, Finland}
\affiliation{Helsinki Institute of Physics, P.O. Box 64, FI-00014 University of Helsinki, Finland}


\begin{abstract}
Event geometry and initial state correlations have been invoked as possible explanations of
long range azimuthal correlations observed in high multiplicity p+p and p+Pb collisions. We study the rapidity dependence of initial state momentum correlations and event-by-event geometry in $\sqrt{s}=5.02~\rm{TeV}$ p+Pb collisions within the 3+1D IP-Glasma model~\cite{Schenke:2016ksl}, where the longitudinal structure is governed by JIMWLK rapidity evolution of the incoming nuclear gluon distributions. We find that the event geometry is correlated across large rapidity intervals whereas initial state momentum correlations are relatively short range in rapidity. Based on our results, we  discuss implications for the relevance of both effects in explaining the origin of collective phenomena in small systems.
\end{abstract}

\maketitle
\section{Introduction}
The collective behavior observed in heavy ion collisions has lead to the discovery of the Quark Gluon Plasma (QGP), and established the behavior of the QGP as a nearly perfect fluid. The main observables associated with this collectivity are the anisotropic flow coefficients $v_n$, which characterize the anisotropies in the transverse momentum distributions of produced particles. Experimental measurements of these coefficients can be described extremely well using relativistic hydrodynamic simulations of heavy ion collisions \cite{Teaney:2009qa,Gale:2013da,Luzum:2013yya,Heinz:2013th,Jeon:2015dfa}. Within the hydrodynamic picture, the final state momentum distributions are explained entirely via the response to the initial state geometry in the transverse (to the beam line) plane. Gradients of the pressure drive the directionally dependent expansion of the system, thus leaving an imprint of the initial shape of the fireball in the final particle spectra. 

More recently, similar signals to those in heavy in collisions have been found in the produced particle spectra of small collision systems, including p/d/$^3$He+A and even p+p \cite{Dusling:2015gta,Loizides:2016tew,Schlichting:2016sqo,Nagle:2018nvi,Schenke:2019pmk,Schenke:2021mxx}  and ultraperipheral Pb+Pb \cite{ATLAS:2021jhn} collisions. 
Such findings have lead to increased research regarding the question how hydrodynamics could possibly be applicable in very small systems that only produce on the order of ten charged hadrons per unit rapidity (see \cite{Florkowski:2017olj} for a review), as well as on the exploration of alternative mechanisms that could generate the observed anisotropies without requiring the creation of a nearly perfect fluid. Examples of the latter include kinetic theory~\cite{He:2015hfa,Greif:2017bnr,Romatschke:2018wgi,Bhaduri:2018iwr,Kersting:2018qvi,Roch:2020zdl,Kurkela:2018qeb,Kurkela:2019kip,Kurkela:2020wwb}, as well as the correlated (multi-)particle production in the color glass condensate framework \cite{Dumitru:2008wn,Kovner:2010xk,Dumitru:2010iy,Kovner:2011pe,Dusling:2012iga,Levin:2011fb,Dusling:2012wy,Dusling:2013qoz,Dumitru:2014dra,Dumitru:2014yza,Schenke:2015aqa,McLerran:2015sva,Schenke:2016lrs,Dusling:2017dqg,Dusling:2017aot,Mace:2018vwq,Mace:2018yvl,Kovner:2018fxj}, where anisotropic momentum distributions result from correlations in the gluon distributions of the incoming nuclei. 

While calculations involving final state effects (e.g. in the hydrodynamic framework) have been rather successful in describing the main features of the momentum anisotropies observed in small collision systems at RHIC and LHC \cite{Bozek:2011if,Bozek:2012gr,Bozek:2013df,Bozek:2013uha,Bozek:2013ska,Bzdak:2013zma,Qin:2013bha,Werner:2013ipa,Kozlov:2014fqa,Schenke:2014zha,Romatschke:2015gxa,Shen:2016zpp,Weller:2017tsr,Mantysaari:2017cni,Schenke:2020mbo}, purely initial state descriptions have so far struggled to fully reproduce quantitative and qualitative features of the data~\cite{Mace:2018yvl,Mace:2018vwq}.
Some have, potentially prematurely, ``ruled out''  initial-stage glasma correlations \cite{Nagle:2021jgy}, however, they should be present and can in principle affect observables, even when geometry driven final state effects dominate.

In the IP-Glasma+\textsc{Music}+UrQMD model \cite{Schenke:2020mbo}, both initial state anisotropies from the Glasma and final state response to the geometry are present. Their relative contributions have been analyzed as functions of multiplicity in \cite{Schenke:2019pmk}, and an observable that should be able to distinguish them as sources of the observed anisotropies, namely the correlation of the elliptic anisotropy with the mean transverse momentum, was analyzed in \cite{Giacalone:2020byk}. The results in these works indicate that while the initial state anisotropy has a non-negligible contribution over a wide range of multiplicities, it starts to be the dominant contribution only for $dN_{\rm ch}/d\eta \lesssim 5-10$, approximately independent of the collision system or energy.

So far, many calculations for proton-nucleus collisions, including the aforementioned ones, were performed under the assumption of boost invariance, which means that correlations of both the transverse geometry and the initial momentum anisotropy extend over arbitrarily large separations in rapidity. In this work we relax the assumption of boost invariance and set out to explore the longitudinal dependence of both the initial state geometry and initial state momentum space correlations. This will provide important input to experimentally distinguish the two types of signals from each other and from short range ``non-flow'' contributions that result e.g. from  mini-jets or resonance decays. 

This paper is organized as follows. We start with a brief description of the 3D IP-Glasma model in Sec.~\ref{sec:two} and subsequently discuss some global event properties in $5.02~{\rm TeV}$ p+Pb collisions in Sec.~\ref{sec:three}. Our main results regarding the longitudinal dependence of the initial state geometry and initial state momentum space correlations are presented in Sec.~\ref{sec:four}. We conclude and present an outlook in Sec.~\ref{sec:five}.


\section{The 3D IP-Glasma model}
\label{sec:two}
We follow the description of~\cite{Schenke:2016ksl}, which is built on the high-energy factorization of the expectation values of sufficiently inclusive quantities~\cite{Gelis:2008rw,Gelis:2008ad}. Based on the Color Glass Condensate effective field theory of high-energy QCD~\cite{Iancu:2003xm},  observables $O(y_{\rm obs})$ at a rapidity $y_{\rm obs}$ can be calculated on an event-by-event basis
\begin{eqnarray}
\label{eq:ObsCl}
O(y_{\rm obs})=O_{\rm cl}\Big(V_{\xT}^{p}(+y_{\rm obs}),V^{Pb}_{\xT}(-y_{\rm obs})\Big)\;,
\end{eqnarray}
as a functional of the light-like Wilson lines $V_{\xT}^{p}(+y_{\rm obs})$ and $V^{Pb}_{\xT}(-y_{\rm obs})$ of the projectile (p) and target (Pb), by solving the classical Yang-Mills (CYM) equations. Starting from initial conditions $V_{\xT}^{p/Pb}(-Y_{\rm max})$ determined by the IP-Glasma model~\cite{Schenke:2012wb,Schenke:2012hg} at the maximal observed rapidity $Y_{\rm max}$, the rapidity evolution of the light-like Wilson lines $V_{\xT}^{p/Pb}(Y)$ is calculated by the JIMWLK evolution equation~\cite{Jalilian-Marian:1996mkd,Jalilian-Marian:1997jx,Jalilian-Marian:1997gr,Iancu:2000hn,Iancu:2001ad}. Based on Eq.~\eqref{eq:ObsCl}, the observables at each rapidity are computed from the solutions to the classical field equations, while the longitudinal (rapidity) structure is governed by the small-$x$ evolution of the Wilson lines. While such factorization, as in Eq.~\eqref{eq:ObsCl}, has been proven only for inclusive quantities which encompass measurements at a single rapidity~\cite{Gelis:2008rw,Gelis:2008ad}, we will use the same prescription to calculate un-equal rapidity correlations on an event-by-event basis. We refer to~\cite{Schenke:2016ksl} for additional discussions of the associated caveats, and provide details of the implementation of the 3D-Glasma model below.

\subsection{IP-Glasma initial condition}
Within the CGC the small-x gluon fields of the incoming nuclei are generated by the moving valence charges according to the Yang-Mills equations 
\begin{equation}\label{eq:YM} 
    [D_\mu,F^{\mu\nu}] = J^\nu\,,
\end{equation}
where $D_\mu=\partial_\mu - igA_\mu$ is the gauge covariant derivative, and $F^{\mu\nu} = \frac{i}{g}[D^\mu,D^\nu] = \partial^\mu A^\nu - \partial^\nu A^\mu - ig[A^\mu,A^\nu]$ is the field strength tensor, with the gluon fields $A^\mu=A^\mu_a t^a$. The $t^a$ are the generators of $SU(N_c)$ (for the number of colors $N_c=3$) in the fundamental representation. The index $a$ is the color index, and runs from 1 to $(N_c^2-1)=8$.
The eikonal currents $J^{\nu}$ on the right hand side of Eq.~\eqref{eq:YM} are given by the sum of the color currents of the two nuclei (the moving large $x$ degrees of freedom)
\begin{equation}
    J^\nu = \delta^{\nu +} \rho_{Pb}({\mathbf x}_\perp) \delta(x^-) + \delta^{\nu -} \rho_{p} ({\mathbf x}_\perp) \delta(x^+)\,.
\end{equation} 
We will use the IP-Glasma model to determine the color charge densities $\rho_{p/Pb}({\mathbf x}_\perp)$ and associated Wilson lines at the initial rapidities (the largest $x$ values). Wilson lines at smaller $x$ then follow from JIMWLK evolution, as discussed in the next subsection.

In IP-Glasma the color charges $\rho_{Pb}({\mathbf x}_\perp)$ and $\rho_p({\mathbf x}_\perp)$ are sampled on an event-by-event basis, assuming local Gaussian correlations as in the McLerran-Venugopalan (MV) model \cite{McLerran:1993ka,McLerran:1994vd}. In practice, one determines the Wilson lines $V_{\xT}$ for each nucleus numerically, approximating the path ordered exponential by the product \cite{Lappi:2007ku}
\begin{equation}
    V_{\xT}^{Pb/p} = \prod_{k=1}^{N_y} \exp \Big( -ig \frac{\rho^k_{Pb/p}(\xT)}{\boldsymbol{\nabla}^2-\tilde{m}^2}\Big)\,,
\end{equation}
where, $\tilde{m}=0.2$~GeV (or $0.8~{\rm GeV}$ as indicated) is an infrared regulator that is used to avoid unphysical Coulomb tails, $N_y=50$ is the number of slices in the longitudinal direction, and, as in the MV model,
the $\rho^k_{Pb}$ and $\rho^k_{p}$ have zero mean and their two-point functions satisfy (suppressing the subscripts $Pb$ and $p$ for clarity)
\begin{equation}
\langle \rho_{i}^a(\bT) \rho_{j}^b(\xT) \rangle = \frac{g^2 \mu^2 (x,\bT)}{N_y} \delta^{ab}\delta^{ij} \delta^{(2)}(\bT-\xT).
\end{equation}
Spatially $(\bT)$ dependent color charge densities, 
$g^2\mu_{Pb/p}(x,\bT)= c_{Q_s}\,Q_s^{Pb/p}(x,T(\bT))$, \footnote{We employ $c_{Q_s}=1.25$ for $\tilde{m}=0.2 {\rm GeV}$  and $c_{Q_s}=1.82$ for $\tilde{m}=0.8 {\rm GeV}$.} are determined using the IPSat model \cite{Bartels:2002cj,Kowalski:2003hm}, which provides the saturation scale $Q_s(x,T(\bT))$ as a function of the nuclear thickness $T(\bT)$ at a given Bjorken $x$.  The nuclear thickness functions $T(\bT)$, which provide the $\bT$ dependence, are determined as in~\cite{Schenke:2020mbo} by sampling the position of individual nucleons from a Woods-Saxon distribution in the case of the Pb nucleus. Subsequently, the position of $N_q=3$ hot spots per nucleon are assigned according to a two-dimensional Gaussian distribution with width $B_p$, and each hot spot is assigned a two-dimensional Gaussian thickness profile of width $B_q$. The parameters $B_{p}=4~{\rm GeV}^{-2}$ and $B_{q}=0.3~{\rm GeV}^{-2}$ of the model are constrained using deeply inelastic scattering data on protons from HERA \cite{Rezaeian:2012ji}. Once the nuclear thickness $T(\bT)$ is determined, we self-consistently determine $Q_s(x,\bT)$ by iteratively solving for
\begin{equation}
    x=x(\bT)=\frac{Q_s(x,T(\bT))}{\sqrt{s_{\rm NN}}} e^{-Y}\,,
\end{equation}
where $\sqrt{s_{\rm NN}}$ is the center of mass energy of the collision. We note that the public IP-Glasma code employed in this study can be found at \cite{ipglasma}, and we refer to \cite{Schenke:2020mbo} for a detailed description of the implementation used in this work.

\begin{figure*}
    \centering
    \includegraphics[width=0.95\textwidth]{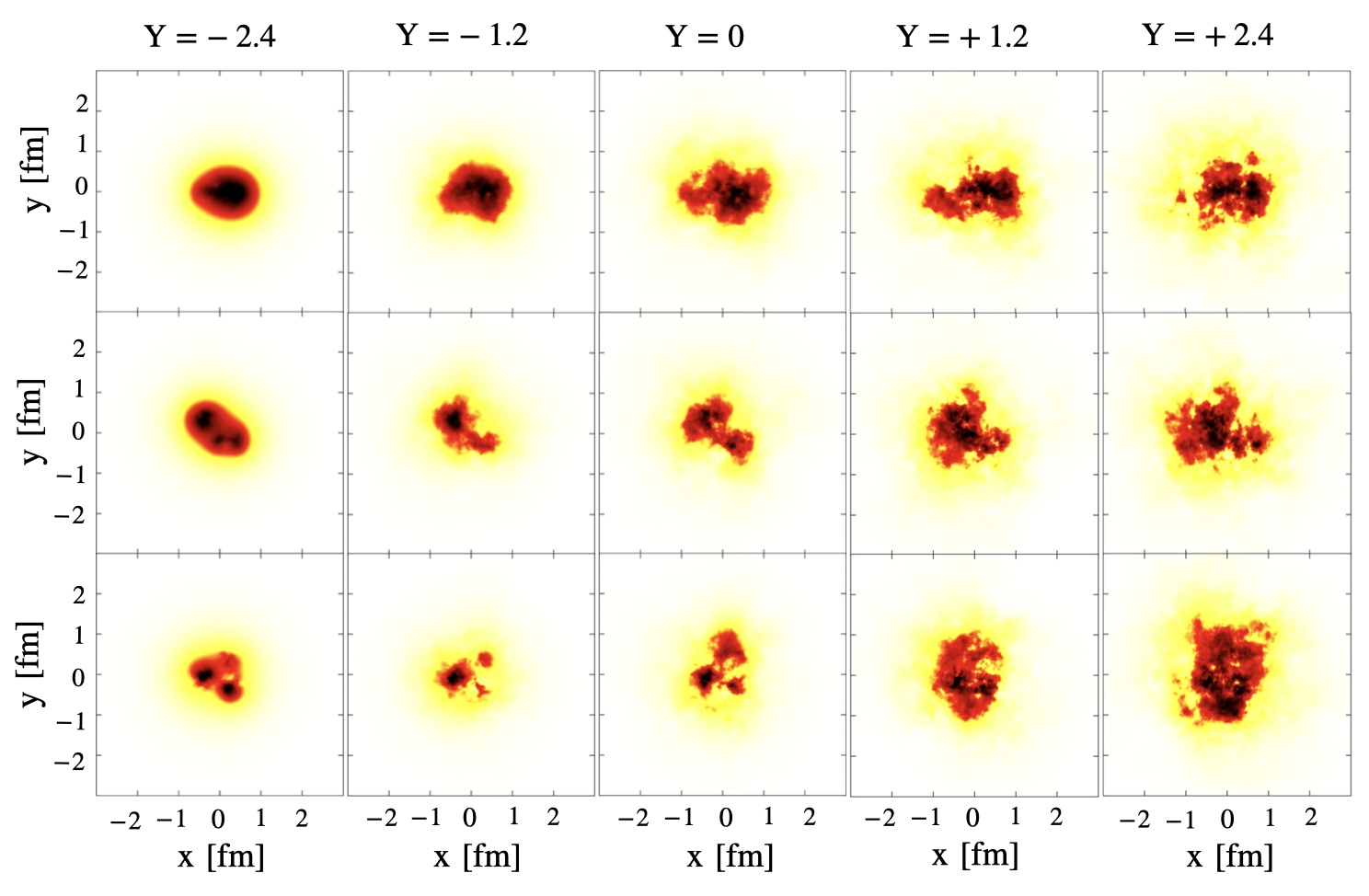}
    \caption{JIMWLK evolution of the gluon fields in three different configurations of the proton for $m=0.2$ GeV and $\alpha_s=0.3$. The trace of Wilson lines
    $1-\rm{Re}[tr(V_{x_\perp})]/N_c$ is shown in the transverse plane for different rapidities $(Y)$ to illustrate the emergence of finer structure and growth of the proton with increasing rapidity.}
    \label{fig:proton_config}
\end{figure*}

\begin{figure*}
    \centering
    \includegraphics[width=0.9\textwidth]{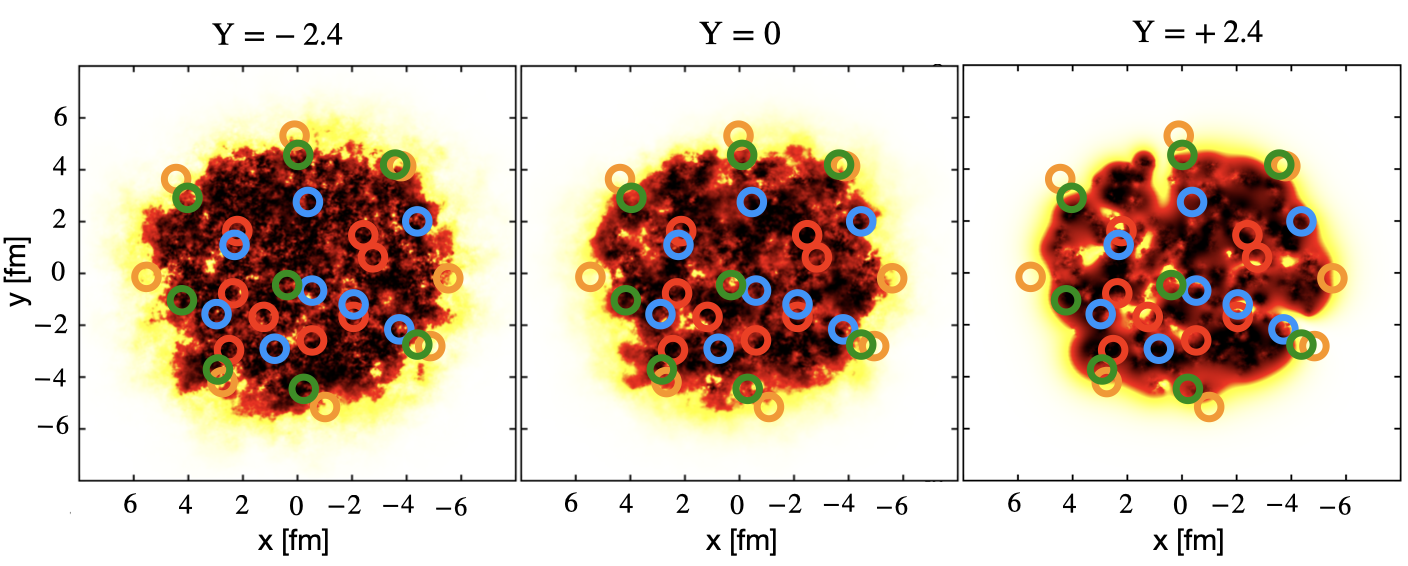}
    \caption{View of the transverse plane for a particular configuration of a right moving lead nucleus at three different rapidities. Circles indicate the collision point of the proton with this lead nucleus for a selection of events. The color coding indicates different centrality classes: red $(0-5)\%$, blue $(40-50)\%$, green $(60-70)\%$ and orange $(80-90)\%$.}
    \label{fig:pOnPb}
\end{figure*}
Based on the above procedure, we generate a total of $N_{p}=32$ and $N_{Pb}=8$ configurations of the Wilson lines $V_{\xT}^{p/Pb}(-Y_{\rm max})$ of the protons and lead nuclei at the largest $x$ value, corresponding to the initial rapidity $Y=-Y_{\rm max}=-2.4$, with transverse coordinates ($\xT$) discretized on a $N_{s}\times N_{s}$ lattice with $N_{s}=1024$ sites and lattice spacing $a_{s}=0.02~{\rm fm}$.

\subsection{JIMWLK evolution}
Starting from the IP-Glasma initial conditions for the Wilson lines $V_{\xT}^{p/Pb}(-Y_{\rm max})$, we perform the JIMWLK \cite{Jalilian-Marian:1996mkd,Jalilian-Marian:1997jx,Jalilian-Marian:1997gr,Iancu:2000hn,Iancu:2001ad}  evolution from $Y=-2.4$ to $Y=+2.4$ for each configuration of the proton and the lead nucleus. We store the configurations for various slices in rapidity, in steps of $Y=0.2$.

The implementation of the JIMWLK solver is equal to that discussed in \cite{Schenke:2016ksl}. Specifically, we express the JIMWLK hierarchy in terms of a functional Langevin equation for the Wilson lines \cite{Weigert:2000gi,Blaizot:2002np}. Each Langevin step can be written as \cite{Lappi:2012vw}
\begin{align}\label{eq:newstep}
  V_{\xT}&(Y+dY) = \notag\\
  &\exp \Big\{-i \frac{\sqrt{\alpha_s dY}}{\pi} \int_{\zT} K_{\xT-\zT}\cdot \left(V_{\zT}\boldsymbol{\xi}_{\zT}
    V^\dag_{\zT}\right)\Big\}\notag\\
    &\times V_{\xT}(Y) \exp \Big\{ i \frac{\sqrt{\alpha_s dY}}{\pi} \int_{\zT} K_{\xT-\zT}\cdot \boldsymbol{\xi}_{\zT}\Big\}\,,
\end{align}
with Gaussian white noise $\boldsymbol{\xi}_{\zT} = (\xi_{\zT,1}^{a} t^a,\xi_{\zT,2}^{a} t^a)$ that is local in transverse coordinate, color, and rapidity, i.e., $\langle \xi_{\zT, i}^b(Y)\rangle = 0$ and
\begin{equation}\label{eq:noise}
  \langle \xi_{\xT, i}^a(Y) \xi_{\mathbf{y}_\perp, j}^b(Y')\rangle = \delta^{ab}\delta^{ij}\delta_{ \xT\mathbf{y}_\perp }^{(2)} \delta(Y-Y')\,.
\end{equation}
Since we are particularly interested in the impact parameter dependence, we
follow \cite{Schlichting:2014ipa} and employ a regularized JIMWLK kernel
\begin{equation}\label{eq:modKernel}
  K_{\xT-\zT} = m|\xT-\zT|~K_{1}(m |\xT-\zT|)~\frac{\xT-\zT}{(\xT-\zT)^2}\,,
\end{equation}
which suppresses emission at large distance scales and limits growth in impact parameter space.
The modified Bessel function of the second kind $K_{1}(x)$ behaves as $x K_{1}(x)=1+\mathcal{O}(x^2)$ for small arguments $x$, leaving the kernel  unmodified at short distance scales. Conversely, for large arguments $K_{1}(x)=\sqrt{\frac{\pi}{2x}} e^{-x}$ decays exponentially, suppressing gluon emissions at large distance scales. This regularization also prevents the unphysical exponential growth of the cross section, which would violate unitarity \cite{Kovner:2001bh}.

We note that the only free parameters controlling the JIMWLK evolution in Eq.~\eqref{eq:newstep} are the (fixed) coupling constant $\alpha_s$ and the infrared regulator $m$, and we will consider variations of both parameters to assess the sensitivity of our results.

We illustrate the JIMWLK evolution of the spatial configuration of three sample protons in Fig.\,\ref{fig:proton_config}, where we plot the trace of the Wilson lines, $1-\rm{Re}[tr(V_{x_\perp})]/N_c$, for five different rapidities. Going left to right, $x$ decreases for the left moving proton. One can see that the average size of the proton grows with the evolution and that shorter scale structures emerge as $Q_s$ grows with decreasing $x$. This is expected as the correlation length in the transverse plane behaves as $\sim 1/Q_s$. Similar features can be observed for the evolution of the lead nuclei, shown in Fig.~\ref{fig:pOnPb}, where for the right moving nucleus, $x$ decreases going from right to left; in addition, the impact parameters of the protons, for events within a given centrality class (see Sec.~\ref{sec:centrality}), are marked by different colored circles.

\subsection{Event generation \& classical Yang-Mills evolution}
Having determined $N_p$ proton configurations and $N_{Pb}$ lead configurations over the entire range of rapidities $-2.4 \leq Y \leq 2.4$, we proceed to generate events, where
for each of the $N_{p}\times N_{Pb}$ combinations of protons and lead nuclei, we perform $N_{\bT}=16$ collisions with different impact parameters $\bT$, sampled according to a two-dimensional uniform distribution with the restriction $0<|\bT|<8~{\rm fm}$.\footnote{Note that in order to avoid interpolation of $SU(N_c)$ matrices, we round the impact parameter $\bT$ to the next lattice site.}

Based on the JIMWLK evolved Wilson lines, the initial conditions for the non-vanishing components of the gauge fields $A^{i}_{\xT}(\tau=0^{+}),E^{\eta}_{\xT}(\tau=0^{+})$ in the forward light-cone at a given rapidity $y_{\rm obs}$ are then given by
\begin{eqnarray}
&&A^{i}_{\xT}(\tau=0^{+},y_{\rm obs})=\frac{i}{g} \left[  \Big(V_{\xT}^{p}(+y_{\rm obs})\partial^i V_{\xT}^{p~\dagger}(+y_{\rm obs})  \Big) \right.  \nonumber \\
&& \qquad \left. + \Big(V_{\xT+\bT}^{Pb}(-y_{\rm obs})\partial^i V_{\xT+\bT}^{Pb~\dagger}(-y_{\rm obs})  \Big) \right]\;, \label{eq:ini1} \\
&&E^{\eta}_{\xT}(\tau=0^{+},y_{\rm obs})=\frac{i}{g} \left[  \Big(V_{\xT}^{p}(+y_{\rm obs})\partial^i V_{\xT}^{p~\dagger}(+y_{\rm obs})  \Big), \right. \nonumber \\ 
&& \qquad \left. \Big(V_{\xT+\bT}^{Pb}(-y_{\rm obs})\partial^i V_{\xT+\bT}^{Pb~\dagger}(-y_{\rm obs})  \Big) \right]\label{eq:ini2}\;.   
\end{eqnarray}
Starting from the lattice discretized version of the initial conditions in Eqs.~\eqref{eq:ini1} and \eqref{eq:ini2}~\cite{Krasnitz:1998ns}, we solve the lattice discretized classical Yang-Mills (CYM) equations of motion up to time $\tau=0.2\,{\rm fm}/c$, at which we determine the energy-momentum tensor 
$T^{\mu\nu}$, gluon spectra $\frac{dN_{g}}{d^2\pT dy}$ and gluon multiplicity $ dN_{g}/dy=\int d^2\pT \frac{dN_{g}}{d^2\pT dy}$ as described in~\cite{Schenke:2015aqa}. \begin{figure}[t!]
    \includegraphics[width=0.48\textwidth]{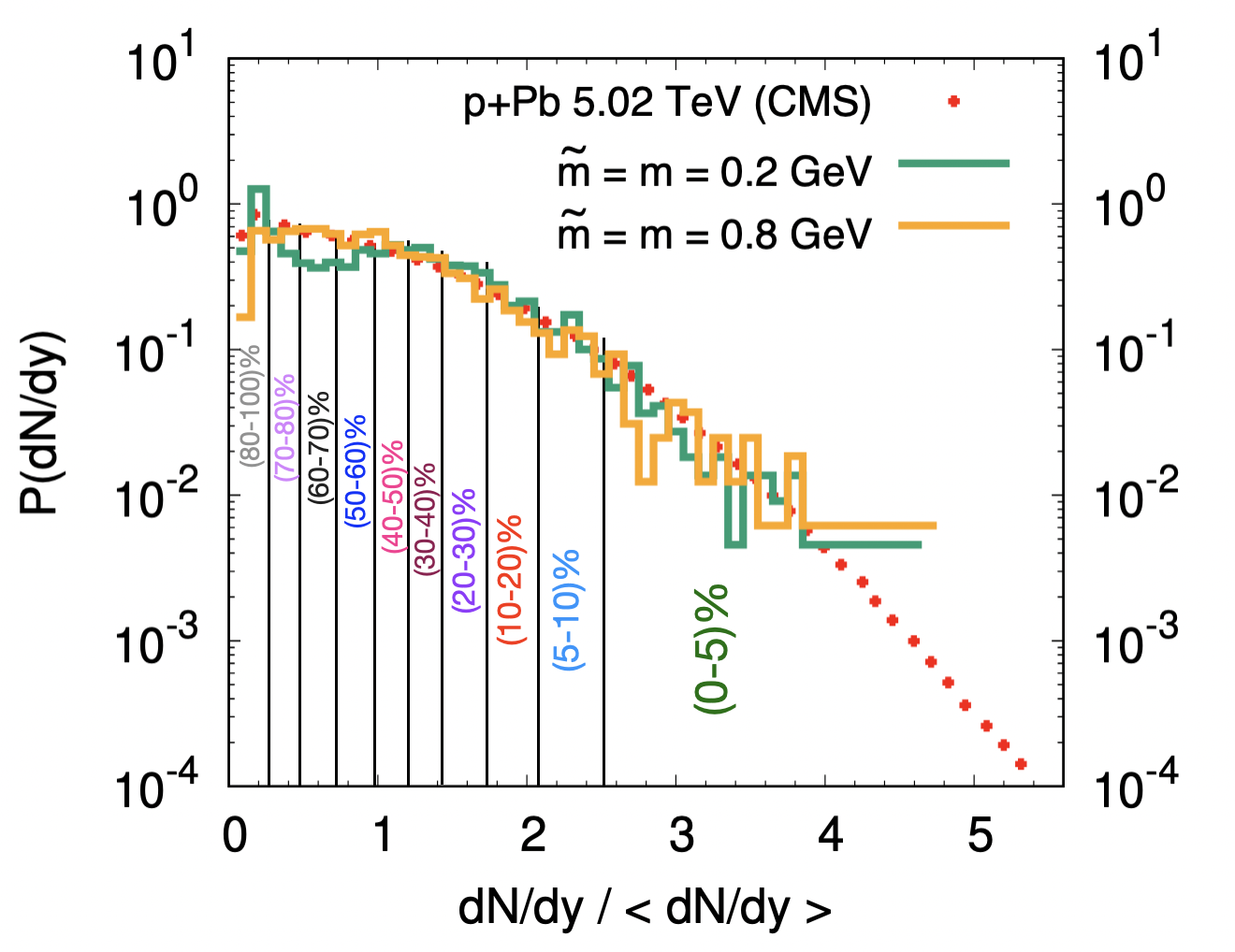}
    \caption{Histogram for the gluon multiplicity distribution $g^2 dN_g/dy$ normalized by its expectation value $\langle g^2 dN_g/dy \rangle$ at mid rapidity (y=0). Simulations are done for two different set of parameters $m=\tilde{m}=0.2~{\rm GeV}$ with $c_{Q_s}=1.25$ and $m=\tilde{m}=0.8~{\rm GeV}$ with $c_{Q_s}=1.82$ along with $\alpha_s=0.15$ 
    Crosses are experimental charged hadron distribution data $dN_{ch}/dy$ for raw reconstructed primary tracks in $\sqrt{s}=5.02~{\rm TeV}$ p+Pb collisions from the CMS collaboration \cite{CMS:2012qk} Centrality classes are indicated by the vertical lines.}
    \label{fig:MD_a015_m02}%
\end{figure}

\begin{figure*}[t!]
\centering
\includegraphics[width=0.45\textwidth]{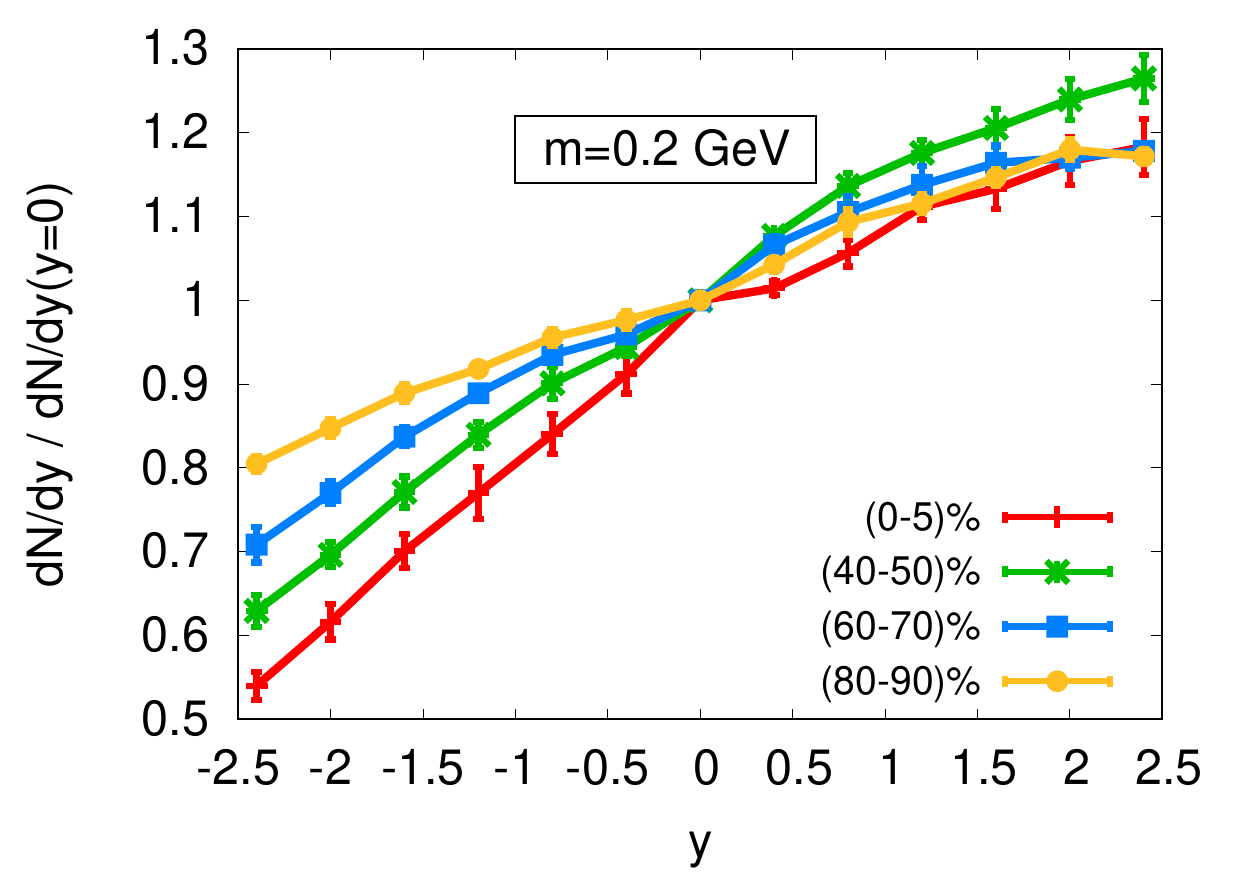}\hspace{0.6em}
\includegraphics[width=0.45\textwidth]{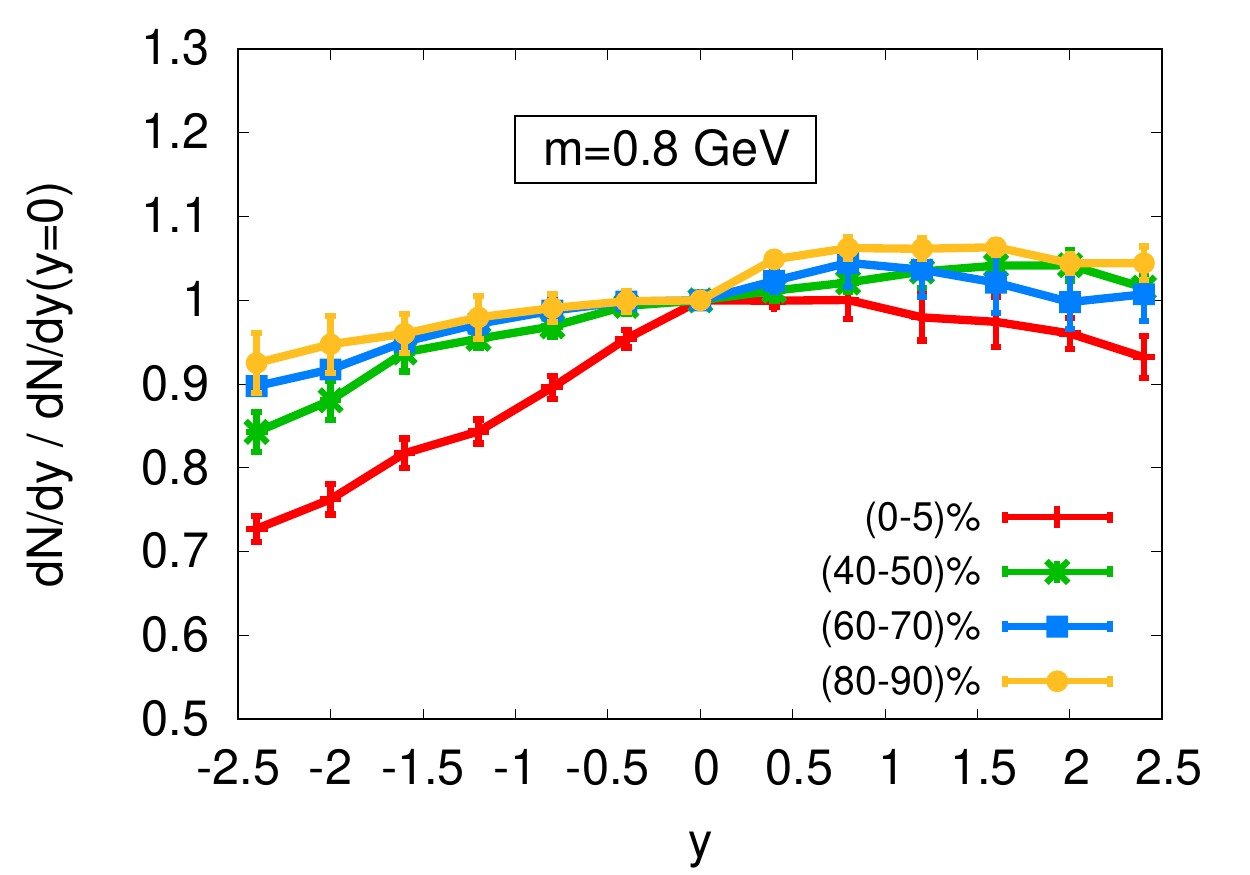}\hspace{0.6em}
\includegraphics[width=0.45\textwidth]{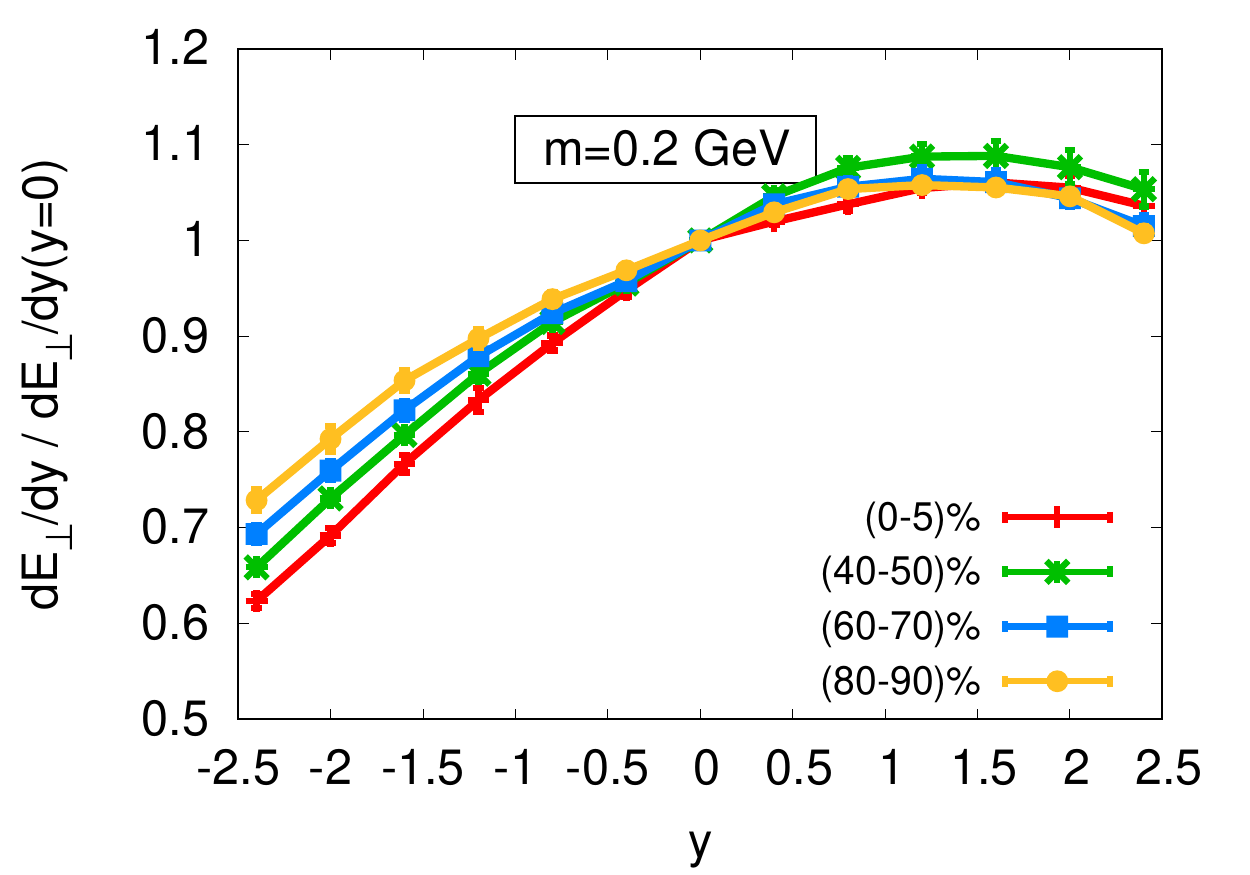}\hspace{0.6em}
\includegraphics[width=0.45\textwidth]{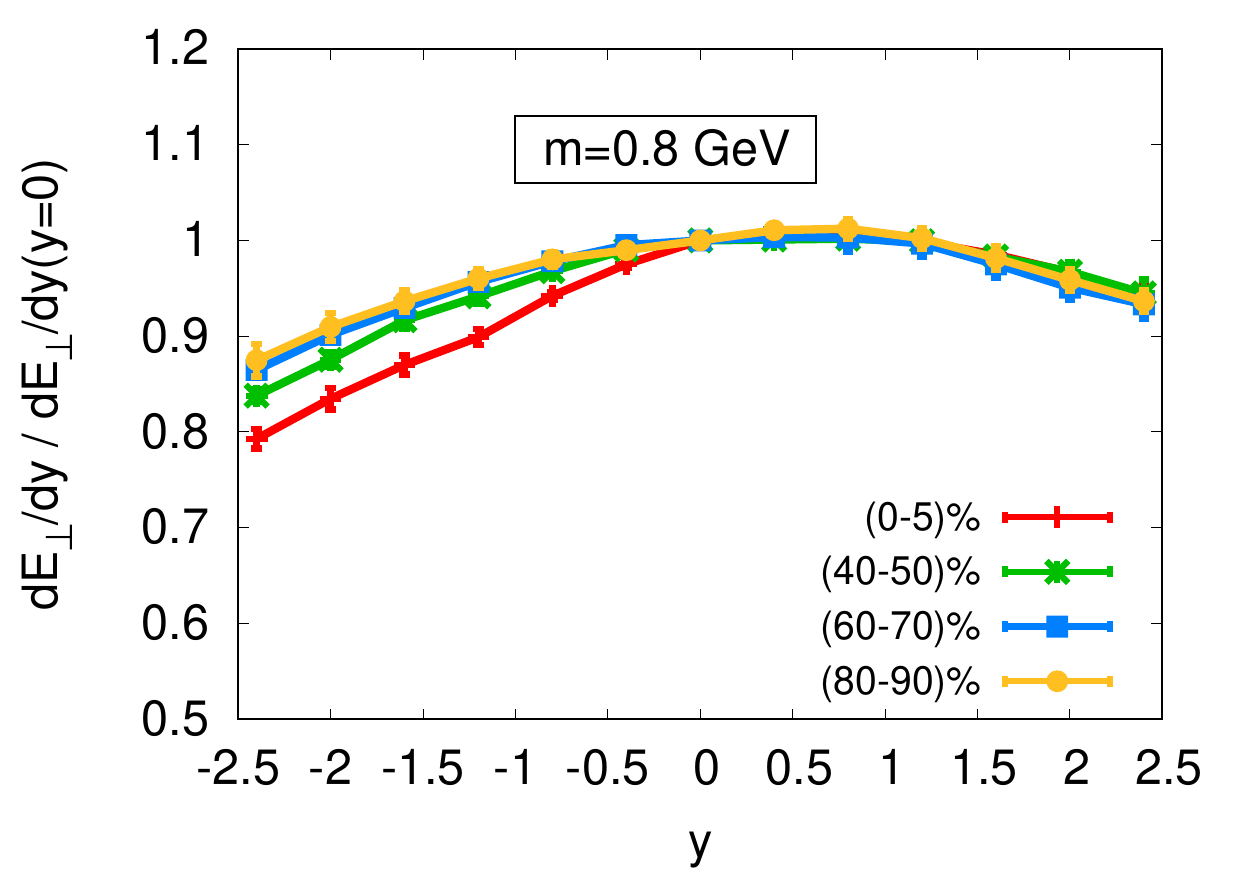}\hspace{0.6em}
\caption{Gluon multiplicity $dN_g/dy$ (top) and transverse energy per unit rapidity $dE_\perp/dy \propto \tau \varepsilon\,(\tau=0.2\,{\rm fm})$ (bottom) relative to their values at mid rapidity for different centrality classes. Simulation parameters: $\alpha_s=0.15$; $\tilde{m} = m=0.2$~GeV (left) and $\tilde{m} = m=0.8$~GeV (right).}
\label{fig:dNbydY}
\end{figure*}

Based on the factorization formula in Eq.\,(\ref{eq:ObsCl}), the rapidity $y_{\rm obs}$ dependence of these observables in each event is then calculated as in~\cite{Schenke:2016ksl} from a series of independent 2+1D CYM simulations, which according to Eqs.~(\ref{eq:ini1},\ref{eq:ini2}) start from the same Wilson lines $V^{p}_{\xT}$ and $V^{Pb}_{\xT}$ evolved up to different rapidities $Y=\pm y_{\rm obs}$. We will consider a rapidity range $y_{\rm obs} \in [-2.4,+2.4]$, where $y_{\rm obs}=-2.4 (+2.4)$ corresponds to no JIMWLK evolution in the proton (lead nucleus), and calculate observables in intervals of $\Delta y=0.4$.

\subsection{Gluon multiplicity and centrality selection}\label{sec:centrality}
Based on the above procedure, we obtain a total of $N_{\rm events}=N_{\bT} \times N_{p}\times N_{Pb}=4096$ events, which we further classify into centrality classes according to their gluon multiplicity $g^2 dN_{g}/dy|_{y_{\rm obs}=0}$ at mid-rapidity $y_{\rm obs}=0$. Since we do not invoke any collision criteria (e.g.~$N_{\rm coll} \geq 1$), we first disregard events with $g^2dN_{g}/dy|_{y_{\rm obs}=0} < 4$ from our event selection and subsequently perform the usual binning. 

We present the gluon multiplicity distribution at mid-rapidity ($y=0$) in Fig.\,\ref{fig:MD_a015_m02}, where we have scaled the distribution by the mean multiplicity, in order to compare to experimental data on the uncorrected reconstructed primary tracks from the CMS Collaboration \cite{CMS:2012qk}. Different curves in Fig.\,\ref{fig:MD_a015_m02}, show the results for two different sets of parameters, namely $m=\tilde{m}=0.2\,{\rm GeV}$ with $c_{Q_s}=1.25$ and $m=\tilde{m}=0.8\,{\rm GeV}$ with $c_{Q_s}=1.82$, which we will continue to investigate in the following. While in both cases the width of the gluon multiplicity distribution agrees well with that of the experimental data on reconstructed tracks, we find that for $m=\tilde{m}=0.2\,{\rm GeV}$ the computed gluon distribution has some peak and dip structure at small multiplicities, which can be attributed to very peripheral events and is not seen in the experimental data. Nevertheless, even in this case, for larger multiplicities (equal or greater than the mean) the data is well described. The figure also indicates the centrality classes as obtained from the gluon distribution.

\section{Global event structure \& nature of high-multiplicity events}
\label{sec:three}
Before we discuss the event-by-event geometry and azimuthal correlations, it is insightful to briefly comment on the general features of low and high multiplicity events in high-energy p+Pb collisions. We first study the rapidity dependence of the multiplicity $dN_{g}/dy$ and transverse energy $dE_{\bot}/dy$.\footnote{We assume that space-time rapidity is equal to the momentum rapidity, $\eta_s=y$, which holds for a system, where the phase-space density of gluons is proportional to $\delta (\eta_s-y)$ \cite{Greif:2017bnr}. Based on this, we will be using $\eta_s$ and $y$ interchangeably.} Different panels in Fig.~\ref{fig:dNbydY} show the rapidity dependence of $dN_{g}/dy$ and $dE_{\bot}/dy$ normalized to their value at mid-rapidity $dN_{g}/dy|_{y=0}$ for different centrality classes $(0-5),(40-50),(60-70)$, and $(80-90)\%$, for the two different sets of parameters $m=\tilde{m}=0.2~{\rm GeV}$ with $c_{Q_s}=1.25$ and $m=\tilde{m}=0.8~{\rm GeV}$ with $c_{Q_s}=1.82$. Absolute values of the multiplicities and transverse energy per unit rapidity at mid-rapidity are provided in Table\,~\ref{tab:EventStatistics}.

\begin{figure}[t!]
    \centering
    \includegraphics[width=0.45\textwidth]{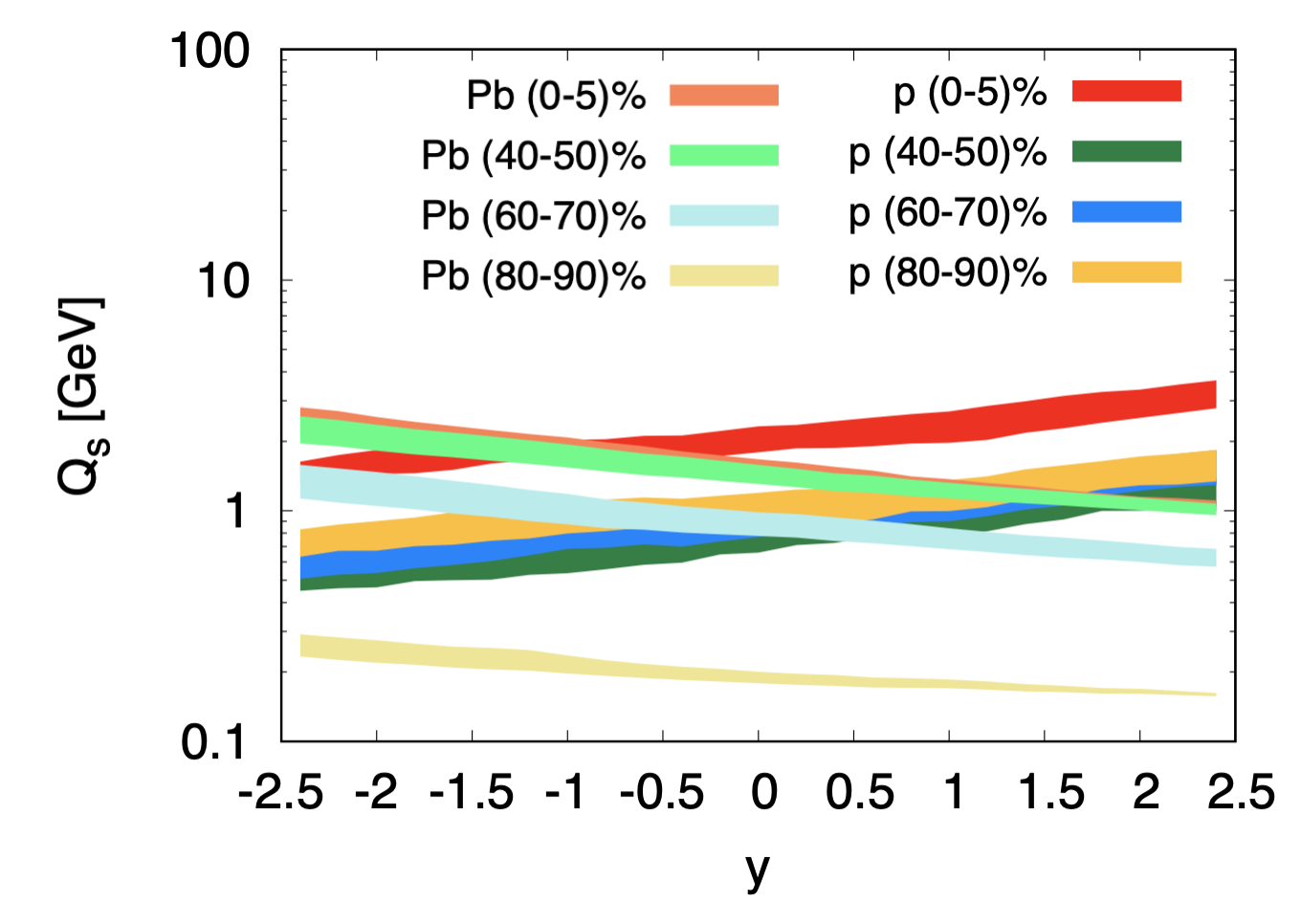}
    \includegraphics[width=0.45\textwidth]{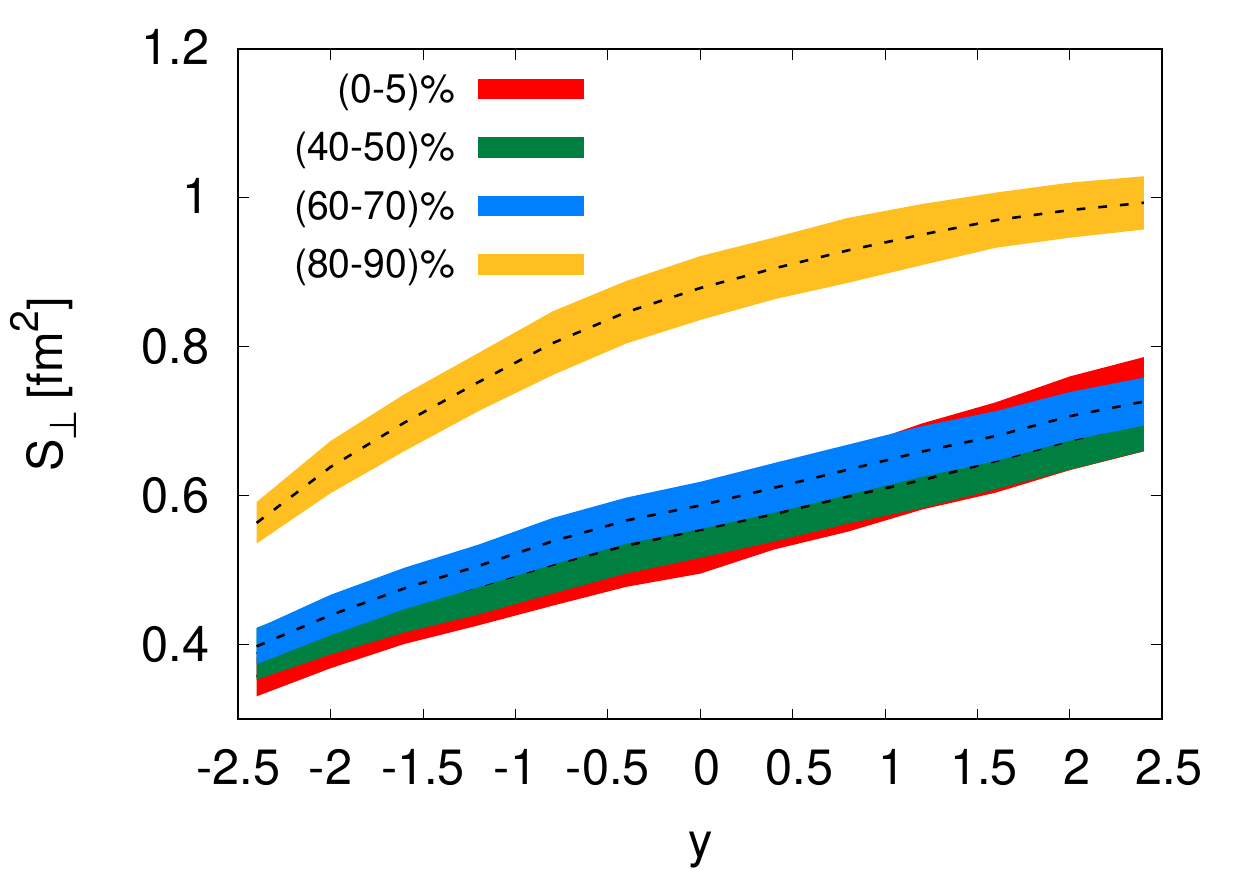}
    \caption{Top: Saturation scale $Q_{s}(y)$ as a function of rapidity $y$ for proton (p) and lead nucleus (Pb) for different centrality classes. Bottom: Rapidity dependence of the system size $S_{\bot}$ 
    for various centrality classes. Results are for $\alpha_s=0.15$ and $m=0.2$~GeV}
    \label{fig:QsST}
\end{figure}

\begin{table}[t]
\centering
\begin{tabular}{|l|l|l|l|l|}
\hline
${\bf{g^2dN_{g}/dy}}$      & $\bf{0-5\%}$  &$\bf{40-50\%}$  &$\bf{60-70\%}$ &$\bf{80-90\%}$ \\ \hline
~$\bf{m=0.2}$~\bf{GeV}~  &~141.1 ~~& ~~~~52.9~ & ~~~~29.2 &~~~~9.2~ \\

~$\bf{m=0.8}$~\bf{GeV}~   &~152.3 ~~  & ~~~~51.2~ & ~~~~33.2~  &~~~~16.6~\\

~$\bf{ALICE~dN_{ch}/d\eta}$~ & ~42.6~~ & ~~~~16.1~& ~~~~9.6~&~~~~4.3\\
  \hline
 $\bf{g^2dE_{\bot}/dy}~\bf{[GeV]} $  &   &   &  &~~~ \\ \hline
~$\bf{m=0.2}$~\bf{GeV}~  &~457.1 ~~& ~~~~162.6~& ~~~~80.1~ &~~~~20.1~ \\
~$\bf{m=0.8}$~\bf{GeV}~   &~697.1 ~~  & ~~~~214.4~ &  ~~~136.7~ &~~~~66.2~\\ \hline
\end{tabular}
\caption{Values for gluon multiplicity $g^2dN/dy$ and transverse energy per unit rapidity $g^2dE_{\bot}/dy$ at $y=0$ for $\alpha_s=0.15$ together with the ALICE data \cite{ALICE:2014xsp} for $dN_{\rm ch}/{d\eta}$. Simulation results are obtained for two different setups, $\tilde{m}=m=0.2$ GeV with $c_{Q_s} = 1.25$, and $\tilde{m}=m=0.8$ GeV with $c_{Q_s} = 1.82$.}
\label{tab:EventStatistics}
\end{table}

\begin{figure*}[ht]
\centering
\includegraphics[width=.33\linewidth]{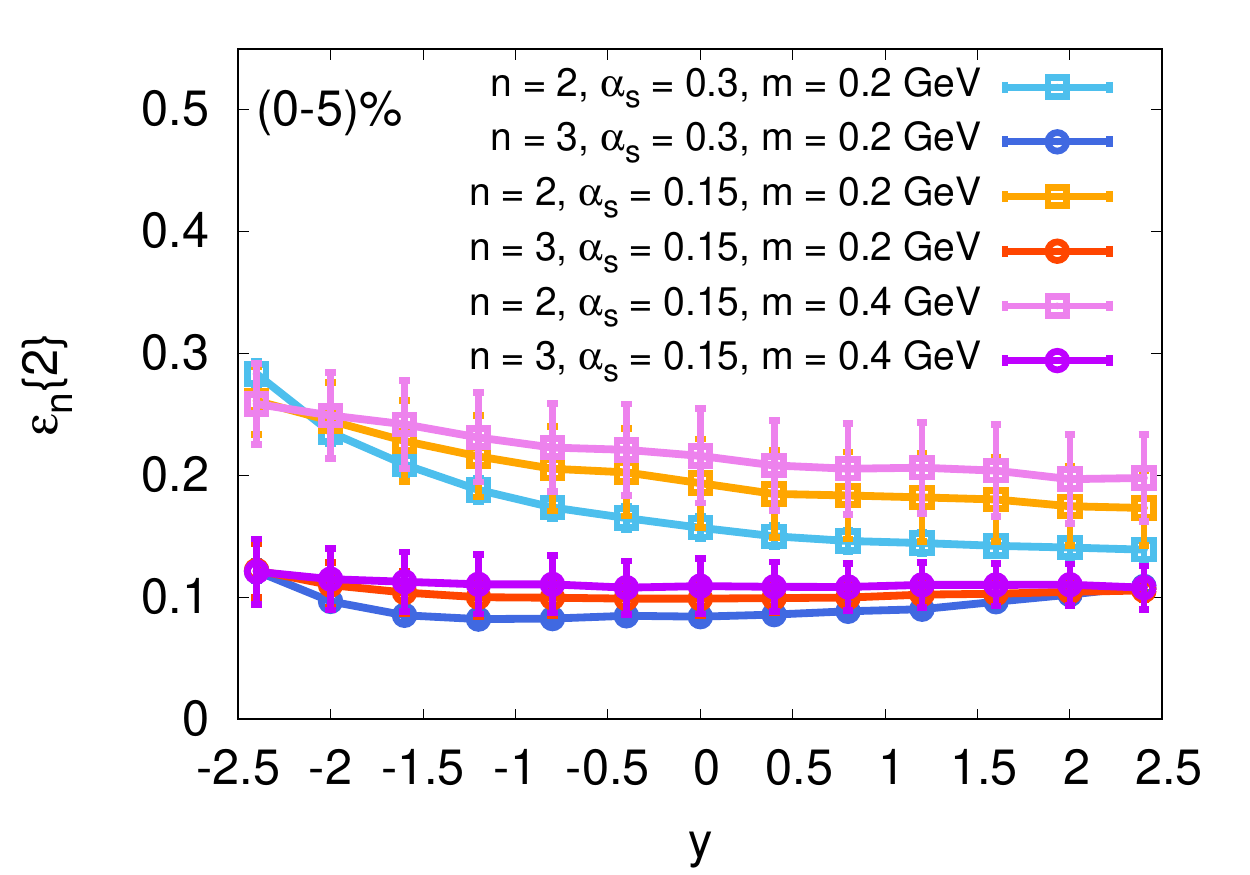}\hfill
\includegraphics[width=.33\linewidth]{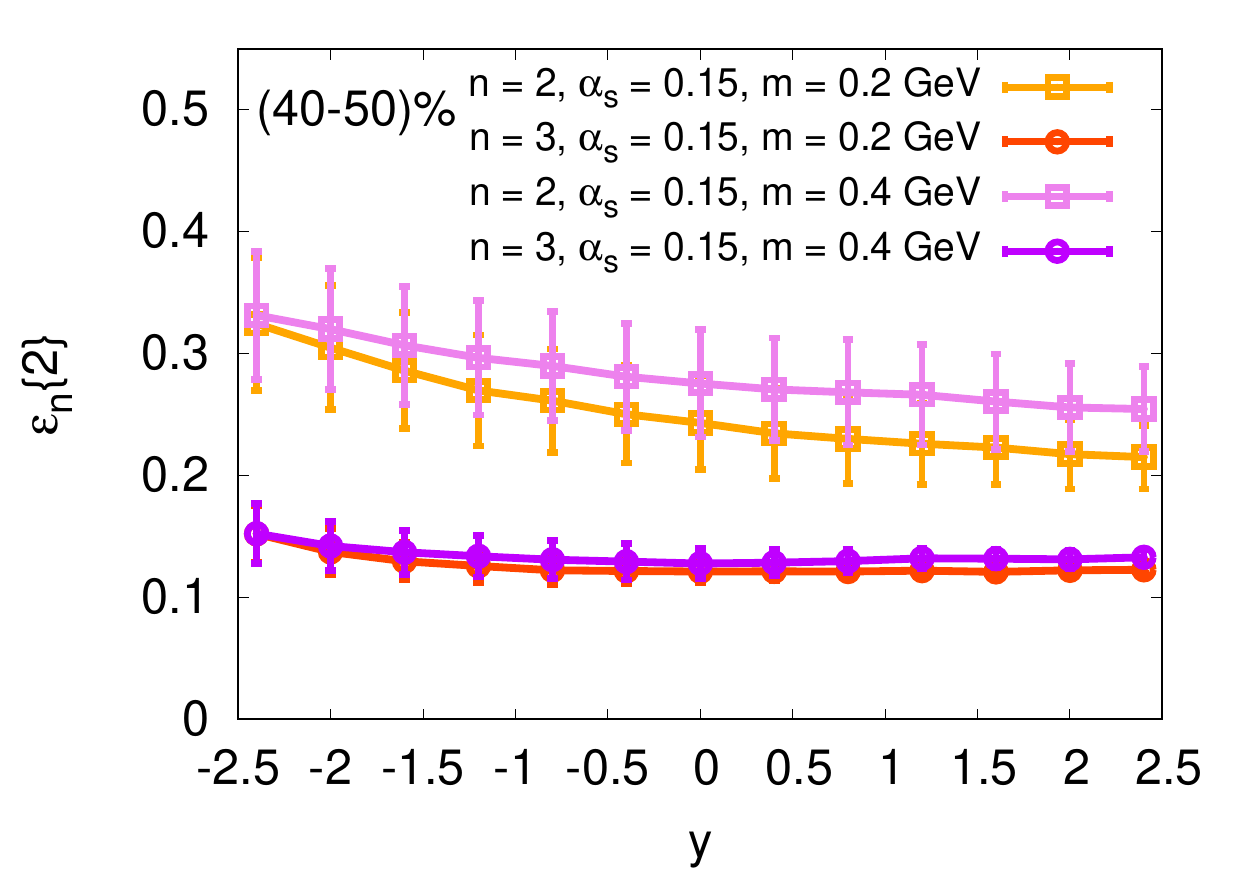}\hfill
\includegraphics[width=.33\linewidth]{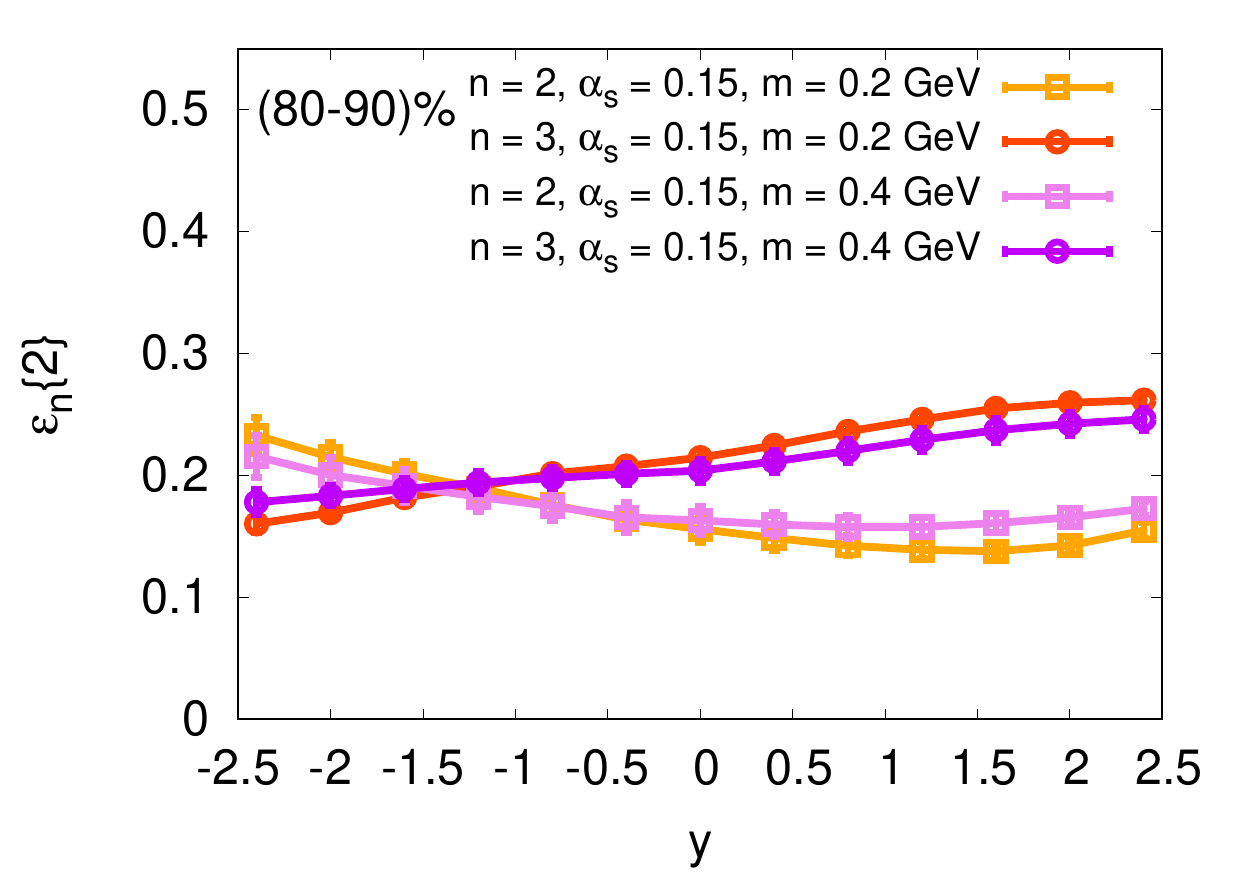}
\includegraphics[width=.33\linewidth]{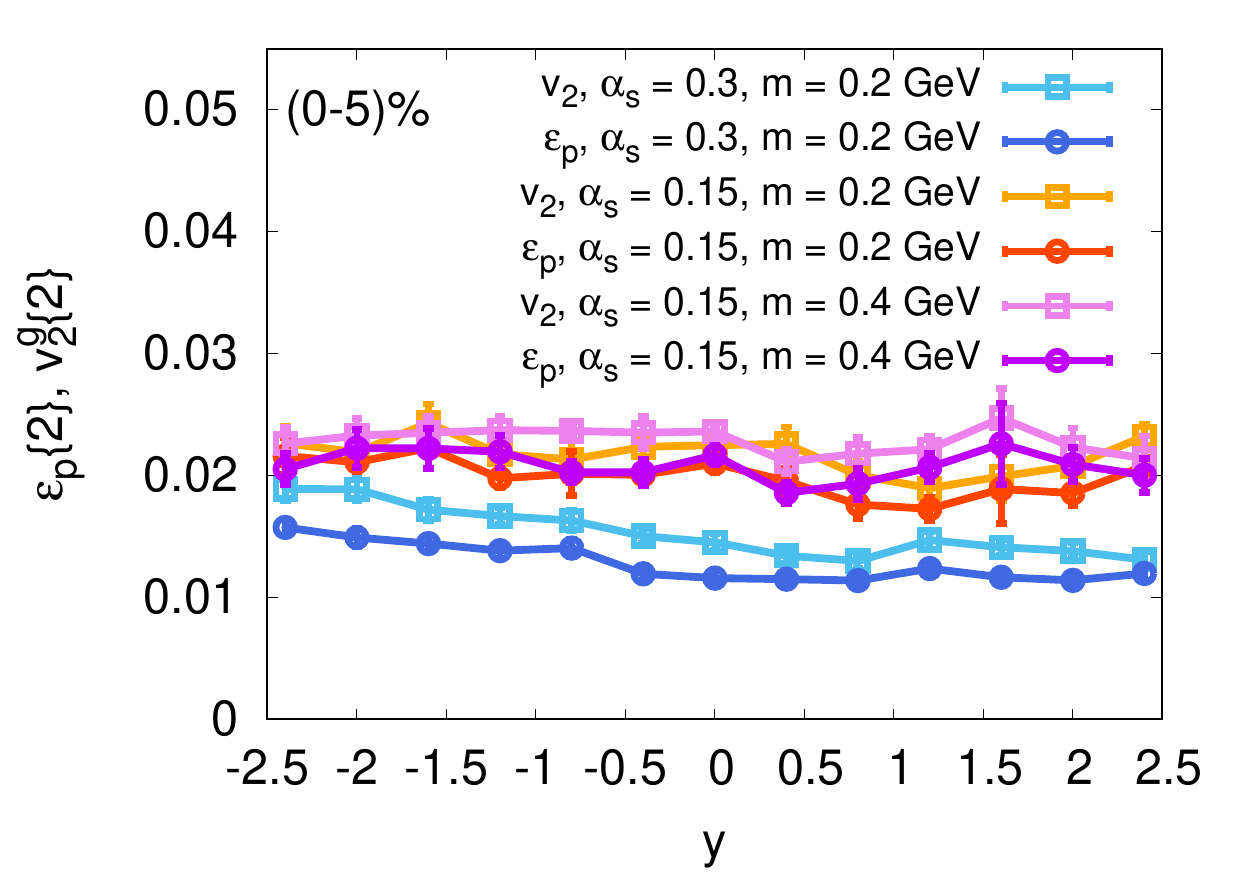}\hfill
\includegraphics[width=.33\linewidth]{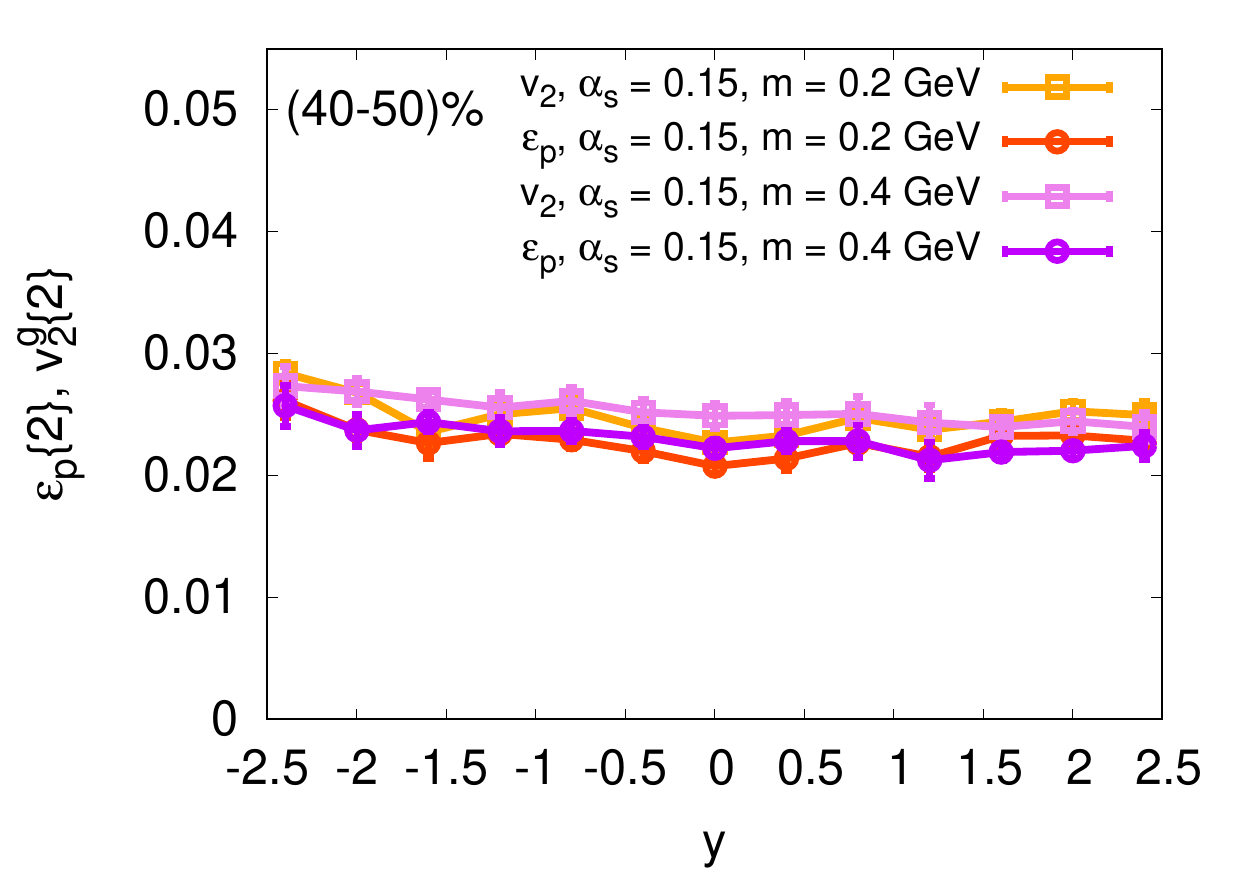}\hfill
\includegraphics[width=.33\linewidth]{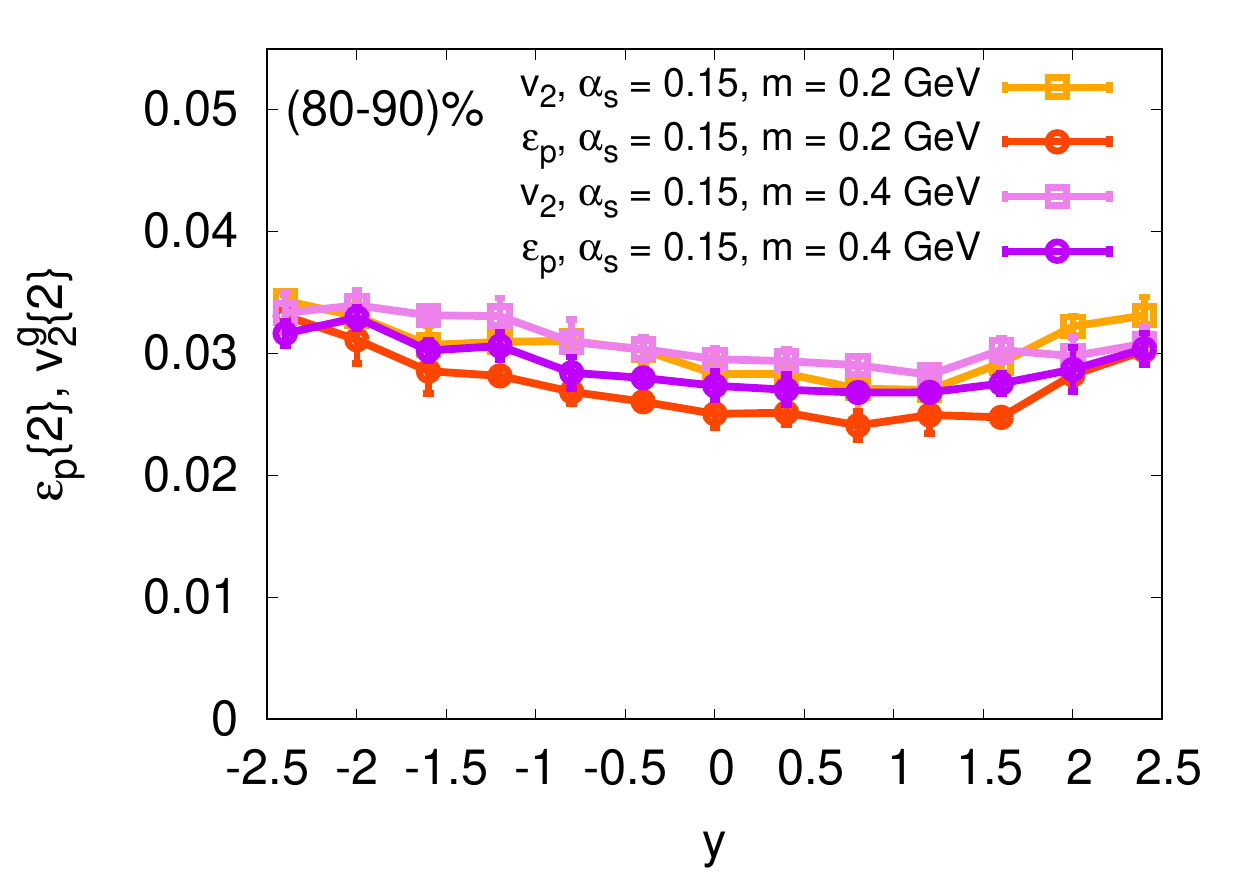}
\caption{Geometric eccentricities $\varepsilon_{n}\{2\}=\sqrt{ \langle |\varepsilon_{n}(y)|^2\rangle}$ (top) and initial momentum anisotropies $\varepsilon_{p}\{2\}=\sqrt{\langle |\varepsilon_p(y)|^2 \rangle}$ together with azimuthal anisotropy $v_2^{g}\{2\}=\sqrt{\langle|v_2^{g}(y)|^2\rangle}$ (bottom) for different centrality classes 0-5\% (left),~40-50\% (center) and 80-90\% (right) as a function of rapidity.}
    \label{fig:Epandv2VsRapidity}
\end{figure*}

Generally, one can see that the rapidity dependence of both the multiplicity $dN_{g}/dy$ and the transverse energy $dE_{\bot}/dy$ flattens as one approaches more peripheral events; however the comparison of the left and right panels indicates that the magnitude of the forward-backward asymmetry in more central events is actually quite sensitive to the value of the infrared regulators $m$ and $\tilde{m}$. Evidently, it would be instructive to compare the results in Fig.~\ref{fig:dNbydY} to experimental measurements, however we are not aware of measurements of $dN/dy$ of identified hadrons in $p+Pb$ collisions. Nevertheless, when comparing our results for the gluon rapidity distribution to $dN_{\rm ch}/d\eta$ of unidentified charged hadrons, we find that the gluon distribution for $m=\tilde{m}=0.2\,{\rm GeV}$ generally shows a steeper rapidity dependence than the experimental data from the ALICE Collaboration \cite{ALICE:2014xsp}, which is essentially symmetric in the (80-100)\% bin and appears to be more in line with the behavior observed for $m=\tilde{m}=0.8\,{\rm GeV}$.


While the self-normalized quantities in Fig.~\ref{fig:dNbydY} emphasize the rapidity dependence, we note that for both parameter sets, the centrality dependence of the absolute yield at mid-rapidity up to $60-70\%$ is approximately in line with that of the experimentally determined charged hadron yield, as can be seen in Table\,\ref{tab:EventStatistics}.

When comparing the top and bottom panels of Fig.~\ref{fig:dNbydY}, one observes that the transverse energy shows a slightly weaker centrality dependence compared to the gluon multiplicity. This is likely a consequence of the transverse energy being more sensitive to the larger of the two saturation scales $Q_s$, as parametrically one has
$dN_{g}/dy \sim Q_{s<}^2 S_\bot$ while $dE_\bot/dy \sim Q_{s>}Q_{s<}^2 S_\bot$~\cite{Dumitru:2001ux,Lappi:2006hq} where $S_{\bot}$ is the transverse area and $Q_{s,>/<}$ denotes the larger/smaller of the two saturation scales.


Next, in order to obtain further insight into the properties of low and high multiplicity events, we will extract the average Pb and p saturation scales $Q_s(y)$ and determine a measure of the system size $S_{\bot}(y)$ for the different centrality classes. Specifically, the saturation scale $Q_{s}(y)$ is extracted from the dipole scattering amplitude
\begin{eqnarray}\label{eq:Dipole_Amplitude}
D(\rT,\dT)=\frac{1}{N_c}\text{tr}\left[ V_{\dT+\rT/2} V^{\dagger}_{\dT-\rT/2}\right]\;,
\end{eqnarray}
averaged over (dipole) impact parameters $|\dT|<0.2~R_{p}$ from the collision point\footnote{ By the collision point we mean the transverse position where the center of mass of the proton hits the lead nucleus. Hence for the proton the dipole amplitude is extracted around its center of mass, while according to Eq.~(\ref{eq:ini1},\ref{eq:ini2}) for the lead nucleus, the collision point is offset from the center of the nucleus and the dipole amplitude is thus measured around the impact parameter $\bT$ of the p+Pb collision.} (see App.~\ref{app:A} for details). By following previous works \cite{Schlichting:2014ipa}, we extract the distance $|\rT|_{c}$ where the dipole amplitude exceeds a value of $c$, i.e.,
\begin{eqnarray}
D(|\rT|_{c},|\dT|<0.2 R_{p})=c\;,
\end{eqnarray}
 and calculate $Q_{s}=2/|\rT|_{c}\log^{1/2}(1/c)$ according to the parametrization $D(\rT)=\exp(-Q_s^2\rT^2/4)$. We employ $c=0.8$ and $0.9$ to estimate the uncertainty of this procedure. While the saturation scale $Q_s$ reflects properties of the gluon distribution of the colliding nuclei, the system size $S_{\bot}$ is determined from the energy momentum tensor $T^{\mu\nu}$ as
\begin{eqnarray}\label{eq:Sperp}
S_{\bot}=\frac{\int d^2\xT~{\mathbf x}_\perp^{2}
~T^{\tau\tau}(\xT)}{\int d^2\xT
~T^{\tau\tau}(\xT)}
\end{eqnarray}
which we evaluate at $\tau=0.2~{\rm fm}/c$ after the collision of the proton and the lead nucleus.

We will focus on the case $m=\tilde{m}=0.2\,{\rm GeV}$, which exhibits a stronger rapidity and centrality dependence of $dN/dy$ and $dE_\perp/dy$. For this case our results for $Q_{s}^{p/Pb}(y)$ and $S_{\bot}(y)$ are compactly summarized in Fig.~\ref{fig:QsST}. With decreasing $x$, which corresponds to increasing rapidity $y$ for the left moving proton and decreasing rapidity $y$ for the right moving lead nucleus, both saturation scales $Q_s$ increase due to the JIMWLK evolution. The proton saturation scale $Q_{s}^{p}$ is similar in the three more peripheral events, while the nucleus' $Q_{s}^{Pb}$ depends more strongly on centrality, indicating that in mid-central and peripheral events the multiplicity is determined by the impact parameter, i.e., the position in the lead nucleus where the proton hits, as well as fluctuations in the lead nucleus. In contrast, the proton saturation scale in the most central bin is significantly larger than for the other centralities, while there is little difference between the lead saturation scale in 40-50\% and 0-5\% centrality classes. This means that the highest multiplicities are reached by upward fluctuations of the proton's gluon density, quantified by $Q_{s}^{p}$.

The size of the interaction region increases approximately linearly (for the three most central centralities studied), which is driven by the growth of the proton size with rapidity \cite{Kovner:2001bh,Schlichting:2014ipa} (see App.~\ref{app:A}). The most peripheral events show a significantly larger area, which may appear counter-intuitive at first sight. However, given the definition of the area measure in \eqref{eq:Sperp}, an overall very small but spread out energy density can lead to a large area, which seems to be the dominant structure of the most peripheral events we studied. For all other centralities, the area is approximately the same, and the difference in multiplicity is driven almost entirely by changes in the $Q_s$ values.


\section{Event geometry \& initial state momentum correlations}
\label{sec:four}
Having established the basic features of the events in different centrality classes, we continue by investigating the longitudinal structure of the event geometry and the initial state momentum anisotropy. We follow standard procedure and characterize the event geometry in terms of the eccentricities 
    \begin{align}\label{eq:en}
        \varepsilon_n(y)=\frac{\int d^2\rT T^{\tau\tau}(y,\rT)~|\rT|^n e^{in\phi_{\rT}}}{\int d^2\rT T^{\tau\tau}(y,\rT)~|\rT|^n}\,,
    \end{align}
where the integer $n$ indicates the harmonic. We will study the cases $n=2$ and $n=3$.

Similarly, following \cite{Schenke:2019pmk,Giacalone:2020byk}, the initial state momentum anisotropy can be characterized in terms of the anisotropic energy flow
\begin{align}\label{eq:ep}
        \varepsilon_p(y)=\frac{\int d^2\rT~ T^{xx}(y,\rT)-T^{yy}(y,\rT)+2iT^{xy}(y,\rT)}{\int d^2\rT~T^{xx}(y,\rT)+T^{yy}(y,\rT)}
\end{align}
or alternatively as in \cite{Schenke:2015aqa} in terms of the azimuthal anisotropy $v_2^{g}$ of the produced gluons\footnote{We note that the additional $|\kT|$ weight is chosen such that in the quasi-particle picture the definitions of $\varepsilon_{p}$ and $v_2^{g}$ agree with each other.}
 \begin{align}\label{eq:v2}
       v_2^{g}(y)=\frac{\int d^2\kT |\kT| \frac{dN_{g}}{dy d^2\kT}(y)e^{2i\phi_{\kT}}}{\int d^2\kT |\kT|~\frac{dN}{dy d^2\kT}(y)}\,.
\end{align}
We evaluate the expressions in Eqs.\,\eqref{eq:en}, \eqref{eq:ep}, and \eqref{eq:v2} at $\tau=0.2\,\rm{fm}/c$ to calculate $\varepsilon_{n},\varepsilon_{p},$ and $v_{2}^{g}$ as functions of rapidity $y$ on an event-by-event basis. Subsequently, to quantify the overall rapidity dependence we compute the correlation functions
\begin{align}
C_\mathcal{O}(y_1,y_2)=\left\langle {\rm Re}\big(\mathcal{O}(y_1)\mathcal{O}^{*}(y_2)\big) \right\rangle
\end{align}
where $\left\langle.\right\rangle$ denotes an event average and $\mathcal{O}$ is any of the above observables. The correlation function $C_\mathcal{O}$ contains information about both the magnitude and rapidity dependence of the correlation function. To focus on the rapidity decorrelation of the transverse geometry and initial state momentum correlations, we will also show results for the normalized rapidity correlation function
\begin{align}\label{eq:CN}
    C^N_{\mathcal{O}}(y_1,y_2)=\frac{C_\mathcal{O}(y_1,y_2)}{\sqrt{\langle |\mathcal{O}(y_1)|^2\rangle\langle |\mathcal{O}(y_2)|^2\rangle}}\,.
\end{align}

\begin{figure}[t]
\centering
\includegraphics[width=0.4\textwidth]{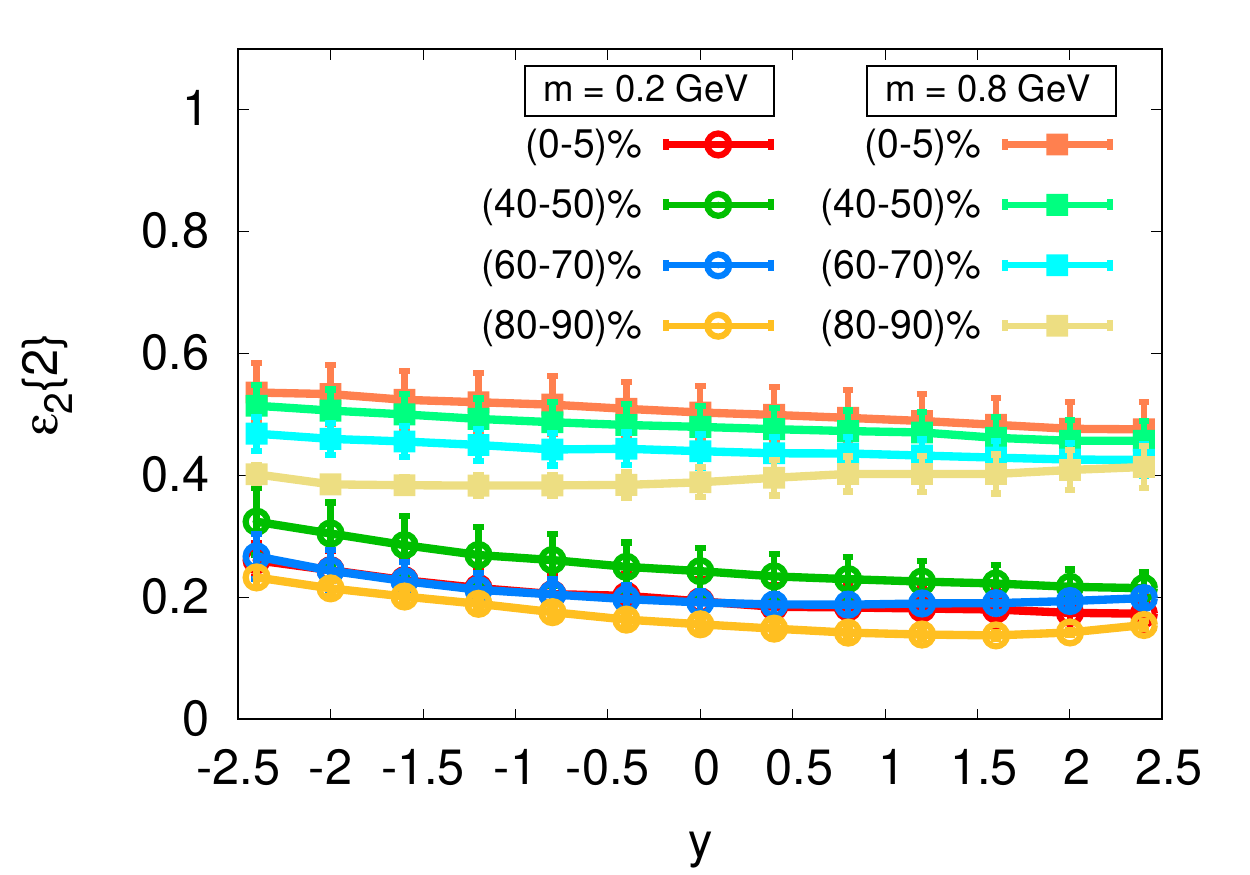}\hfill
\includegraphics[width=0.4\textwidth]{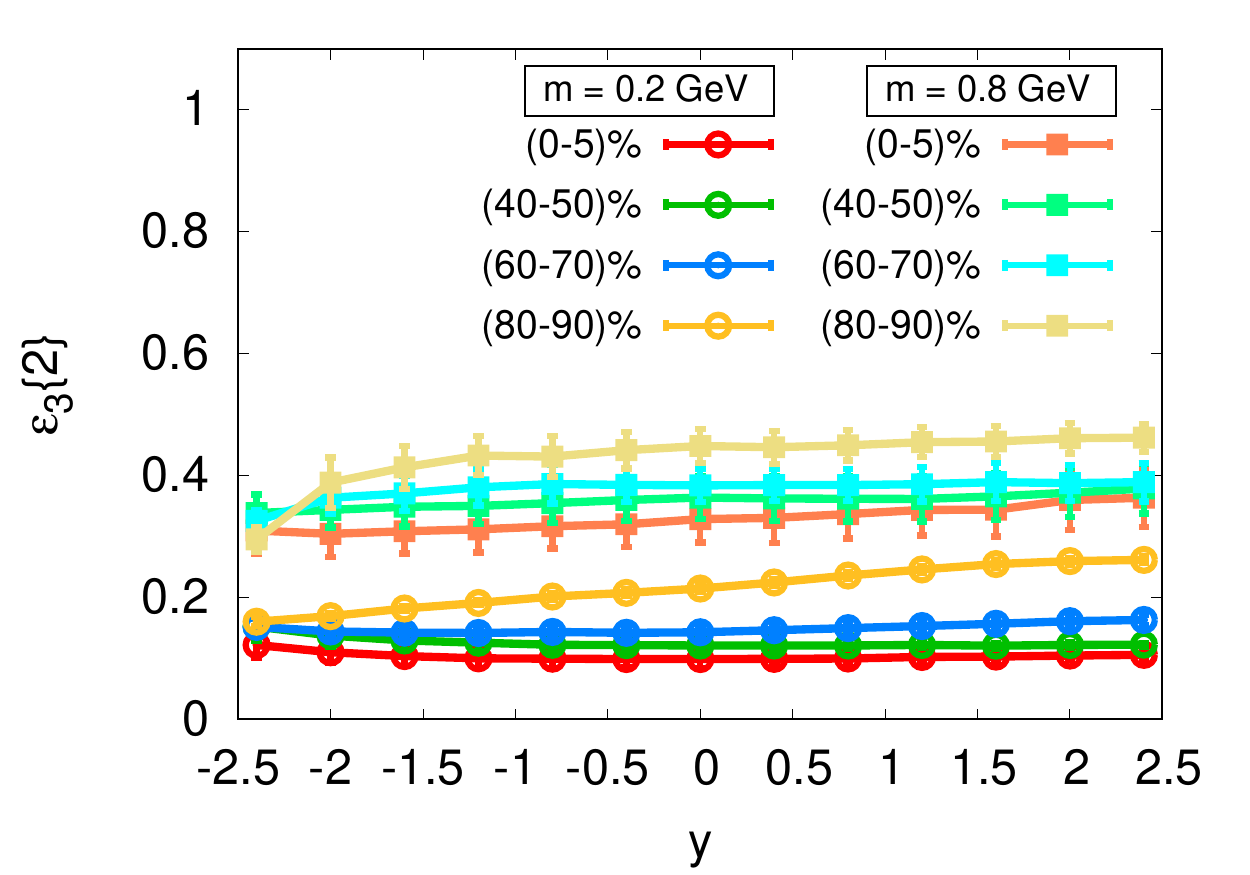}\hfill
\includegraphics[width=0.4\textwidth]{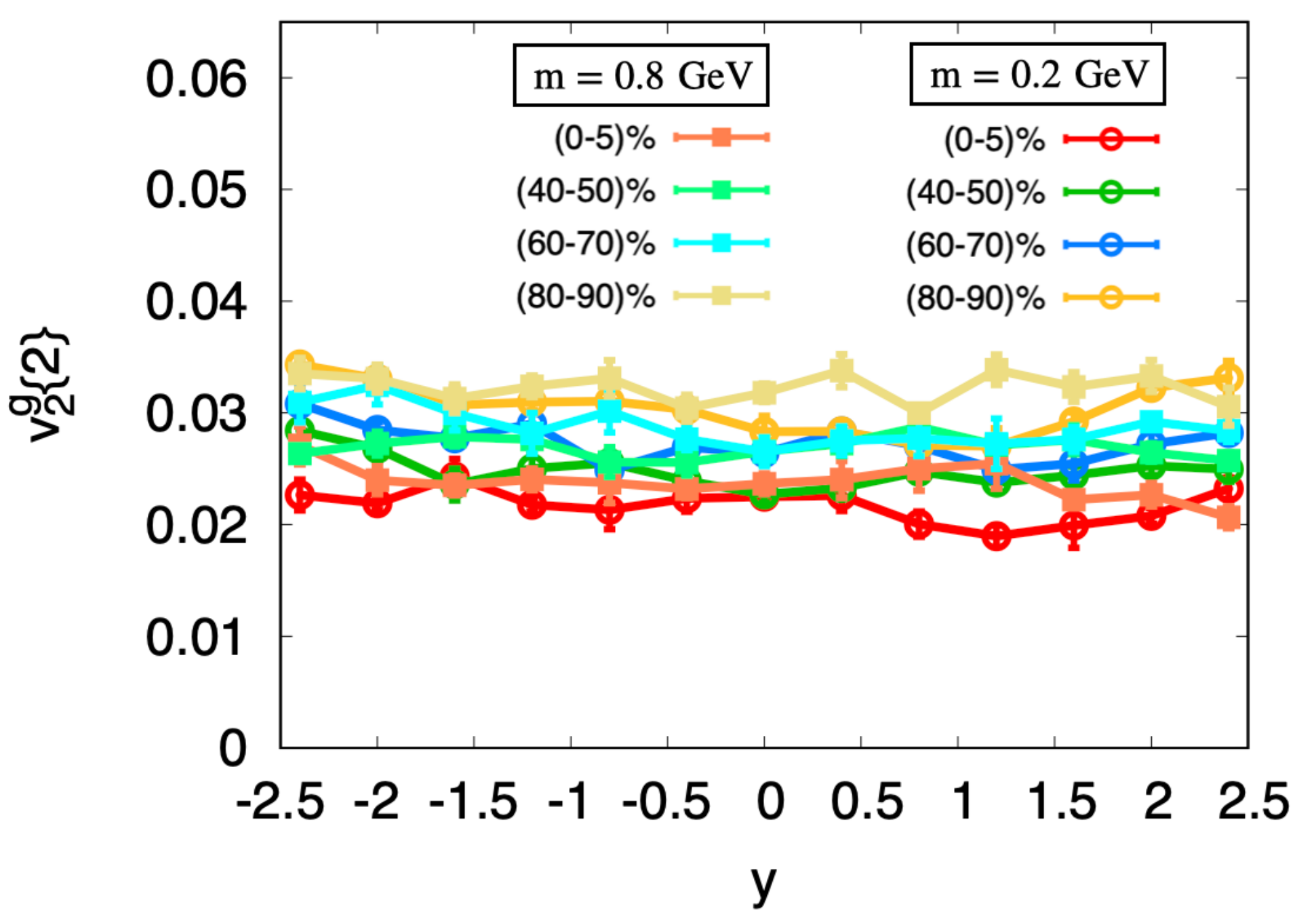}\hfill
\caption{Comparison of the rapidity dependence of $\sqrt{ \langle |\varepsilon_{2}(y)|^2\rangle}$ (top), $\sqrt{ \langle |\varepsilon_{3}(y)|^2\rangle}$ (middle) and $\sqrt{\langle|v_2^{g}(y)|^2\rangle}$ (bottom) for different centrality classes for $\alpha_s=0.15$ and distinct IR regulators such that $m=\tilde{m}$.}
    \label{fig:E2andv2_Centrality_Plots}
\end{figure}

\subsection{Rapidity dependence of event geometry and momentum anisotropy}
In Fig.\,\ref{fig:Epandv2VsRapidity} (top) we show the rapidity dependence of eccentricities $\varepsilon_2\{2\}(y)=\sqrt{\langle|\varepsilon_{2}(y)|^2\rangle}$ and $\varepsilon_3\{2\}(y)=\sqrt{\langle|\varepsilon_{3}(y)|^2\rangle}$ for different parameters and centrality classes. In most cases $\varepsilon_2$ decreases with increasing rapidity, and does so more rapidly for larger $\alpha_s$ and smaller $m$, as expected by how these parameters affect the JIMWLK evolution speed. For our standard parameters of $\tilde{m}=m=0.2\,{\rm GeV}$ and $\alpha_s=0.15$ the rapidity dependence is rather weak.
For the most peripheral bin, $\varepsilon_2$ has a shallow minimum as a function of rapidity. The triangularity $\varepsilon_3$ has an even weaker rapidity dependence than $\varepsilon_2$ in the two more central bins, and increases with increasing rapidity in the most peripheral bin. Given the comparable size of $\varepsilon_2$ and $\varepsilon_3$ in this bin, one might expect the observed anti-correlation between the two quantities, as it is difficult geometrically to generate a large $\varepsilon_2$ and $\varepsilon_3$ at the same time (This can be seen most easily when arranging just three hot spots. A maximal triangularity goes along with a reduced ellipticity and vice versa.)

\begin{figure*}[ht]
    \centering
    \begin{subfigure}
        \centering
        \includegraphics[width=0.4\textwidth]{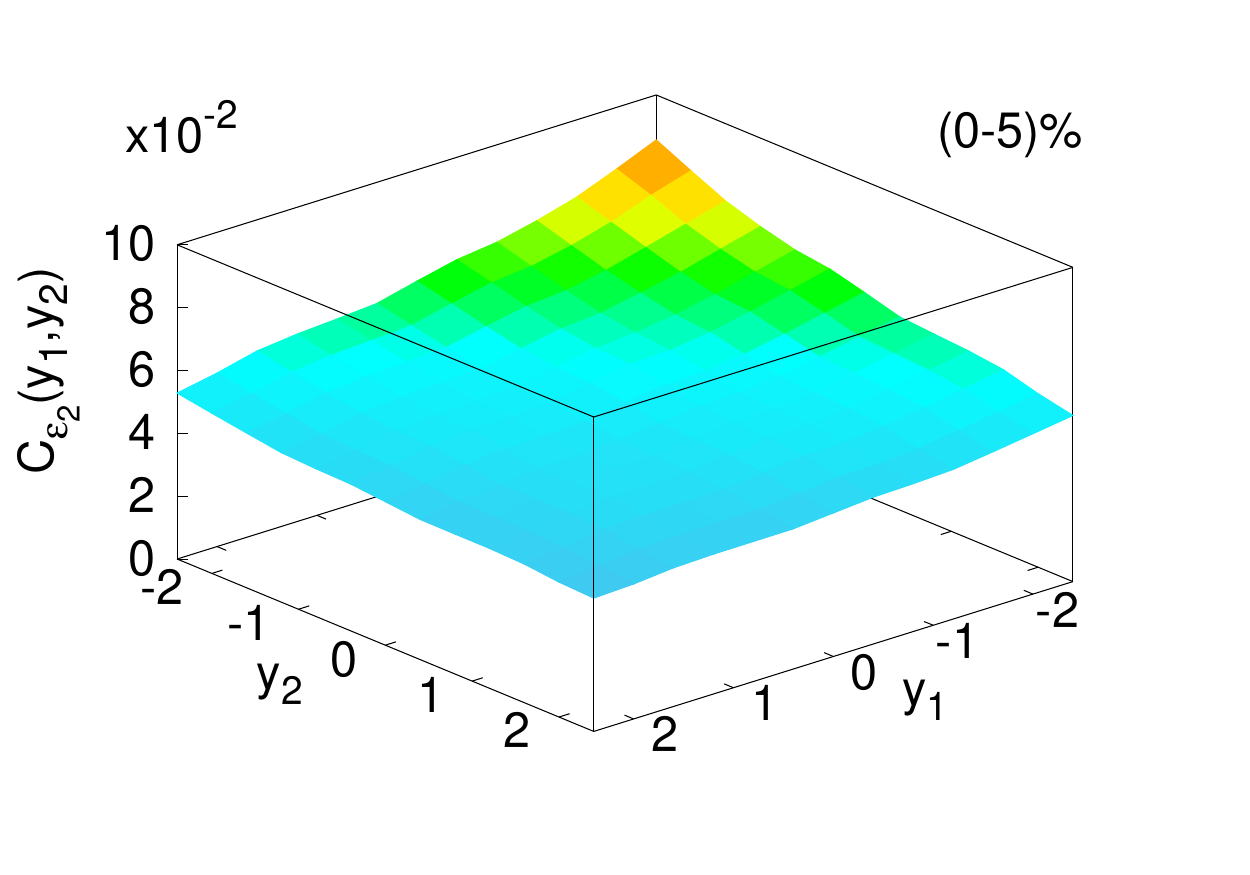}
    \end{subfigure}
    \begin{subfigure}  
        \centering 
        \includegraphics[width=0.4\textwidth]{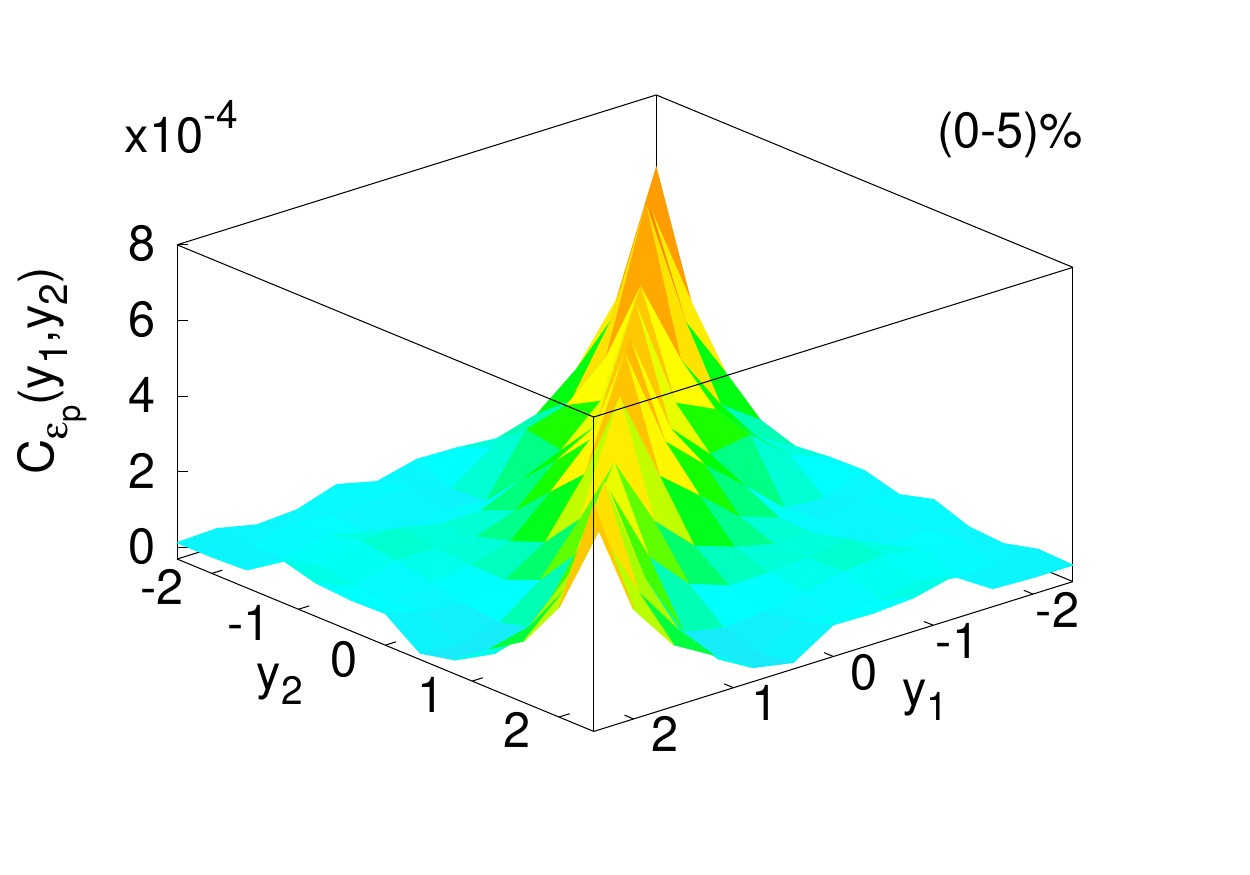}
    \end{subfigure}
    \vskip\baselineskip
    \begin{subfigure}  
        \centering 
        \includegraphics[width=0.4\textwidth]{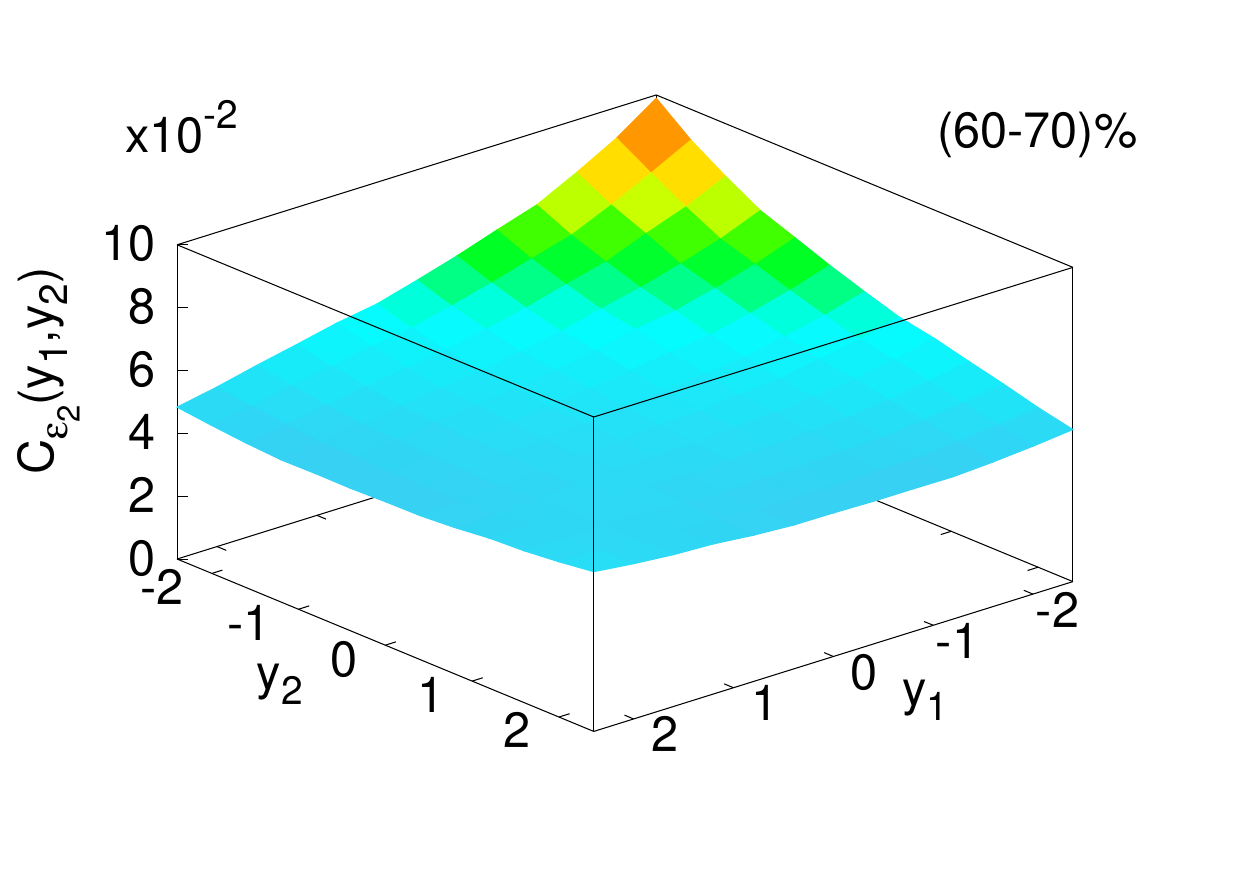}
    \end{subfigure}
    \begin{subfigure}
        \centering 
        \includegraphics[width=0.4\textwidth]{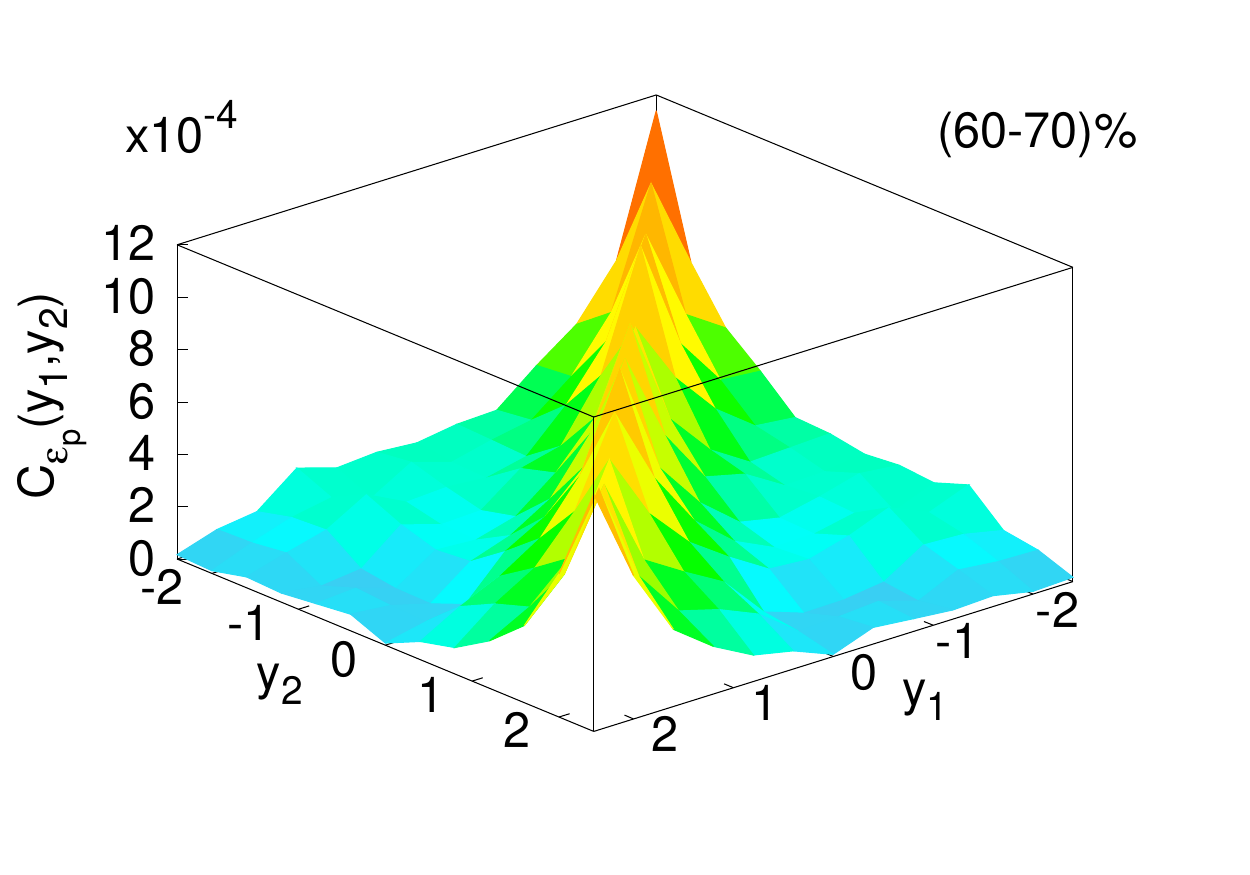}
    \end{subfigure}
    \caption{Two point correlation function for second order eccentricity $\varepsilon_2$ (top-left) and momentum anisotropy $\varepsilon_p$ (top-right) for $(0-5)\%$ centrality class for $\alpha_s=0.15$ and $m=0.2$ GeV. Bottom panel demonstrates the same observable for $(60-70)\%$ centrality class.
    }
    \label{fig:3d_unnormalised_plots}
\end{figure*}

\begin{figure*}
\centering
\includegraphics[width=.31\linewidth]{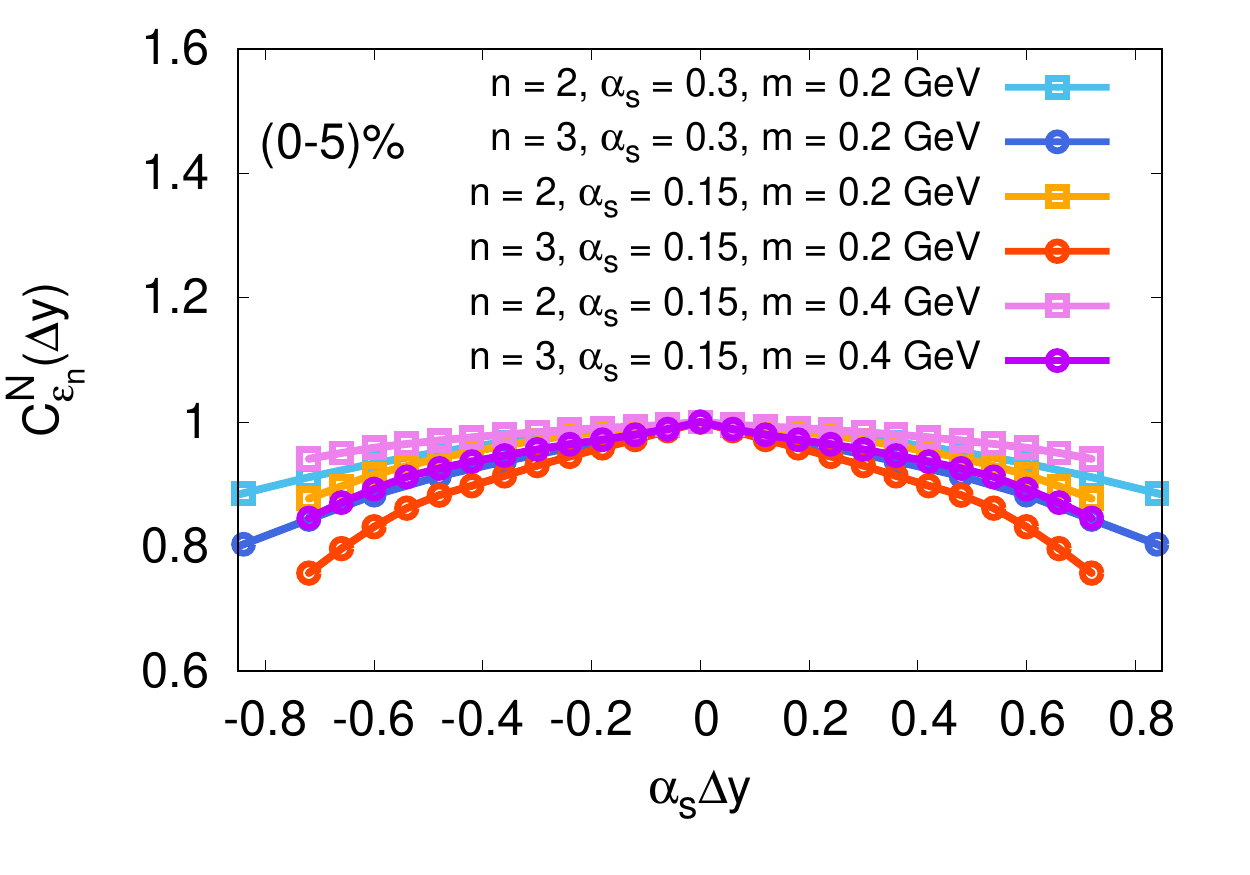}\hfill
\includegraphics[width=.31\linewidth]{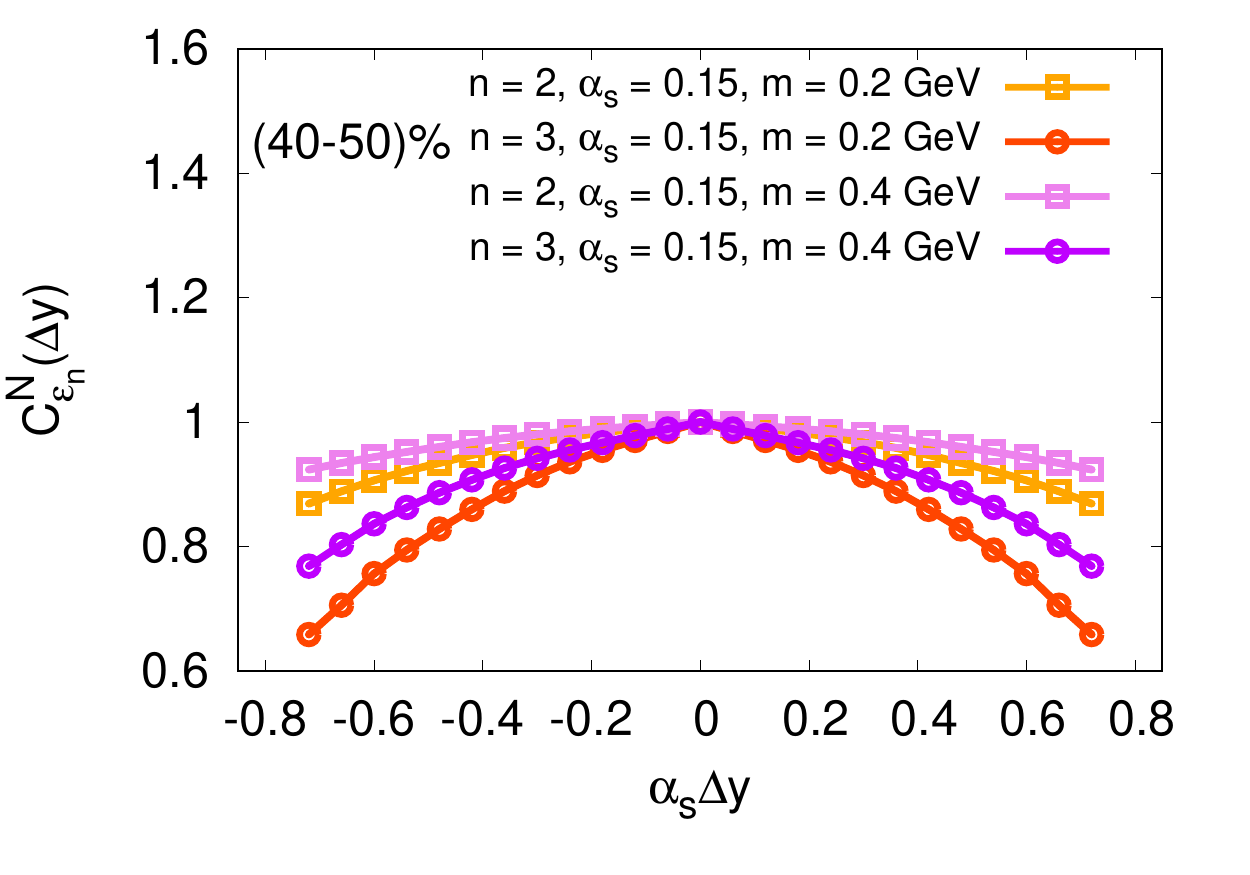}\hfill
\includegraphics[width=.31\linewidth]{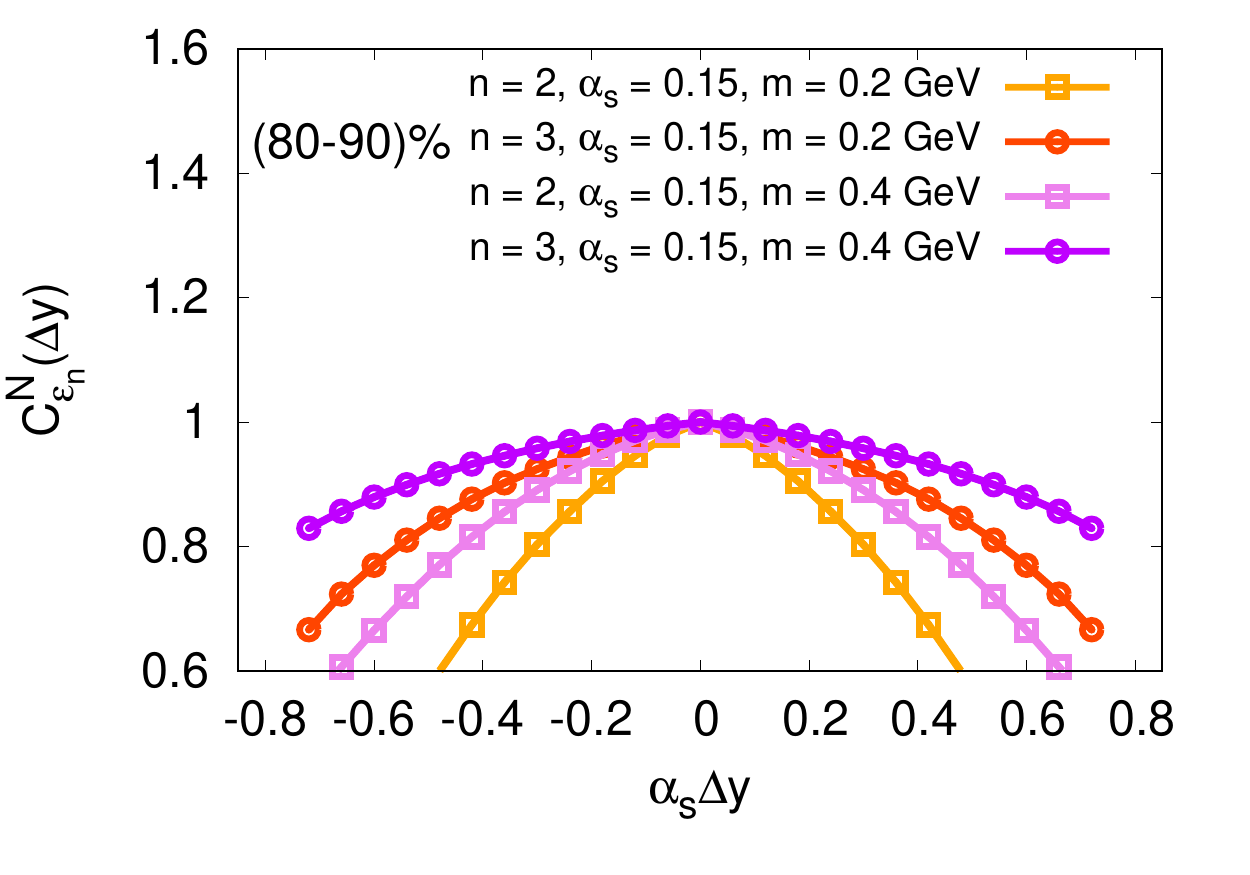}\hfill
\includegraphics[width=.31\linewidth]{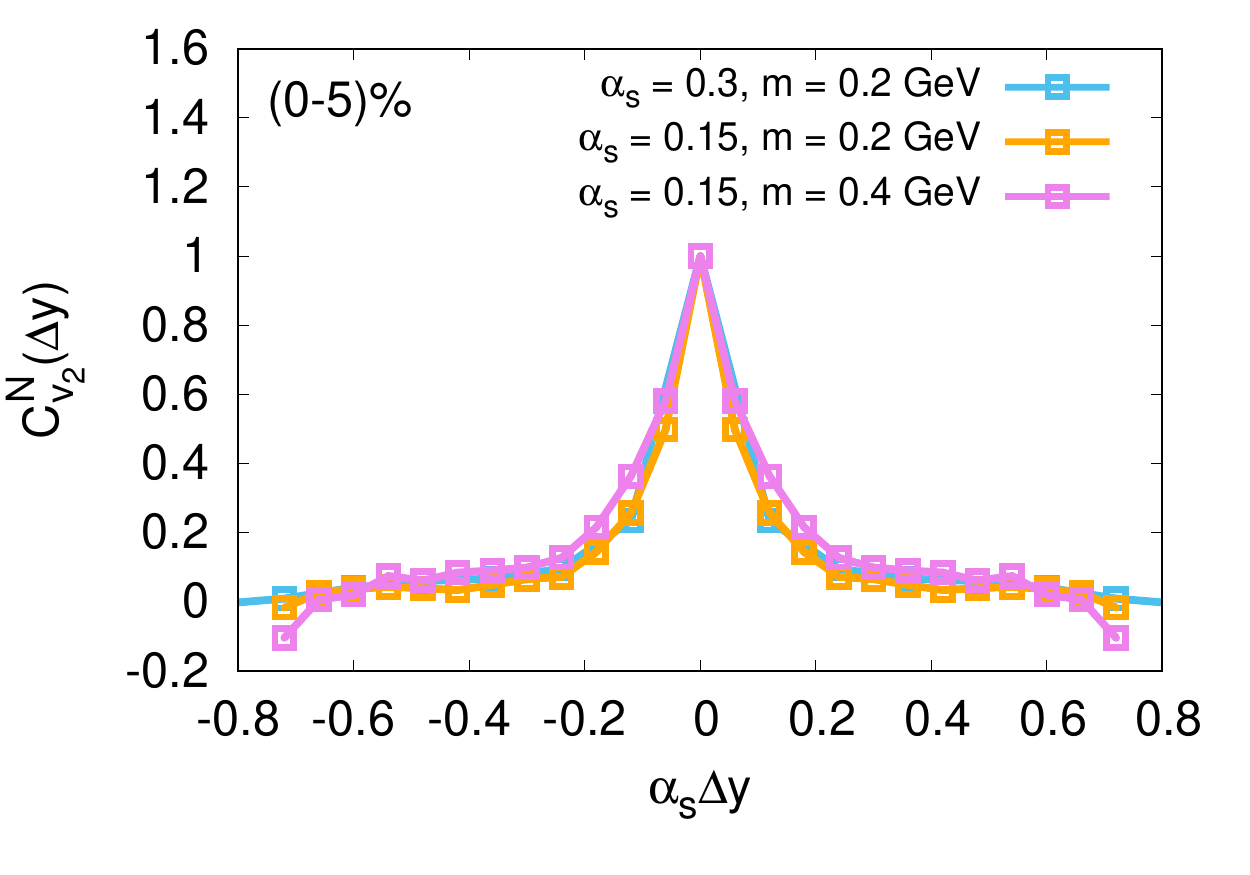}\hfill
\includegraphics[width=.31\linewidth]{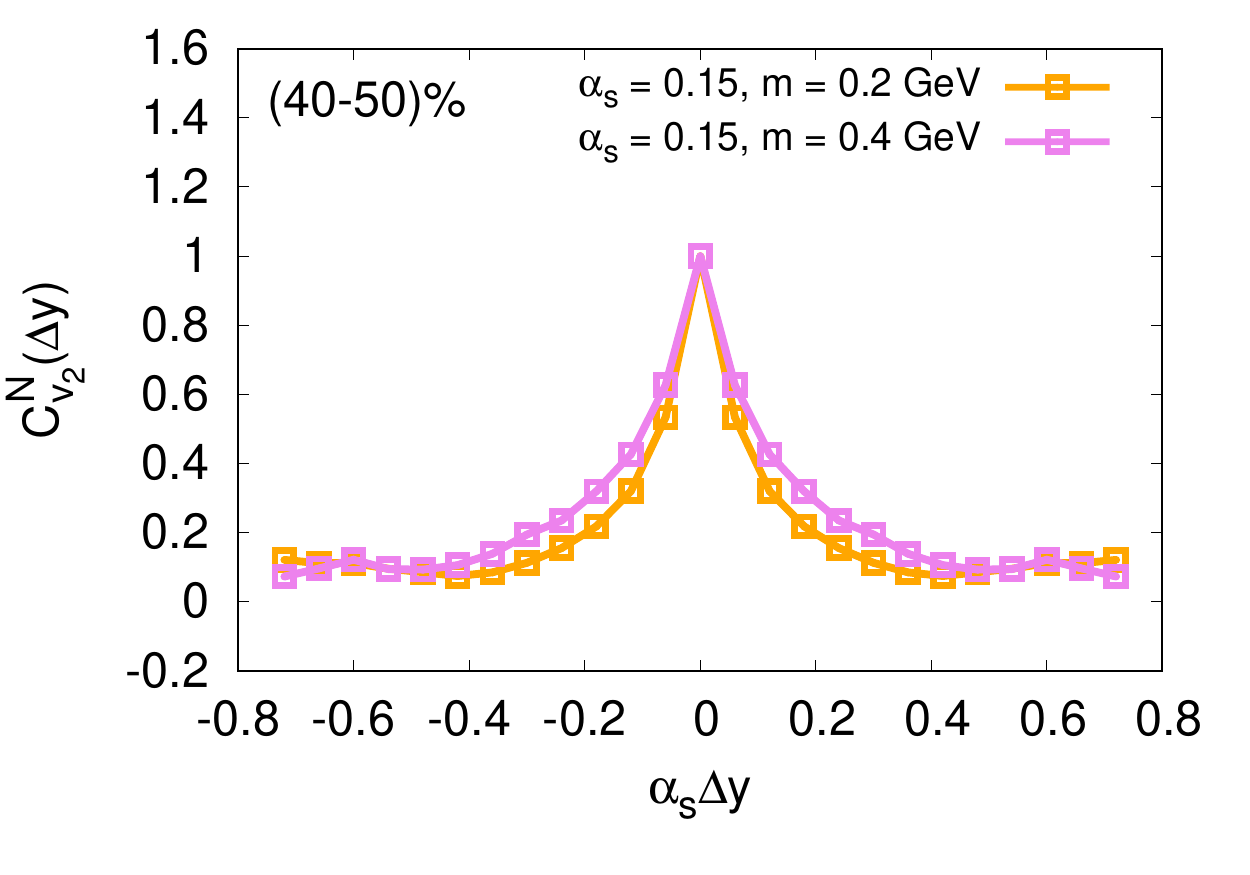}\hfill
\includegraphics[width=.31\linewidth]{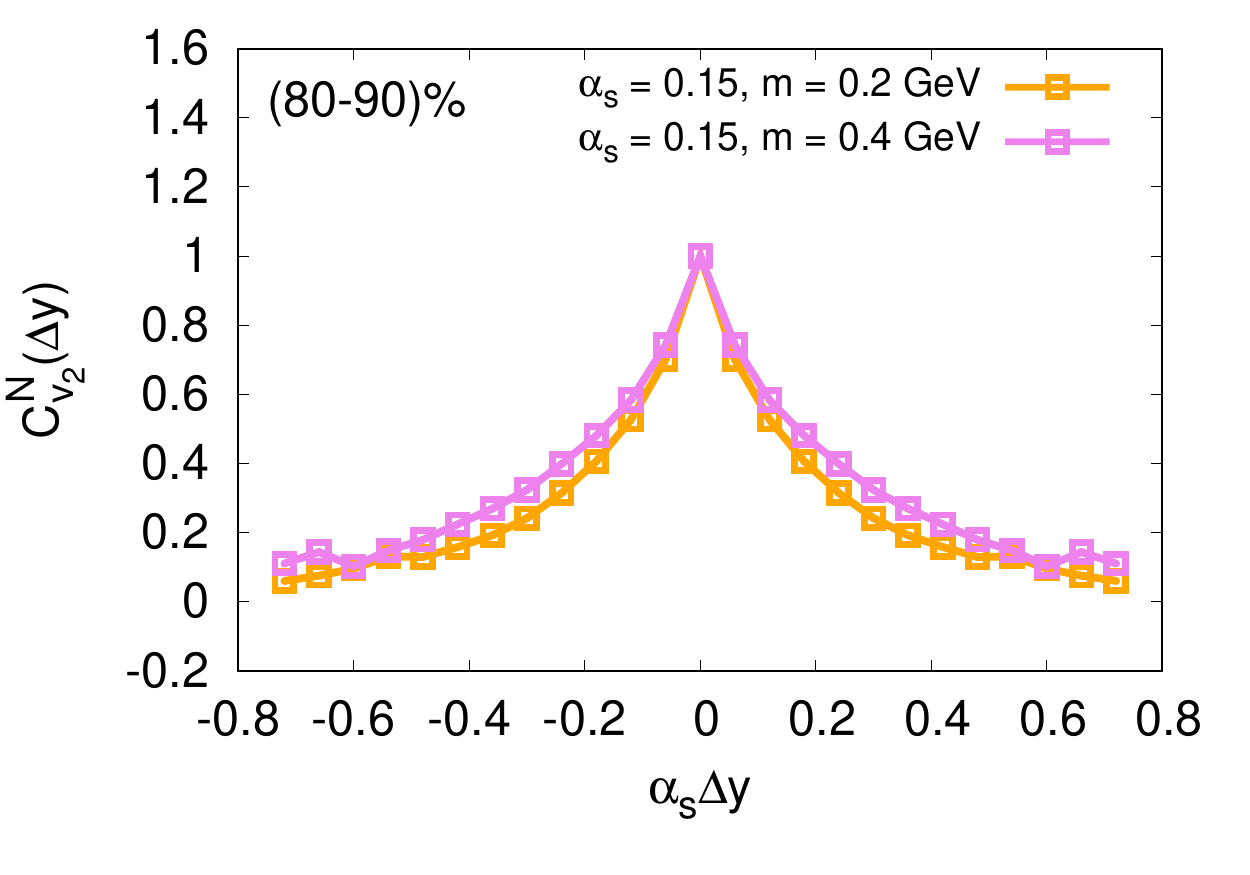}\hfill
\caption{Normalized two point correlation functions $C^{N}(\Delta y)$ for geometric eccentricities $\varepsilon_{2},\varepsilon_{3}$ (top) and initial state momentum anisotropies $v_2$ (bottom) for different centrality classes 0-5\% (left),~40-50\% (center) and 80-90\% (right) as a function of the rapidity separation $\alpha_{s} \Delta y$.
}
    \label{fig:Norm_CorrelationsEpandv2}
\end{figure*}

The bottom panel in Fig.\,\ref{fig:Epandv2VsRapidity} shows the rapidity dependence of the initial state anisotropy $\varepsilon_p\{2\}(y)=\sqrt{\langle|\varepsilon_{p}(y)|^2\rangle}$ and the gluon elliptic momentum anisotropy $v_2^{g}\{2\}(y)=\sqrt{\langle|v_{2}^{g}(y)|^2\rangle}$ for the same centrality classes and parameter sets as the eccentricities above. First, it is clear to see that both quantities follow each other closely. The anisotropy of the energy momentum tensor is thus a good predictor of the gluon momentum anisotropy in the situation that strong final state interactions are not included. 
Comparing the three panels, we can see that the initial momentum anisotropy increases with decreasing gluon multiplicity. The rapidity dependence of $\varepsilon_{p}$ and $v_{2}^{g}$ is negligible in most cases, with the case using $\alpha_s=0.3$ (shown only for the most central bin) showing the strongest decrease with increasing rapidity. In the most peripheral bin the two quantities show a minimum around $y=1$, which is where the transverse energy is maximal.

In Fig.\,\ref{fig:E2andv2_Centrality_Plots} we focus on the centrality dependence of the rapidity dependent $\varepsilon_2$, $\varepsilon_3$, and $v_2^{g}$ and compare results for the two different parameter sets $m=\tilde{m}=0.2 \,{\rm GeV}$ and $m=\tilde{m}=0.8 \,{\rm GeV}$ with $\alpha_s=0.15$ in both cases.\footnote{Since $\varepsilon_p$ and $v_2^{g}$ are essentially identical, we only show the centrality dependence of $v_2^g$.} Generally, the sharper profiles for $m=\tilde{m}=0.8,{\rm GeV}$ lead to larger geometric eccentricities $\varepsilon_{2}$ and $\varepsilon_{3}$ across all rapidities and centrality classes, as pointed out previously in \cite{Demirci:2021kya}. While for $m=\tilde{m}=0.8\, {\rm GeV}$ both $\varepsilon_{2}$ and $\varepsilon_{3}$ exhibit a monotonic behavior as a function of centrality, we find that for smaller values of the infrared regulator $m=\tilde{m}=0.2\, {\rm GeV}$, the eccentricity $\varepsilon_2$ is maximal for 40-50\% central collisions, and minimal in the most peripheral bin, and $\varepsilon_3$ increases monotonically towards more peripheral events and shows the strongest centrality dependence on the lead going side. 

On the other hand, the magnitude and centrality dependence of the gluon momentum anisotropy $v_2^{g}$ is rather insensitive to the infrared regulator and only very weakly dependent on the rapidity. However, as has been observed previously \cite{Schenke:2015aqa, Schenke:2019pmk}, the initial momentum anisotropy driven $v_2^{g}$ increases monotonically with decreasing multiplicity (towards more peripheral events). We show here that this is true for all studied rapidities. Furthermore, the value of $v_2^g$ is largely independent of rapidity in all centrality bins, which is also a new result.

\begin{figure*}[t]
\centering
\includegraphics[width=.49\textwidth]{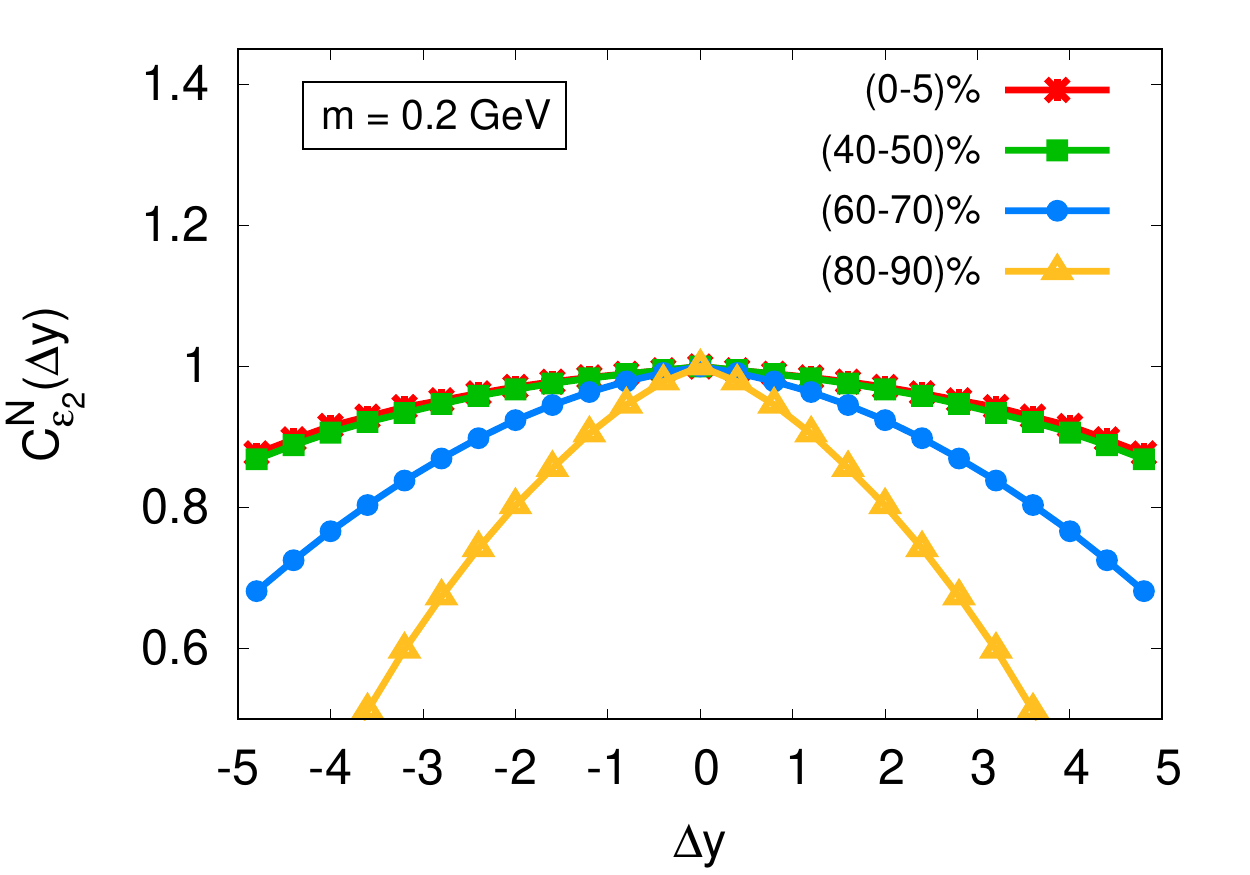}\hfill
\includegraphics[width=.49\textwidth]{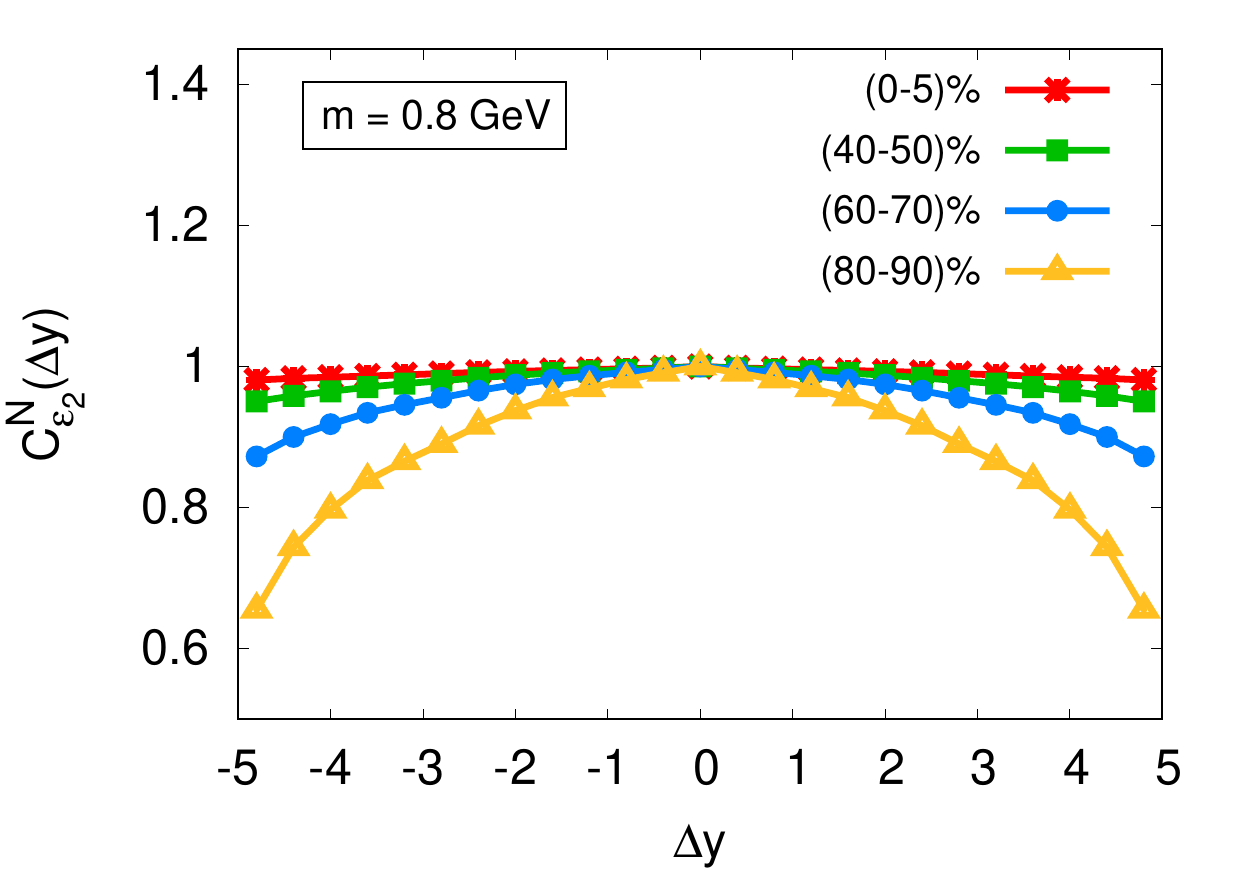}\hfill
\includegraphics[width=.49\textwidth]{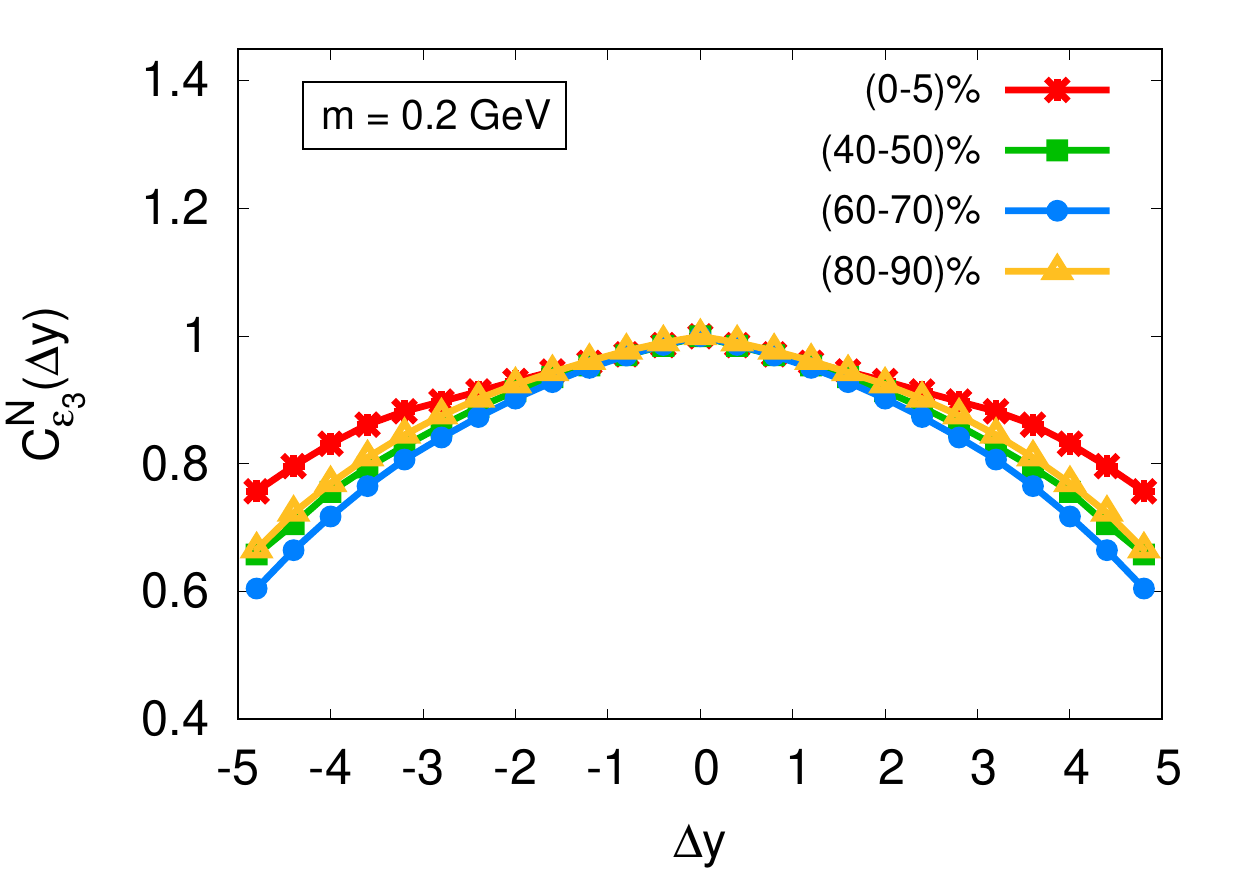}\hfill
\includegraphics[width=.49\textwidth]{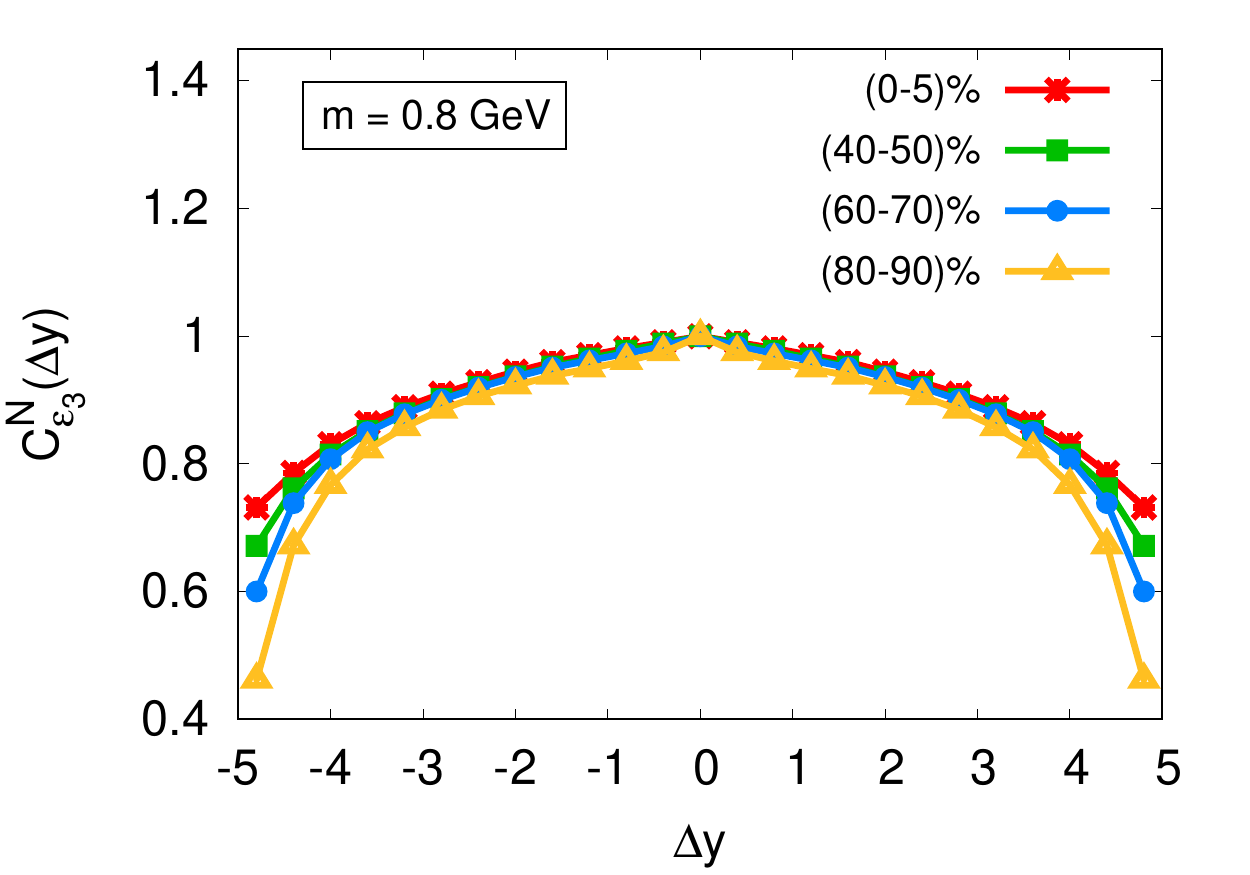}\hfill
\includegraphics[width=.49\textwidth]{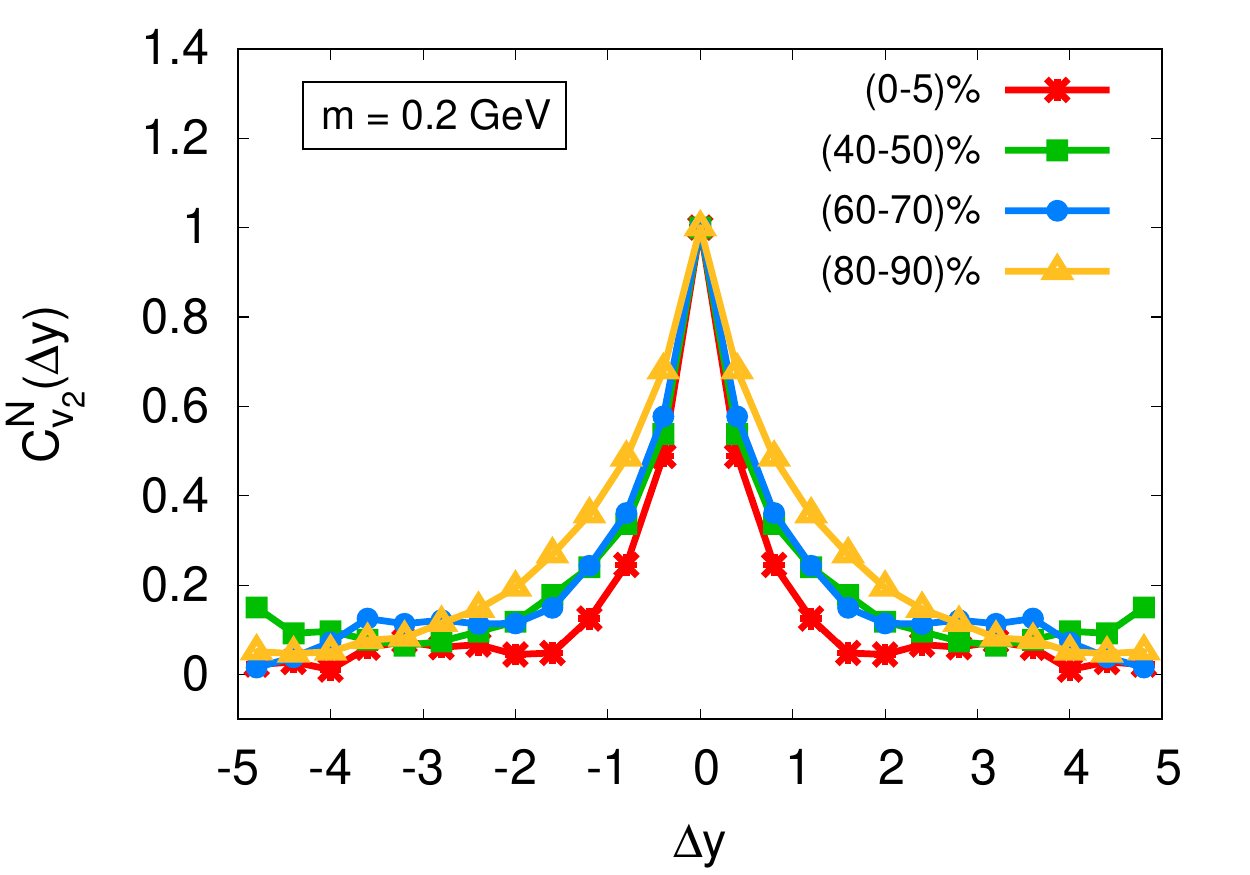}\hfill
\includegraphics[width=.49\textwidth]{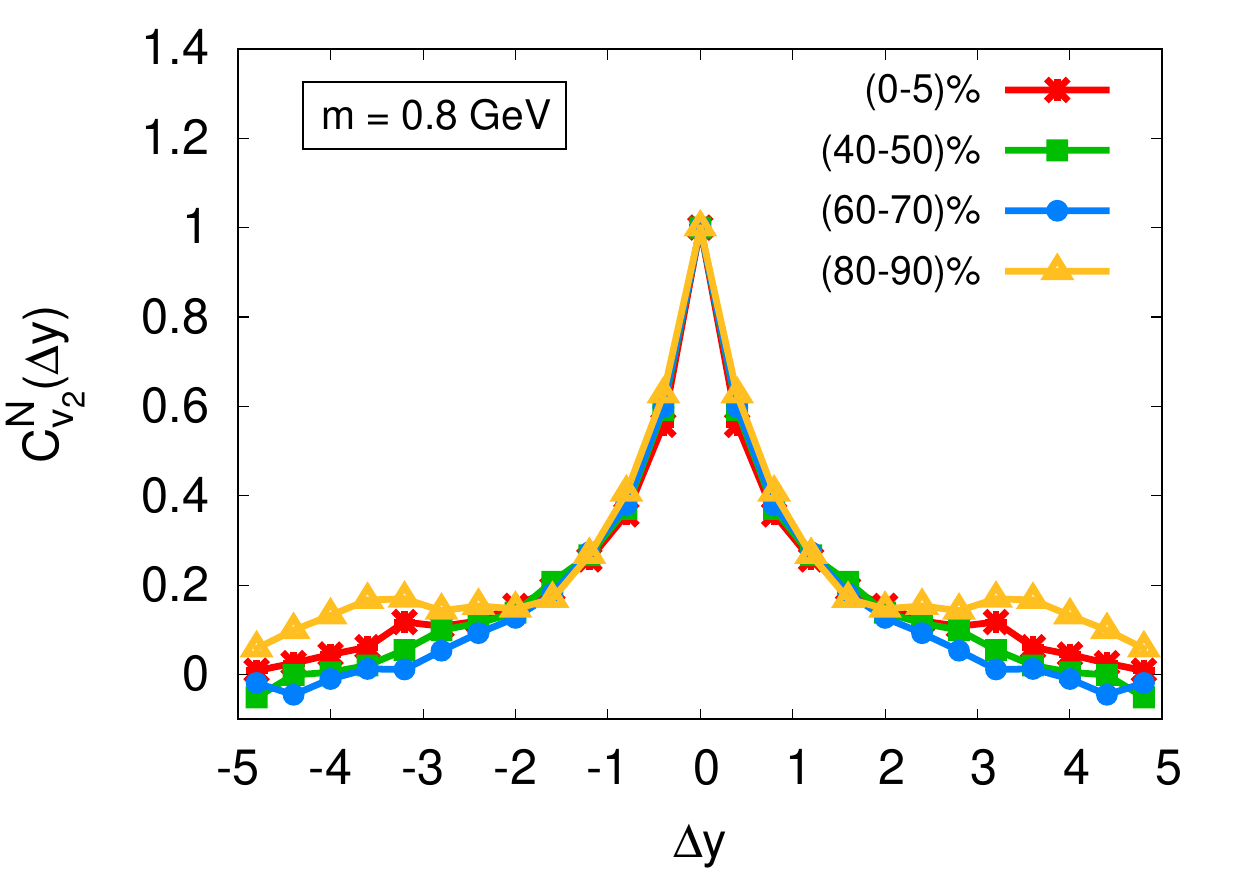}\hfill
\caption{Comparison of the normalized correlation function $C^{N}_{\mathcal{O}}(\Delta y)$ of the geometric eccentricity $\varepsilon_2$ (top), $\varepsilon_3$ (middle), and initial state gluon momentum anisotropy $v_2$ (bottom) for different centrality classes using $\alpha_s=0.15$, and $m=\tilde{m}=0.2\,{\rm GeV}$ (left) or $0.8\,{\rm GeV}$ (right).}
    \label{fig:Compare_CorrelationFunction}
\end{figure*}

\subsection{Decorrelation of event geometry and momentum anisotropy}
Now that we have established the overall rapidity dependence of the initial state geometry and momentum anisotropy, we will investigate the correlation across different rapidities, as quantified by the correlation functions $C_{\varepsilon_2}(y_1,y_2)$ and $C_{\varepsilon_p}(y_1,y_2)$ shown in Fig.\,\ref{fig:3d_unnormalised_plots}. The top panel shows results for (0-5)\% central collisions, the bottom for (60-70)\% central collisions. The overall magnitude of this correlator is related to the size of $\varepsilon_2$ and $\varepsilon_p$, as $C_{\varepsilon _2}(y,y)=\big(\varepsilon_{2}\{2\}(y)\big)^2$ and similarly for $C_{\varepsilon_p}$. We see that $C_{\varepsilon_2}(y_1,y_2)$ is maximal for both rapidities being most negative, where the $\varepsilon_2$ is largest. Fixing one rapidity, we can see the decorrelation when varying the other rapidity. $C_{\varepsilon_p}(y_1,y_2)$ is maximal for $y_1=y_2$ and does not vary much along this diagonal, as $\varepsilon_p$ (or $v_2^g$) is approximately constant as a function of rapidity. However, when comparing the results for $\varepsilon_{2}$ and $\varepsilon_{p}$, we can already see that the decorrelation of the initial state momentum anisotropy in $C_{\varepsilon_p}(y_1,y_2)$ is much faster than the decorrelation of the event geometry in $C_{\varepsilon_2}(y_1,y_2)$. One also observes that
$C_{\varepsilon_2}(y_1,y_2)$ is only weakly dependent on centrality, while $C_{\varepsilon_p}(y_1,y_2)$ shows some increase when going to more peripheral events, related to the increase of the initial state momentum anisotropy ($v_2^{g}$) for lower multiplicity.

In Fig.\,\ref{fig:Norm_CorrelationsEpandv2}, we show the normalized correlation functions $C^N_{\varepsilon_{n}}(\alpha_s \Delta y)$ for $n=2,3$, and $C^N_{v_2}(\alpha_s \Delta y)$ as functions of the scaled rapidity difference $\alpha_s \Delta y$. They are obtained from Eq.\,\eqref{eq:CN} as $C^{N}_{\mathcal{O}}(\Delta y)=\frac{1}{2y_{\rm max}-\Delta y}\int_{-y_{\rm max}+|\Delta y|/2}^{+y_{\rm max}-|\Delta y|/2} dY ~C^{N}_{\mathcal{O}}(Y+\Delta y/2,Y-\Delta y/2)$. For the geometric correlators we find that the decorrelation with rapidity is stronger for $n=3$ than $n=2$. This is consistent with experimental observations in heavy ion collisions \cite{CMS:2015xmx, ATLAS:2020sgl}. The decorrelation scales only approximately with $\alpha_s$ as we see small differences between the $\alpha_s=0.15$ and $\alpha_s=0.3$ case. As expected, smaller $m$ leads to a faster decorrelation. The centrality dependence shown in the three top panels demonstrates how the  rapidity decorrelation becomes faster towards more peripheral events.

\begin{figure*}
    \centering
    \includegraphics[width=0.99\linewidth]{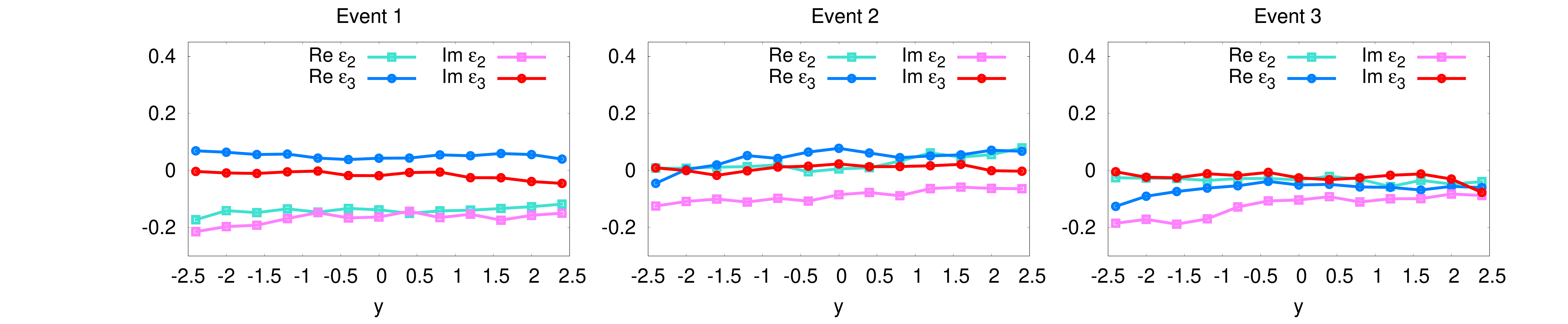}
    \includegraphics[width=0.99\linewidth]{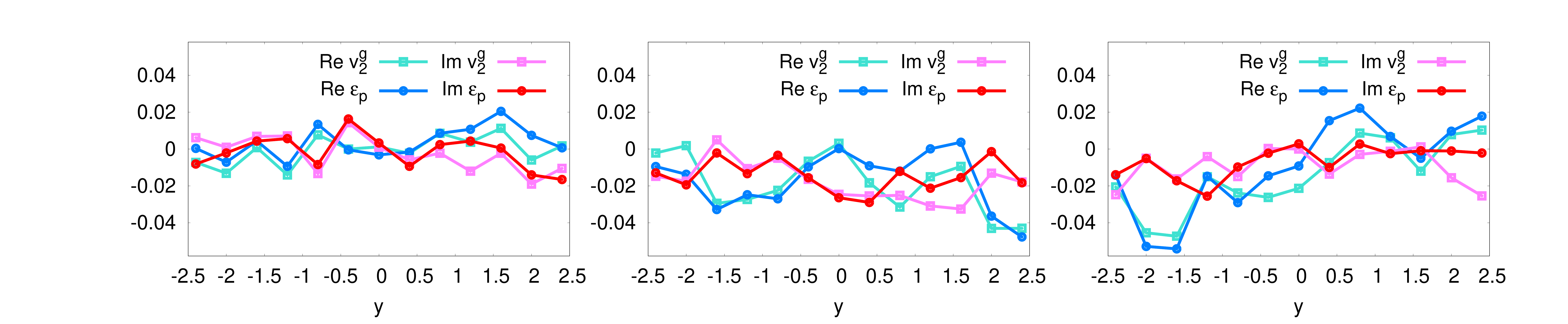}
    \caption{Rapidity dependence of the real and imaginary parts of the $2^{\rm nd}$ and $3^{\rm rd}$ order spatial eccentricities (top-panel) for three different events in the $(0-5)\%$ centrality class (top-panel). Similar result are given for the azimuthal anisotropy of initial state gluon $v_2^g$ and initial state momentum anisotropy $\epsilon_p$ in the bottom panel. Simulation parameters: $\alpha_s=0.15$ and $m=\tilde{m}=0.2$~\rm{GeV}.}
    \label{fig:MagnitudeComparisons}
\end{figure*}
\begin{figure}
    \centering
    \includegraphics[width=0.45\textwidth]{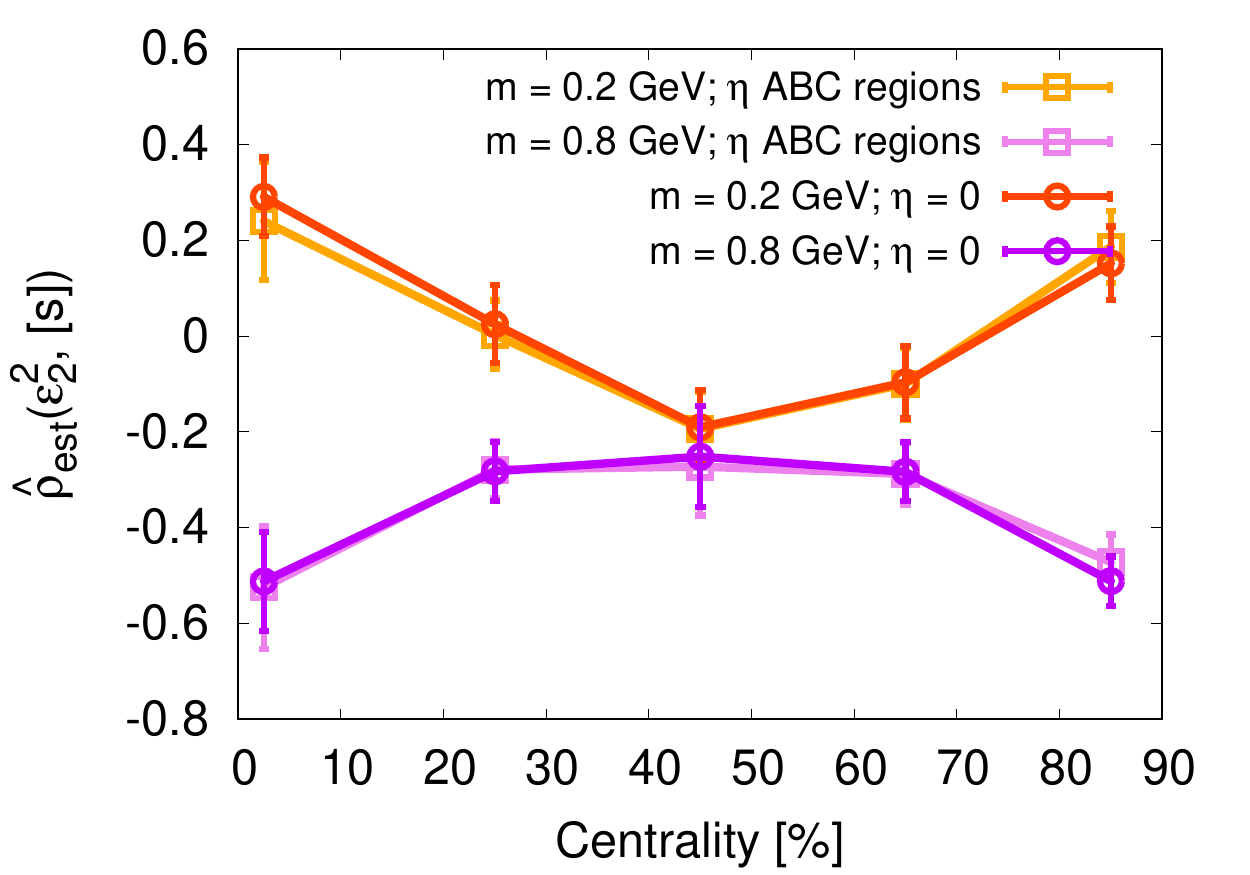}
    \includegraphics[width=0.45\textwidth]{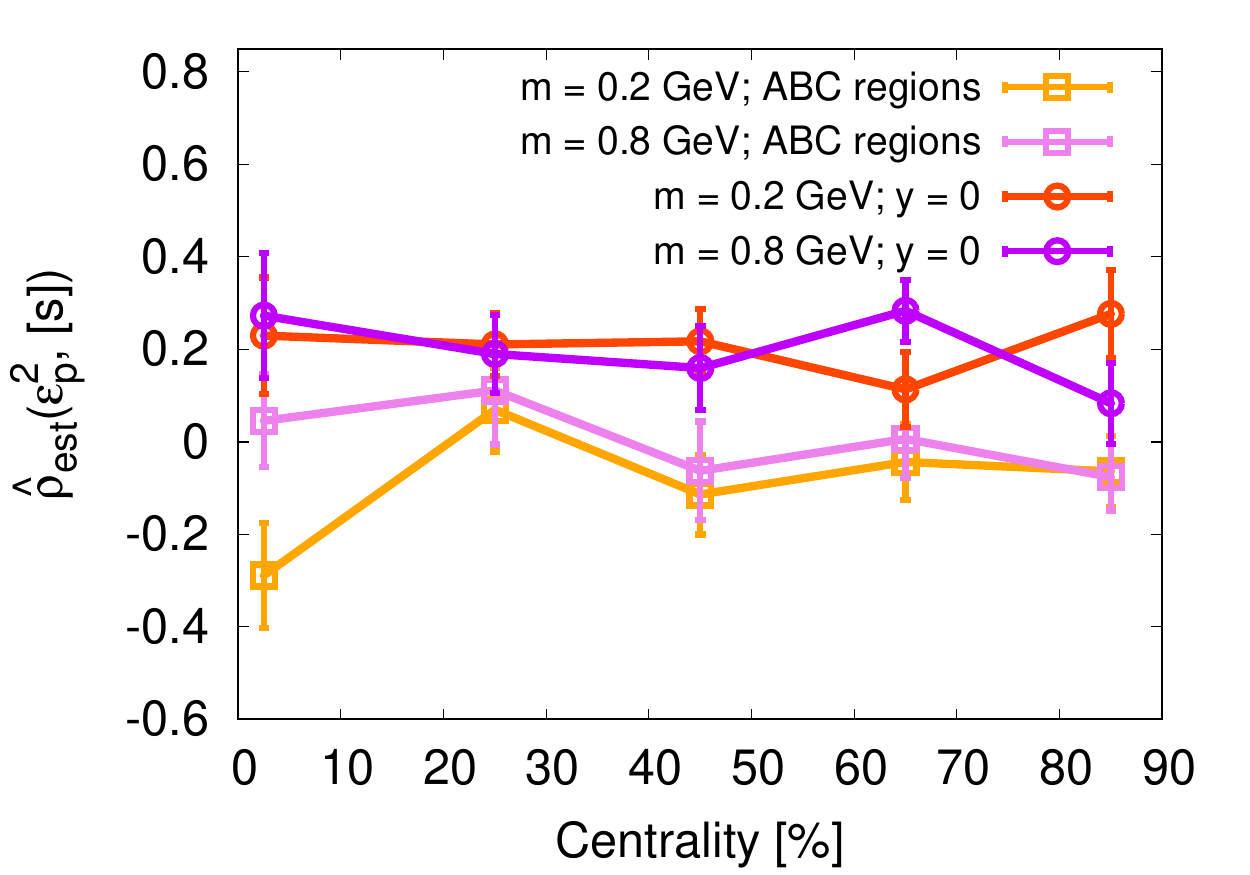}
    \caption{The estimators  based on the initial geometry $\hat{\rho}_{\rm est}(\varepsilon_2^2,[s])$ (top) and initial state momentum anisotropy $\hat{\rho}_{\rm est}(\epsilon_p^2,[s])$ (bottom) as a function of centrality for two different values of the infrared regulator $(m=\tilde{m})$. The correlation measure denoted as ABC is obtained for different rapidity regions: region A with $-2.4<y<-0.8$, central region B with $|y|<0.8$ and region C with $0.8<y<2.4$}
    \label{fig:rhoest}
\end{figure}
In the bottom panel of Fig.\,\ref{fig:Norm_CorrelationsEpandv2}, we present the correlator for the gluon momentum anisotropy $C^N_{v_2}(\alpha_s \Delta y)$, which shows a much more rapid decorrelation than the geometric quantities, but the opposite centrality dependence, with the most peripheral bin showing the broadest correlation in rapidity. The scaling with $\alpha_s$ works more accurately in this case, and smaller $m$ leads only to a slightly faster decorrelation. The quick decorrelation in the initial momentum anisotropy with JIMWLK evolution, compared to the geometric case, can be expected based on the fact that every gluon emission in the evolution leads to a color decorrelation, quickly scrambling information of color domains at the initial rapidity. Conversely, the larger scale geometric structures are much more robust to the evolution, as they are not sensitive to the color structure. 

The centrality dependence of these results is highlighted again in Fig.\,\ref{fig:Compare_CorrelationFunction}. For the geometric quantities, the width of the correlation function decreases with increasing centrality, while it increases for the initial momentum anisotropy. This can be understood as follows: The geometry of the more dilute peripheral events can be changed more easily by additional gluon emissions in the evolution (dominantly via the modification of the proton's shape). Denser protons are more robust to changes of the geometry by the same amount of emissions. Regarding the momentum anisotropy, it is maximal in the most peripheral bins. Consequently it takes more evolution to destroy it. 

We find that in the most peripheral events, where the initial momentum anisotropy can potentially dominate the observed charged hadron anisotropy \cite{Giacalone:2020byk}, the correlation drops by 50\% within approximately one unit of rapidity (for the preferred JIMWLK evolution speed with $\alpha_s=0.15$). 

In order to better understand the decorrelation that is observed in the event averaged quantities, we study both real and imaginary parts of the spatial and the momentum anisotropies for three individual events in Fig.\,~\ref{fig:MagnitudeComparisons}. In the top panel, we show real and imaginary parts of $\varepsilon_n$ as functions of rapidity. As expected from the slow decorrelation observed above, the plotted quantities vary smoothly and weakly with rapidity. 


In the bottom panel, we show real and imaginary parts of $\varepsilon_p$ and $v_2^g$ as functions of rapidity. For all events (columns), we observe rather quick variations of the preferred direction of anisotropy with rapidity even though the magnitude of anisotropy given by the absolute value does not change too rapidly, even in a single event. These rapid variations explain the quick decrease of the correlator with rapidity, the main driver being fluctuations in the angle. We note that even in a single event, $\varepsilon_p$ resembles $v_2^g$ closely.

\subsection{Estimators for the correlation between mean transverse momentum and elliptic anisotropy}
Finally, we consider estimators for the correlation of mean transverse momentum and the elliptic anisotropy, which has been suggested as an observable to distinguish between geometry and initial momentum anisotropy as the origin of the observed anisotropy \cite{Giacalone:2020byk}. The relevant correlator studied experimentally is defined as
\begin{equation}\label{eq:v2PT}
    \hat{\rho}(v_2^2,[p_T])=\frac{\langle \deltan v_2^2 \,\deltan [p_T]\rangle}{\sqrt{\langle(\deltan v_2^2)^2\rangle\langle(\deltan [p_T])^2\rangle}}\,,
\end{equation}
where $v_2$ is the measured elliptic anisotropy and $[p_T]$ is the mean transverse momentum in a given event, and the event-by-event deviation for observable $O$ at fixed multiplicity is defined as~\cite{Olszewski:2017vyg} 
\begin{align}\label{eq:fixmult}
\deltan O &\equiv \delta O  - \frac{\langle \delta O \delta N \rangle }{ \sigma_N^2 } \delta N \,,
\end{align}
where $\delta O = O - \langle O \rangle$, $N$ is the multiplicity and $\sigma_N$ the variance of $N$ in a given centrality bin.

Because we are considering initial state quantities in this work, we compute estimators for $\hat{\rho}$ by replacing $v_2$ with $\varepsilon_2$ (or $\varepsilon_p$) and $[p_T]$
by the average entropy density $[s]=[e^{3/4}]$ where $e$ is the energy density, approximated as $T^{\tau\tau}$. The average [f] is computed as
\begin{align}
    [f]=\frac{\int d^2 \mathbf x_\perp e(\mathbf x_\perp) f(\mathbf x_\perp)}{\int d^2 \mathbf x_\perp e(\mathbf x_\perp)}\,.
\end{align}
The estimator using the ellipticity $\varepsilon_2$, $\hat{\rho}_{\rm est}(\varepsilon_2^2, [s])$, is shown as a function of centrality in top panel of Fig.\,\ref{fig:rhoest} for two different ways of choosing the rapidity bins where the different components of $\hat{\rho}$ are computed. One takes all quantities at rapidity zero ($y=0$), the other uses three different rapidity bins (ABC regions) for the different components of $\hat{\rho}$, following the prescription used by the ATLAS Collaboration \cite{ATLAS:2019pvn}. We find that for the larger $m=\tilde{m}$ the geometry estimator is always negative, as can be expected from geometric considerations \cite{Giacalone:2020byk}. Since the infrared regulators $m,\tilde{m}$ have a strong effect on the event geometry, this also affects the $\hat{\rho}$ estimator. When considering the smaller $m=\tilde{m}$, we even find positive values for most central and most peripheral events. 
While this is at odds with calculations of this estimator in the IP-Glasma model without JIMWLK evolution \cite{Giacalone:2020byk}, it is conceivable that the JIMWLK evolution, which has greater effects on the geometry for smaller $m$, causes this difference in the most central and most peripheral events. The appearance of positive values for smaller $m$, which leads to larger systems, is in line with findings in a previous work, where the geometric $\hat{\rho}$ correlator turned positive when increasing the system size \cite{Bozek:2020drh}.

Most importantly, for the geometric estimator, we do not see a large dependence on the choice of rapidity bins, which is related to the weak decorrelation of the geometry observed. Hence our results justify the use of the boost-invariant approximation to compute the correlator in geometry driven models \cite{Bozek:2020drh,Giacalone:2020byk}.

When replacing $\epsilon_2$ by the initial state momentum anisotropy $\varepsilon_p$, we observe a positive correlation in the $\hat{\rho}_{\rm est}(\varepsilon_p^2, [s])$ estimator when considering both quantities at mid-rapidity, which again is in line with the findings in \cite{Giacalone:2020byk}. However, due to the rapid decorrelation of $\varepsilon_p$ in rapidity, this signal does not appear to survive the rapidity gap, as the correlator $\hat{\rho}_{\rm est}(\varepsilon_p^2, [s])$ is consistent with zero when considering the selection in different rapidity intervals (ABC).




\section{Conclusions \& Outlook}
\label{sec:five}
We have presented results for rapidity dependent quantities in p+Pb collisions, computed within the color glass condensate framework, which involves the calculation of classical gluon fields in the proton and lead nucleus in IP-Glasma, their leading quantum corrections via JIMWLK evolution of the corresponding Wilson lines, and computation of production and time evolution of the gluon fields generated by the collision at different rapidities. 

We showed results for the rapidity dependence of gluon production $dN_g/dy$ and the transverse energy $dE_\perp/dy$ for different centralities, and analyzed the role of the saturation scale $Q_s$ in the proton and nucleus as well as that of the overlap area for gluon production as a function of centrality.

We studied the transverse geometry, quantified by the eccentricities $\varepsilon_2$ and $\varepsilon_3$ as a function of centrality and rapidity, finding rather mild dependencies. The initial momentum anisotropy, quantified by either the anisotropy of the energy momentum tensor $\varepsilon_p$ or that of the gluon distribution $v_2^g$, showed a weak rapidity dependence for all centralities, and increased when increasing centrality from central to peripheral events.

We computed the unequal rapidity correlations of both the geometric and initial momentum anisotropy vectors and observed very different behavior between the two. The geometry decorrelates much more slowly as a function of the rapidity difference, compared to the initial momentum anisotropy. For the latter, the correlation is widest in the most peripheral centrality bin, but still drops to about half its maximal value for a rapidity difference $\Delta y = 1$. This result implies that when using large rapidity gaps to measure flow harmonics or the $\hat{\rho}$ correlator experimentally, the initial momentum anisotropy may play a smaller role than previously assumed. In order to access this contribution, smaller rapidity gaps need to be employed, which will make the separation from other non-flow effects difficult.

Regarding the geometry, we find a faster decorrelation for $\varepsilon_3$ than for $\varepsilon_2$, which is in line with observations in heavy ion collisions \cite{CMS:2015xmx, ATLAS:2020sgl}. The fast decorrelation of $\varepsilon_3$ can play an important role for the difference between different $v_3$ measurements in small asymmetric systems at RHIC \cite{PHENIX:2018lia,Lacey:2020ime,PHENIX:2021bxz,Nagle:2021jgy}.
 
For all these observables, we studied in detail the dependence on the infrared regulators employed in the calculation, as well as that on the strong coupling constant $\alpha_s$, which controls the evolution speed of the JIMWLK equations. We assumed a fixed coupling constant. Running coupling effects have been included in leading logarithmic JIMWLK evolution calculations \cite{Lappi:2011ju,Dumitru:2011vk,Lappi:2012vw}. Generally, their inclusion should lead to a faster rapidity evolution of the long range geometric structures, and a slower evolution of short range momentum correlations in the transverse plane.

Finally, we computed an initial state estimator for the correlation between the elliptic anisotropy and the average transverse momentum at fixed multiplicity. For the larger infrared regulator, this quantity is always negative, in line with previous findings using the IP-Glasma model (without JIMWLK evolution) \cite{Giacalone:2020byk}. For the smaller regulator, positive values are found in the most central and most peripheral bins, which could be attributed to effects from the JIMWLK evolution on the details of the geometry at large length scales.

We conclude that even at collision energies available at the LHC, for small systems the rapidity dependence is not to be neglected. When rapidity gaps are employed experimentally, the theoretical description will not get away with the assumption of boost invariance in most cases. Rapidity dependent calculations are required and the experimental procedures should be matched as closely as possible. Already the centrality selection is affected by the rapidity dependence, and we recommend for the purpose of an easier comparison to theoretical calculations to perform centrality classifications using measurements around midrapidity. 

We note that from a theoretical point of view, our description calculates  mean-field type correlations and propagates them using 
JIMWLK. The sub-leading correction to the limit of a large number of small $x$ constituents includes the absence of particles that already scattered and conditional constraints on the small-x evolution. We are not aware of how to include these effects in dense-dense calculations, and their potential effects on the observables studied are unclear. However, by considering the dilute limit of the projectile, the authors of \cite{Iancu:2013uva,Iancu:2014qnj} developed a framework to study multi–particle production with rapidity correlations, which may provide a way to assess these effects in future studies.

In the future, it will also be interesting to couple the computed rapidity dependent initial state to hydrodynamics, possibly via an intermediate kinetic theory stage \cite{Kurkela:2018vqr,Kurkela:2018wud}. Also, a construction of a fully 3 dimensional Wilson line configuration followed by 3+1D Yang-Mills evolution, as explored in \cite{McDonald:2020oyf,Schlichting:2020wrv,Ipp:2021lwz}, will be desirable.


\section*{Acknowledgments} S.S. and P.S. are supported by the Deutsche Forschungsgemeinschaft (DFG, German Research Foundation) through the CRC-TR 211 ’Strong interaction matter under extreme conditions’– project
number 315477589 – TRR 211. P.S. is also supported by the Academy of Finland, project 321840 and under the European Union’s Horizon 2020 research and innovation programme by the STRONG-2020 project (grant
agreement No 824093). B.P.S. is supported by the U.S. Department of Energy, Office of Science, Office of Nuclear Physics, under Contract Number DE-SC0012704.
This research used resources of the National Energy Research Scientific Computing Center, a DOE Office of Science User Facility supported by the Office of Science of the U.S. Department of Energy
under Contract No. DE-AC02-05CH11231. 

\appendix
\section{Dipole amplitude \& Saturation scale}\label{app:A}
\begin{figure*}[t!]
\centering
\includegraphics[width=.33\linewidth]{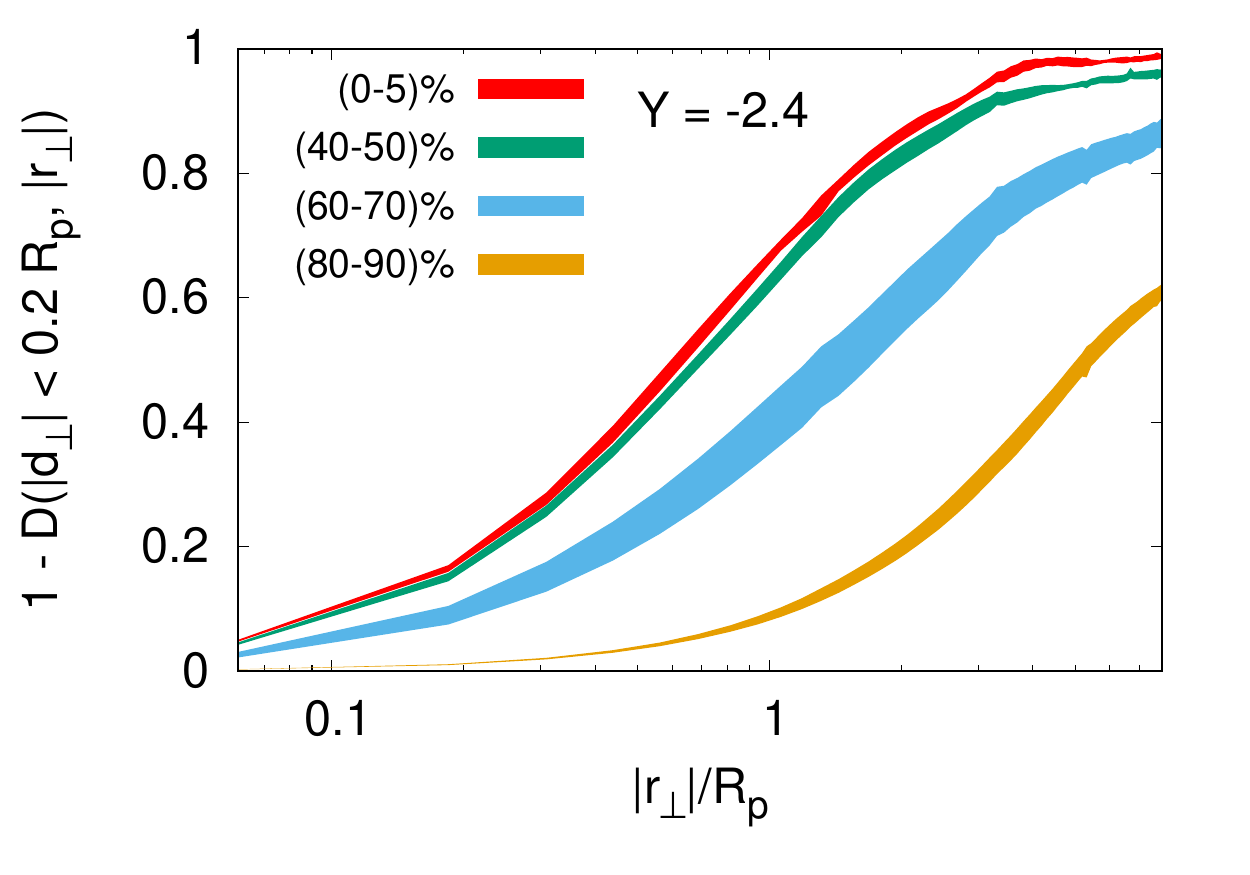}\hfill
\includegraphics[width=.33\linewidth]{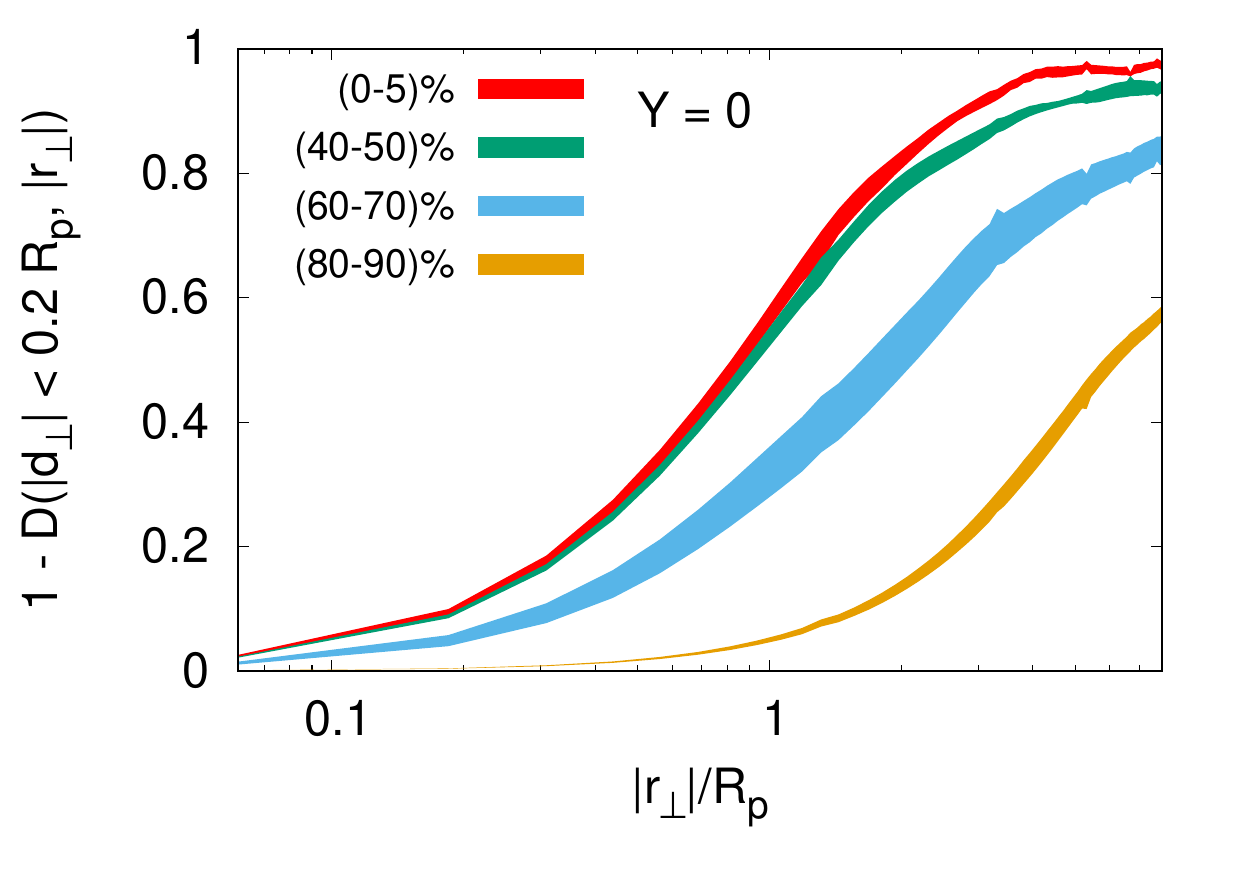}\hfill
\includegraphics[width=.33\linewidth]{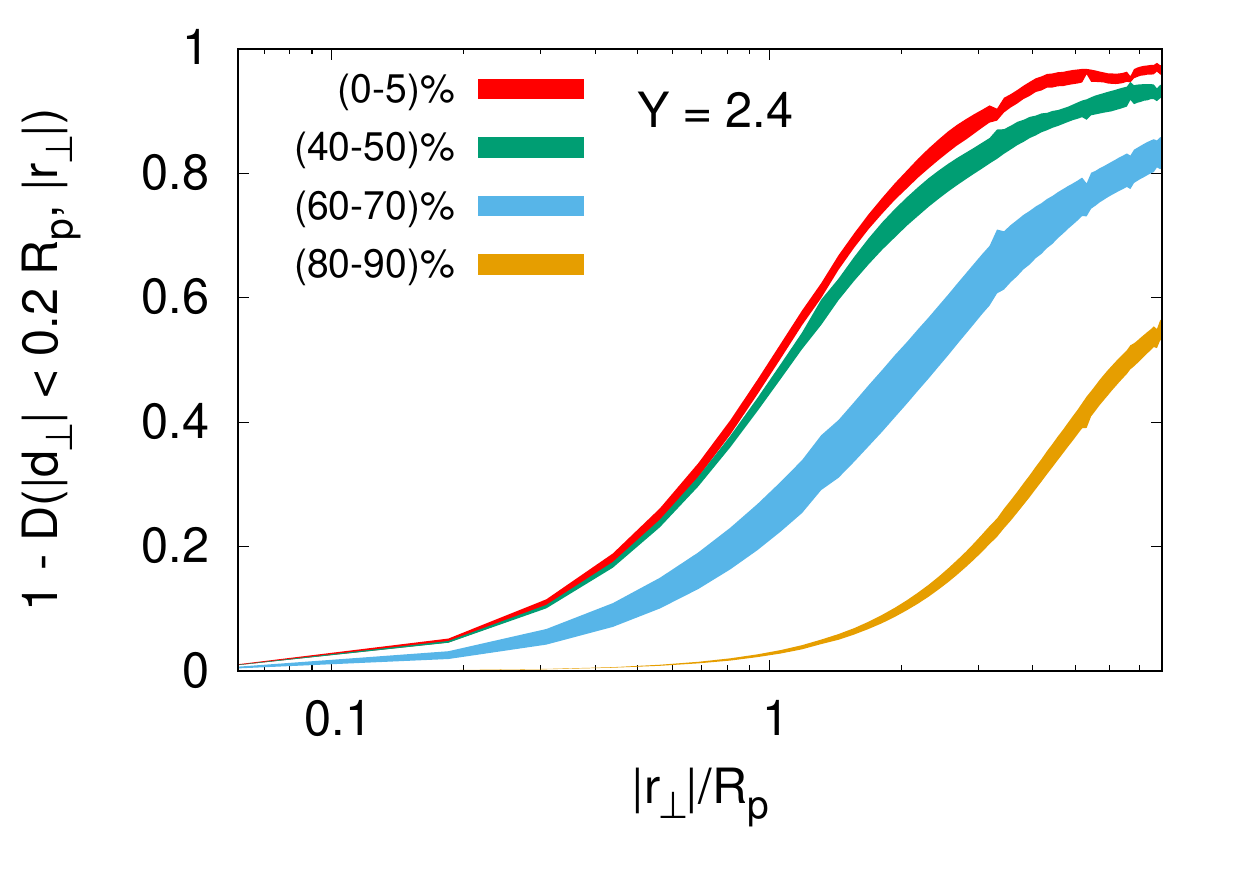}\hfill
\includegraphics[width=.33\linewidth]{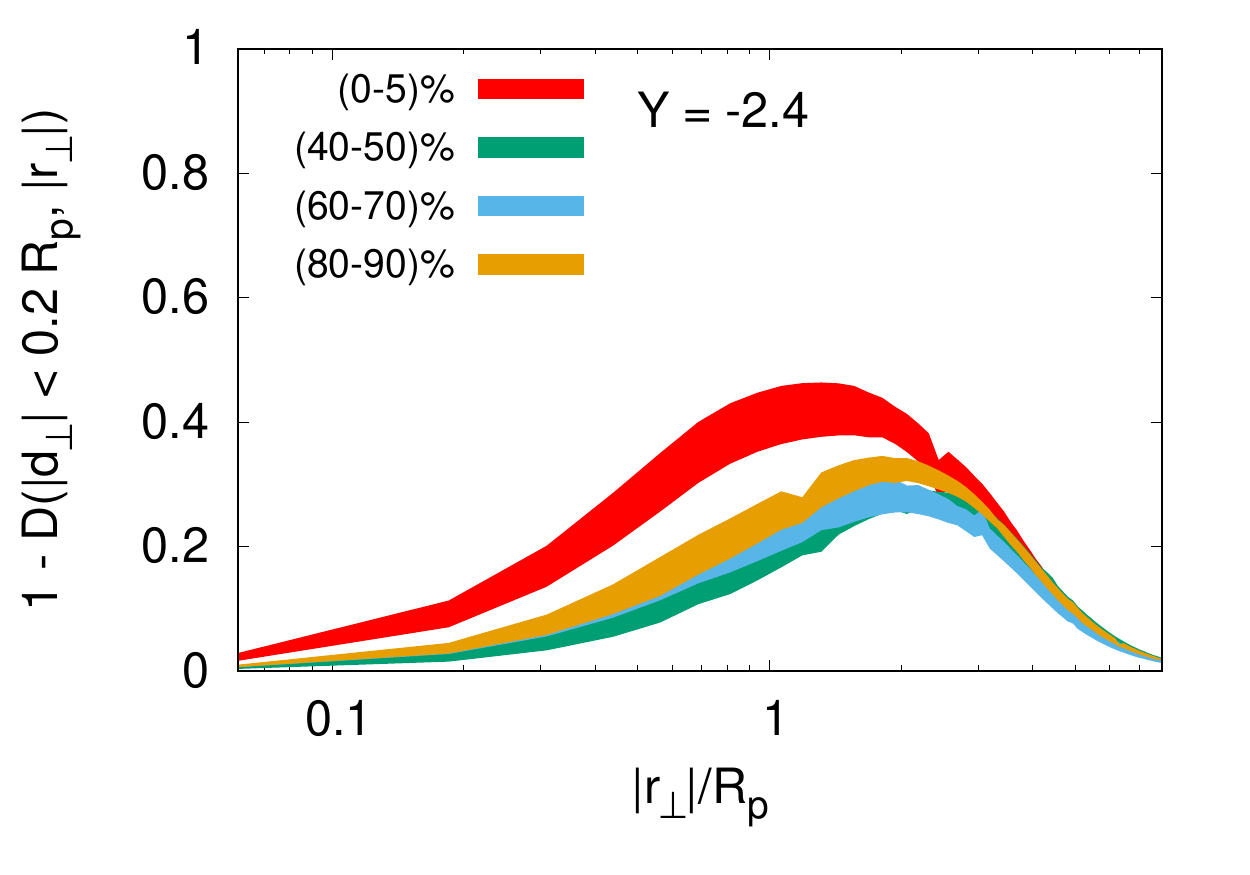}\hfill
\includegraphics[width=.33\linewidth]{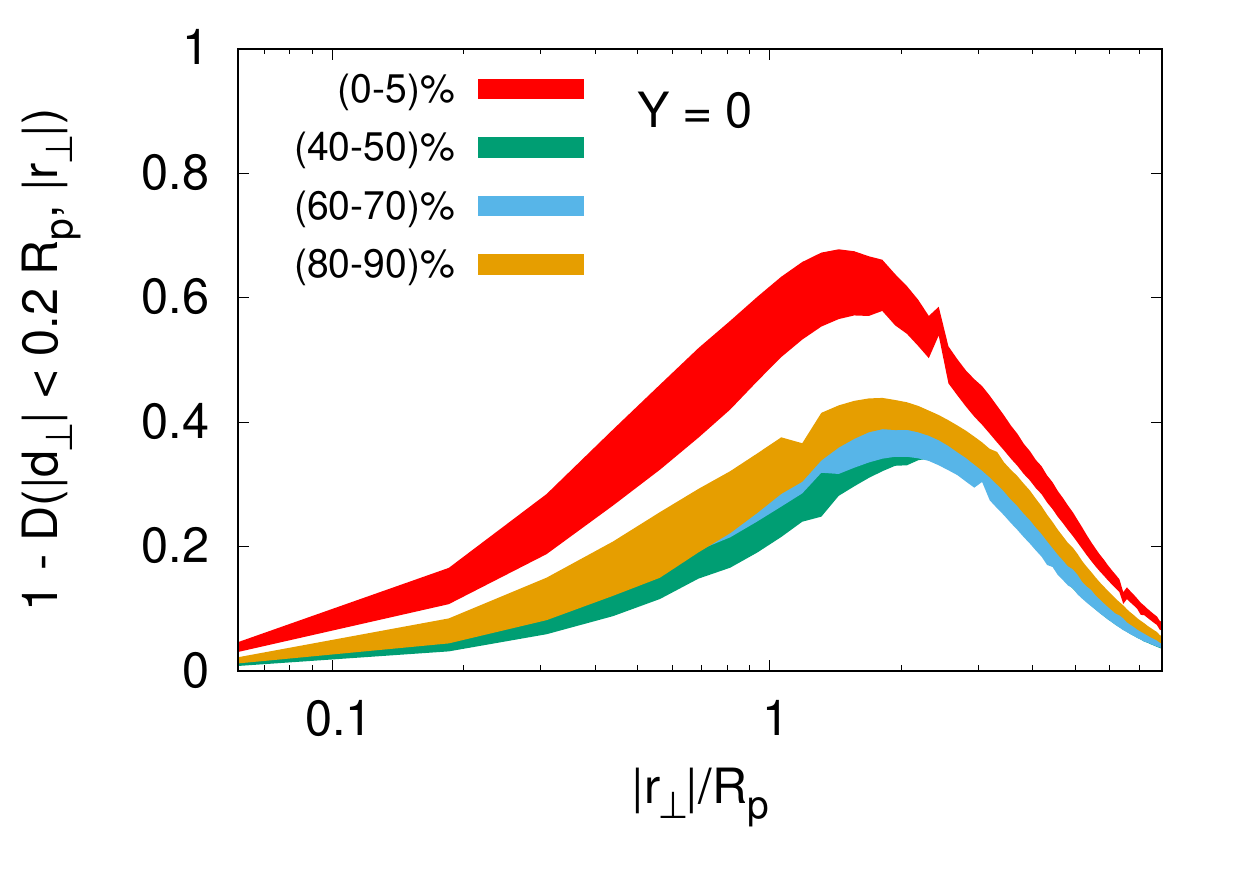}\hfill
\includegraphics[width=.33\linewidth]{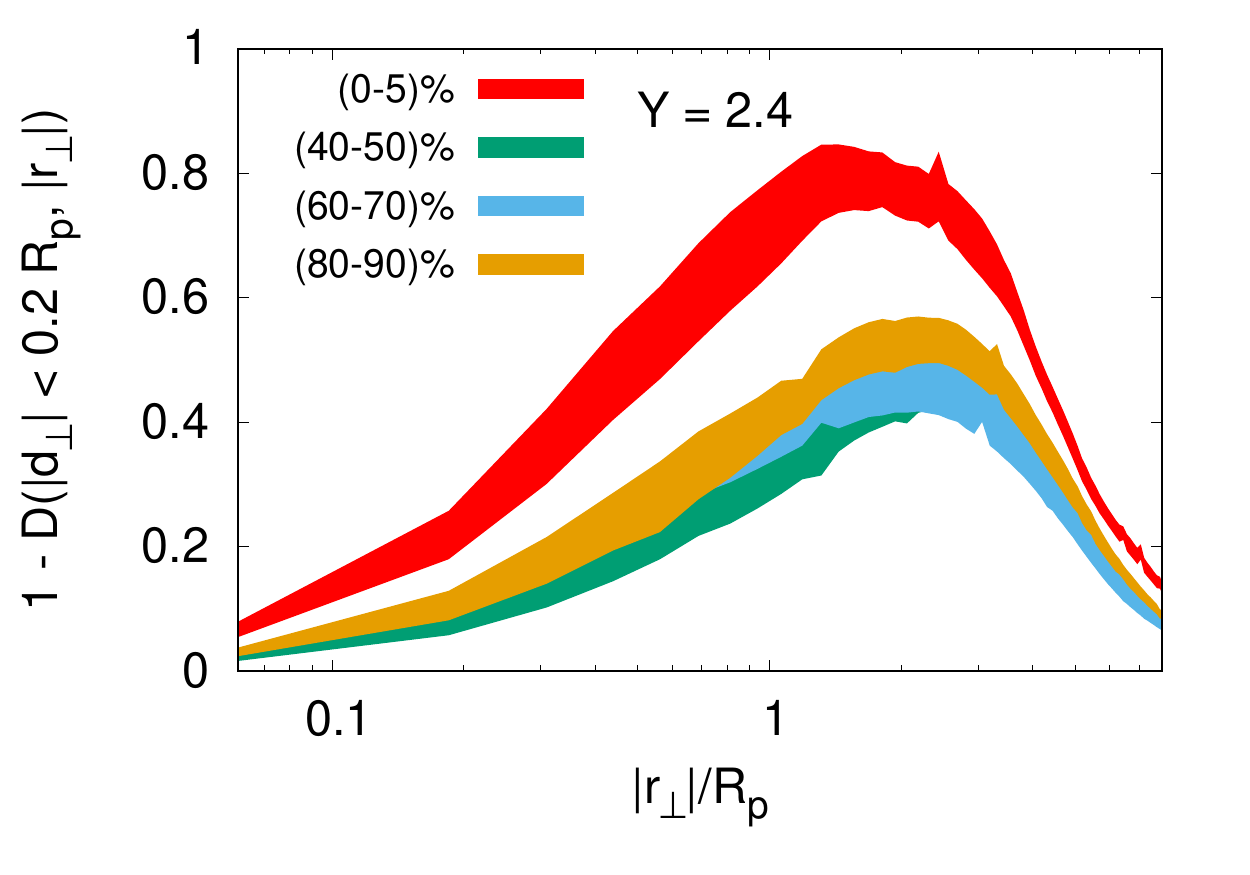}\hfill

\caption{Dipole scattering amplitudes $1-D(\rT,|\dT|<0.2R_{p})$ of the lead nucleus (top) and proton (bottom) at three different rapidites $Y=-2.4,0,+2.4$ as a function of dipole size $|\rT|$ in units of the proton radius $R_{p}$.}
    \label{fig:DA_for_p_Pb}
\end{figure*}
In order to characterise the gluon distribution of the proton and the Pb nucleus, we use the dipole scattering amplitude, Eq.~\eqref{eq:Dipole_Amplitude}, where $\dT=(\xT+\yT)/2$ is the impact parameter and $\rT=\xT-\yT$ is the size of a color singlet dipole with color charges at positions $\xT$ and $\yT$. We show the dipole scattering amplitude $1-D(\rT,|\dT|<0.2R_{p})$ for the Pb nucleus (top) and proton (bottom) as a function of dipole size $r_\perp=|\rT|$ for a fixed range of impact parameter $|\dT|<0.2R_{p}$ measured at three different rapidities $y=-2.4,0,+2,4$ in various centrality classes in Fig.~\ref{fig:DA_for_p_Pb}. This choice of $\dT$ is based on \cite{Schlichting:2014ipa} where the dominant support of $D(|\dT|,|\rT|)$ dwells in the region of small impact parameter. 


Due to color transparency the dipole scattering amplitude $1-D$ vanishes  at $r_\perp=0$ and then gradually rises and reaches a maximum at $r_\perp/R_p \sim 1$. For the Pb nucleus we observe that the scattering amplitude saturates for $(0-5)\%$ and $(40-50)\%$ centrality classes, while the other two centrality classes are dilute even for $Y=-2.4$ which corresponds to the smallest $x$. For proton (bottom), the dipole amplitude is much below the saturation level, even after full rapidity evolution $(Y=2.4)$, and starts to fall when the separation between the dipole exceeds the size of the proton $r_\perp \gg R_p$ because the dipole no longer hits the target, as previously observed in \cite{Schlichting:2014ipa}. We also note that for protons the shape of the dipole amplitude as a function of $r_\perp$ does not change much with centrality, in particular for the three more peripheral bins.

\begin{figure*}
    \centering
    \includegraphics[width=0.45\textwidth]{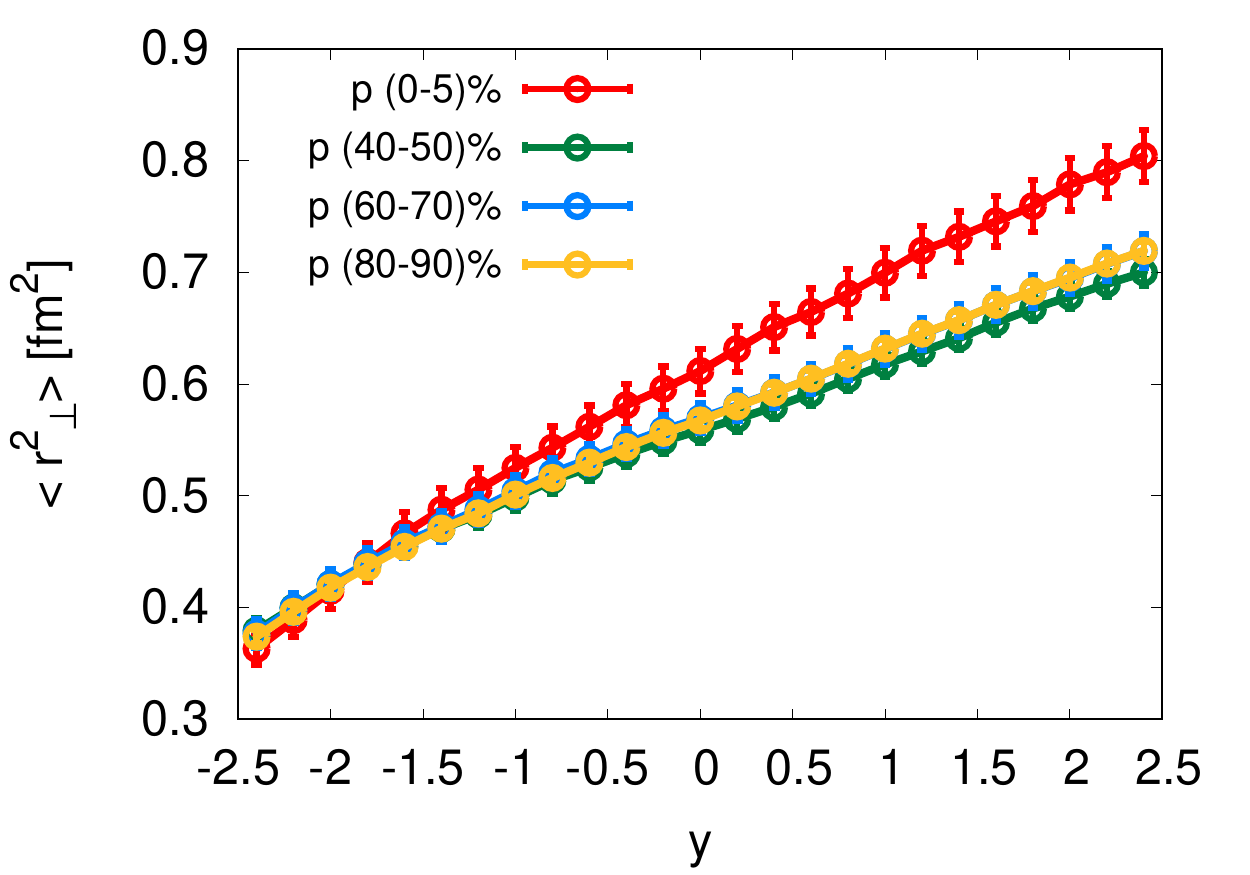}
    \includegraphics[width=0.45\textwidth]{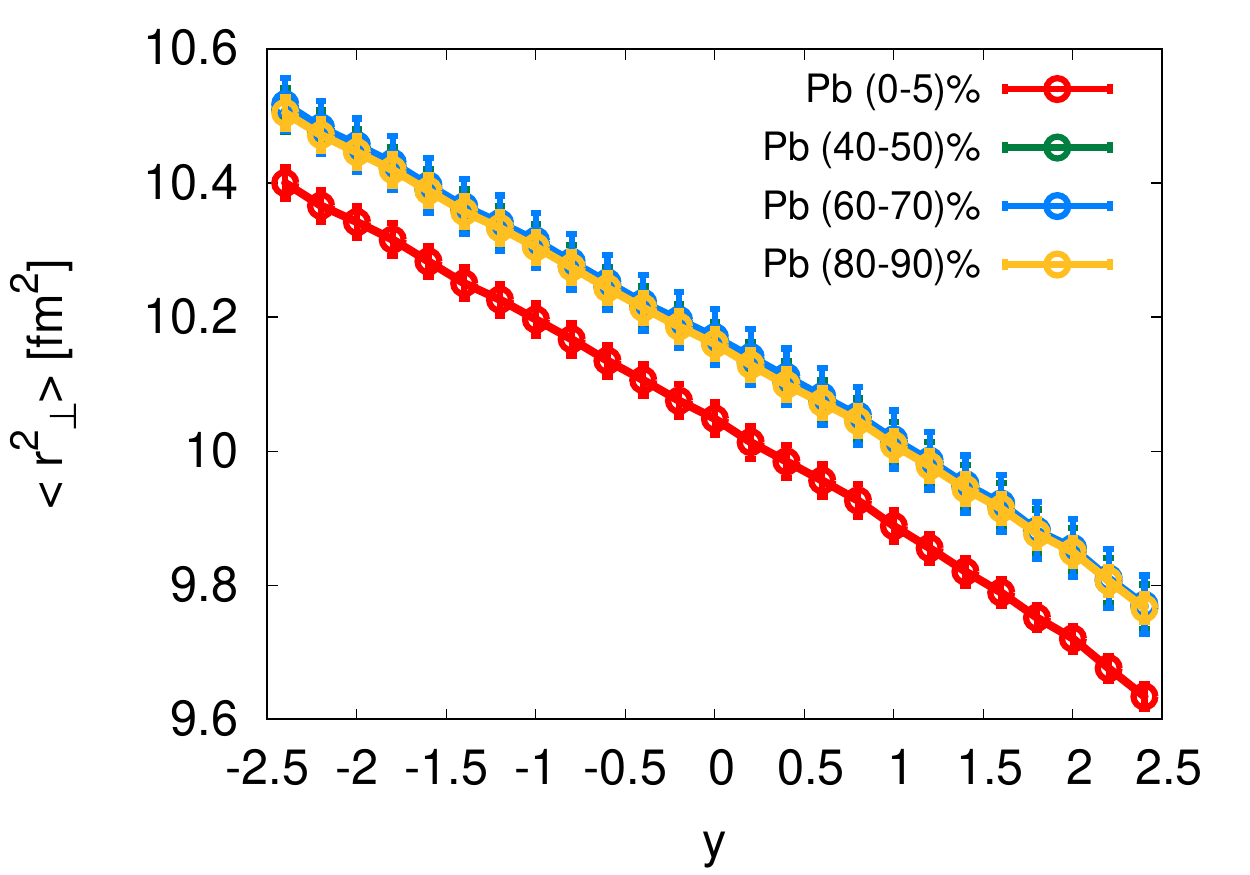}
    \includegraphics[width=0.45\textwidth]{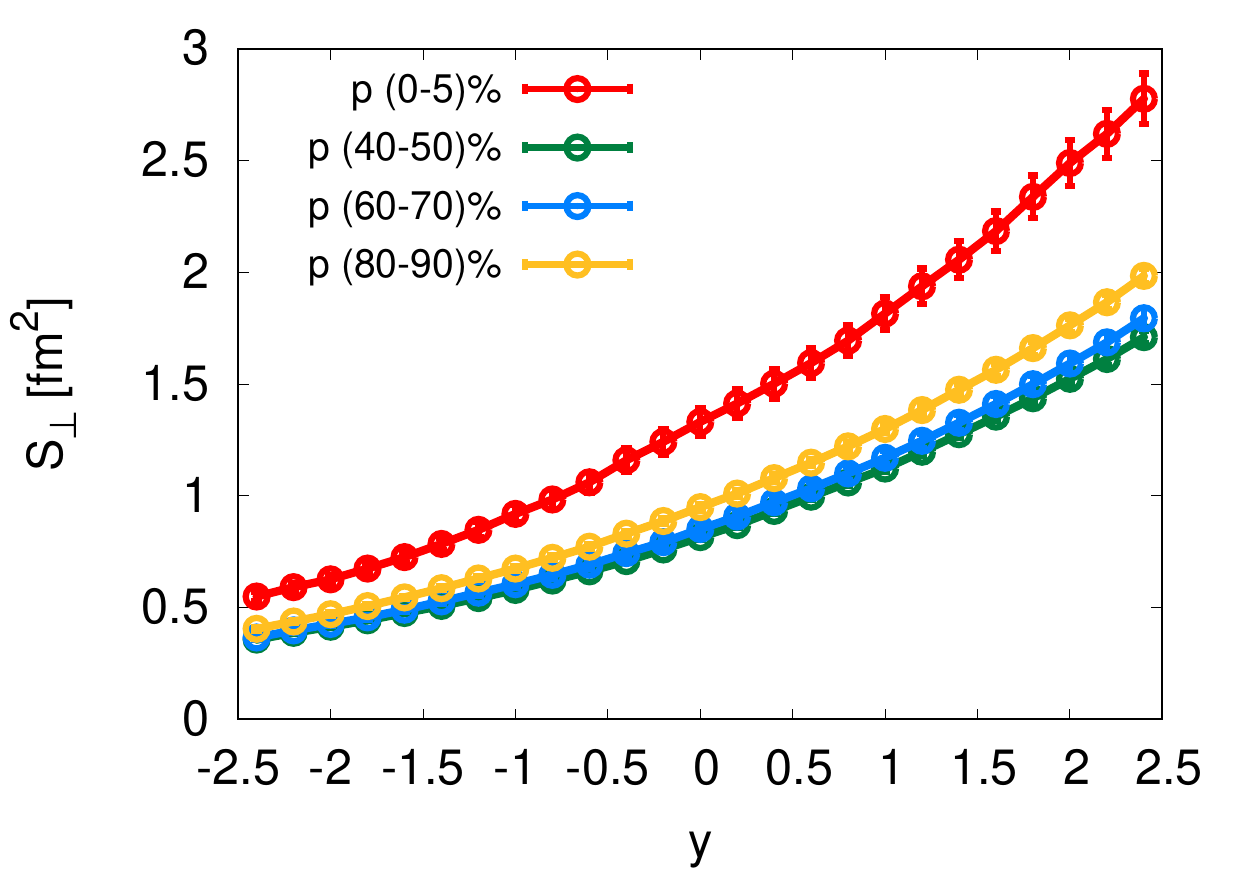}
    \includegraphics[width=0.45\textwidth]{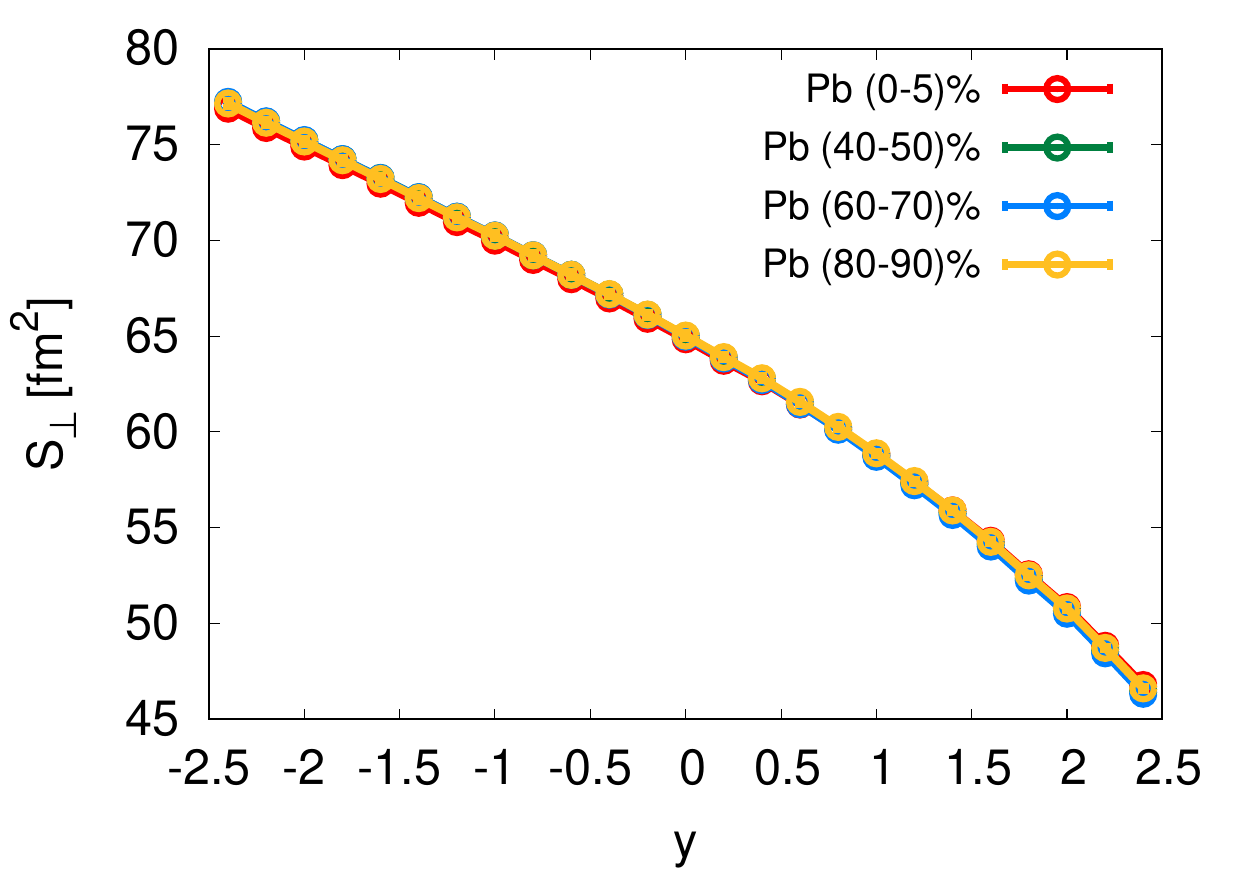}
    \caption{Mean radius squared (top) and transverse area (bottom) for proton (left) and Pb nucleus (right) as a function of rapidity for different centrality classes. Simulation parameter: $\alpha_s=0.15$ and $m=\tilde{m}=0.2~\rm GeV$.
    }
    \label{fig:size_radius}
\end{figure*}

In order to investigate the system size, we use the Weizs\"acker Williams fields $E^-_{\mu}$, which are represented by light-like Wilson lines $V_{p/Pb}$ on a two dimensional lattice with transverse coordinates as:
\begin{align}
    E^-_{j,\mathbf x}&{}=\frac{i}{4}\Big[V^\dagger_{\mathbf x +j}V_{\mathbf x }+V^\dagger_{\mathbf x }V_{\mathbf x -j}-V^\dagger_{\mathbf x }V_{\mathbf x +j}-V^\dagger_{\mathbf x -j}V_{\mathbf x }\Big]-\notag\\
&{}\frac{i}{4N_c}\rm{Tr}\Big[V^\dagger_{\mathbf x +j}V_{\mathbf x }+V^\dagger_{\mathbf x } V_{\mathbf x -j}-V^\dagger_{\mathbf x }V_{\mathbf x +j}-V^\dagger_{\mathbf x -j}V_{\mathbf x }\Big]
\end{align}
The mean radius squared is then determined from $E^2(\mathbf x_\perp)=\rm{Tr} [E^-_x(\mathbf x_\perp)E^-_x(\mathbf x_\perp)+E^-_y(\mathbf x_\perp)E^-_y(\mathbf x_\perp)]$ as
\begin{align}
    \langle \mathbf r_\perp^2(y) \rangle =\frac{\int d^2 \mathbf r_\perp \mathbf r_\perp^2 E^2(\mathbf r_\perp,y)}{\int d^2 \mathbf r_\perp E^2(\mathbf r_\perp,y)}\,,
\end{align}
Similarly, the transverse area $S_\perp$ is obtained as
\begin{align}
    S_\perp=\int \Theta(E^2(\mathbf x_\perp)-\Lambda^2) d^2\mathbf x_\perp
\end{align}
where the Heaviside function implies that only regions with field strength $E^2(\mathbf x_\perp)$ larger than the cut-off scale $\Lambda^2=0.02~ \rm{GeV/fm}^3$ contribute to the integral. The results for $ \langle r_\perp^2 \rangle$ and $S_\perp$ as a function of rapidity for different centrality classes for proton (left) and Pb nucleus (right) are summarised in Fig.~\ref{fig:size_radius}. We find that similar to Fig.~\ref{fig:QsST}, the mean radius squared grows almost linearly in the direction of increasing (decreasing) rapidity for the proton (Pb nucleus), and is considerably independent of centrality, except for the most central bin being different from the other three.



We observe that for the left moving proton, the transverse area $S_{\bot}$ grows quadratically with decreasing x, in agreement with the observation made in \cite{Schlichting:2014ipa}, where the similarly defined proton radius grows linearly with the rapidity evolution. Since the transverse size $S_\bot$ of the Pb nucleus is significantly larger to begin with, the effect of Gribov diffusion on the Pb nucleus is smaller, leading to a slower increase of the area with decreasing $x$ (decreasing rapidity). With regards to the centrality dependence, one finds that due to the larger overall field strength the transverse area $S_{\bot}$ of protons is somewhat larger for the most central events, while for the Pb nucleus no significant centrality dependence is observed.

\bibliography{bib}
\end{document}